\documentclass[aps,prd,reprint,secnumarabic,superscriptaddress,preprintnumbers,amsmath,amssymb,longbibliography,nofootinbib]{revtex4-2}
\usepackage{latexsym}
\usepackage{textcomp}
\usepackage{gensymb}
\usepackage{graphicx}
\usepackage{verbatim}
\usepackage{amsthm}
\usepackage{float}
\usepackage{amsmath}
\usepackage{booktabs}
\usepackage{graphicx,subfigure}
\usepackage{epstopdf}
\usepackage{indentfirst}
\usepackage{epsfig}
\usepackage{epsfig,subfigure,psfrag}
\usepackage{dcolumn}
\usepackage[T1]{fontenc}
\usepackage{bm}
\usepackage{slashed}
\usepackage{cancel}
\usepackage[table]{xcolor}
\usepackage{tabularx}
\usepackage{hyperref}
\usepackage[normalem]{ulem}
\usepackage{braket}
\usepackage{enumitem}
\usepackage{multirow}
\usepackage{makecell}
\usepackage{colortbl}

\hypersetup{
    colorlinks=true,       
    linkcolor=blue,          
    citecolor=blue,        
    filecolor=blue,      
    urlcolor=blue           
}

\renewcommand\thesection{\Roman{section}}

\makeatletter
\renewcommand*{\p@subsection}{\thesection.}
\makeatother

\newcommand{\beq}{\begin{equation}}
\newcommand{\eeq}{\end{equation}}
\newcommand{\beqnarray}{\begin{eqnarray}}
\newcommand{\eeqnarray}{\end{eqnarray}}

\newcommand{\Eu}{Eu$_5$In$_2$Sb$_6$}
\newcommand{\Eua}{Eu$_5$In$_2$Sb$_6$}
\newcommand{\Eub}{EuZn$_2$P$_2$}

\begin{document}

\title{\texorpdfstring{SPLENDOR:~a novel detector platform to search for\\light dark matter with narrow-gap semiconductors}{SPLENDOR}}

\author{P. Abbamonte}
\affiliation{Department of Physics, University of Illinois Urbana-Champaign, Urbana, IL 61801, USA}
\affiliation{Materials Research Laboratory, University of Illinois, Urbana, IL 61801,USA}

\author{A. Albert}
\affiliation{Los Alamos National Laboratory, Los Alamos, NM 87545, USA}

\author{D.\,S.\,M. Alves}
\email[]{spier@lanl.gov}
\affiliation{Los Alamos National Laboratory, Los Alamos, NM 87545, USA}

\author{J. Anczarski}
\affiliation{Department of Physics, Stanford University, Stanford, CA 94305, USA}
\affiliation{SLAC National Accelerator Laboratory, Menlo Park, CA 94025, USA}

\author{T. Aralis}
\affiliation{SLAC National Accelerator Laboratory, Menlo Park, CA 94025, USA}
\affiliation{Kavli Institute for Particle Astrophysics and Cosmology, Stanford University, Stanford, CA 94035, USA}

\author{T.\,U. B\"ohm}
\affiliation{Los Alamos National Laboratory, Los Alamos, NM 87545, USA}
\affiliation{Leibniz Institute for Solid State and Materials Research Dresden, Helmholtzstr. 20, D-01069 Dresden, Germany}

\author{C. Boyd}
\affiliation{Department of Physics, University of Illinois Urbana-Champaign, Urbana, IL 61801, USA}

\author{J. Chen}
\affiliation{Department of Physics, University of Illinois Urbana-Champaign, Urbana, IL 61801, USA}
\affiliation{Materials Research Laboratory, University of Illinois, Urbana, IL 61801,USA}

\author{P.-H. Chu}
\affiliation{Los Alamos National Laboratory, Los Alamos, NM 87545, USA}

\author{M.\,S. Cook}
\affiliation{Los Alamos National Laboratory, Los Alamos, NM 87545, USA}
\affiliation{Oak Ridge National Laboratory, Materials Science and Technology Division, Oak Ridge, TN 37831, USA}

\author{C.\,W.~Fink}
\email[]{cwfink@syr.edu}
\affiliation{Los Alamos National Laboratory, Los Alamos, NM 87545, USA}
\affiliation{Department of Physics, Syracuse University, Syracuse, NY 13244, USA}

\author{M.\,L. Graesser}
\affiliation{Los Alamos National Laboratory, Los Alamos, NM 87545, USA}

\author{Y. Kahn}
\affiliation{Department of Physics, University of Illinois Urbana-Champaign, Urbana, IL 61801, USA}
\affiliation{Department of Physics, University of Toronto, Toronto, ON M5S 1A7, Canada}

\author{C.\,S. Kengle}
\affiliation{Department of Physics, University of Illinois Urbana-Champaign, Urbana, IL 61801, USA}
\affiliation{Materials Research Laboratory, University of Illinois, Urbana, IL 61801,USA}
\affiliation{Los Alamos National Laboratory, Los Alamos, NM 87545, USA}

\author{T. Kucinski}
\affiliation{Los Alamos National Laboratory, Los Alamos, NM 87545, USA}

\author{N.\,A. Kurinsky}
\affiliation{SLAC National Accelerator Laboratory, Menlo Park, CA 94025, USA}
\affiliation{Kavli Institute for Particle Astrophysics and Cosmology, Stanford University, Stanford, CA 94035, USA}

\author{C. Lane}
\affiliation{Los Alamos National Laboratory, Los Alamos, NM 87545, USA}

\author{A. Leder}
\affiliation{Los Alamos National Laboratory, Los Alamos, NM 87545, USA}

\author{R. Massarczyk}
\affiliation{Los Alamos National Laboratory, Los Alamos, NM 87545, USA}

\author{A. Mazumdar}
\affiliation{Los Alamos National Laboratory, Los Alamos, NM 87545, USA}
\affiliation{Department of Physics and Astronomy, University of North Carolina, Chapel Hill, NC 27599, USA}
\affiliation{Triangle Universities Nuclear Laboratory, Durham, NC 27708, USA}

\author{S.\,J. Meijer}
\affiliation{Los Alamos National Laboratory, Los Alamos, NM 87545, USA}

\author{W. Nie}
\affiliation{Los Alamos National Laboratory, Los Alamos, NM 87545, USA}
\affiliation{Department of Physics, SUNY University at Buffalo, Buffalo, NY 14260, USA}

\author{E.\,A. Peterson}
\affiliation{Los Alamos National Laboratory, Los Alamos, NM 87545, USA}

\author{A.  Phipps}
\affiliation{Department of Physics, California State University, East Bay, Hayward, CA 94542, USA}

\author{F. Ronning}
\affiliation{Los Alamos National Laboratory, Los Alamos, NM 87545, USA}

\author{P.\,F.\,S. Rosa}
\email[]{pfsrosa@lanl.gov}
\affiliation{Los Alamos National Laboratory, Los Alamos, NM 87545, USA}

\author{I. Rydstrom}
\affiliation{Department of Physics, Santa Clara University, Santa Clara, CA 95053, USA}
\affiliation{Department of Physics, University of California Berkeley, Berkeley, CA 94720, USA}

\author{N.\,S. Sirica}
\affiliation{Los Alamos National Laboratory, Los Alamos, NM 87545, USA}
\affiliation{U.S. Naval Research Laboratory, Washington, DC 20375, USA}

\author{Z.\,J. Smith}
\affiliation{Department of Physics, Stanford University, Stanford, CA 94305, USA}
\affiliation{SLAC National Accelerator Laboratory, Menlo Park, CA 94025, USA}

\author{K. Stifter}
\affiliation{SLAC National Accelerator Laboratory, Menlo Park, CA 94025, USA}
\affiliation{Kavli Institute for Particle Astrophysics and Cosmology, Stanford University, Stanford, CA 94035, USA}

\author{S.\,M. Thomas}
\email[]{smthomas@lanl.gov}
\affiliation{Los Alamos National Laboratory, Los Alamos, NM 87545, USA}

\author{S.\,L. Watkins}
\affiliation{Los Alamos National Laboratory, Los Alamos, NM 87545, USA}
\affiliation{Pacific Northwest National Laboratory, Richland, WA 99352, USA}

\author{B.\,A. Young}
\affiliation{Department of Physics, Santa Clara University, Santa Clara, CA 95053, USA}

\author{J.-X. Zhu}
\affiliation{Los Alamos National Laboratory, Los Alamos, NM 87545, USA}

\collaboration{SPLENDOR Collaboration}
\noaffiliation

\date{\today}

\begin{abstract}
We present the design and current status of SPLENDOR, a novel detector
platform that combines narrow-gap semiconductor targets with low-noise charge readout
to achieve sensitivity to dark matter energy deposits well below the eV scale.
SPLENDOR is designed to be a modular and scalable system able to accommodate different target materials and signal readout technologies.  
SPLENDOR's present strategy entails: (i) the use of strongly correlated $f$-electron semiconductors
with anisotropic electronic structures to enable not only sub-eV energy thresholds, but also directional sensitivity to the incoming dark matter flux, allowing for signal-background discrimination via daily modulation, and (ii) custom charge readout based on cryogenic high-electron-mobility transistor (cryoHEMT) amplifiers approaching single-electron resolution.
We report on the selection and characterization of \Eu~as the target material for SPLENDOR's first prototype detector, as well as the development and calibration of the prototype amplifier chain, achieving a measured charge resolution of $20\pm7$ electrons in silicon test samples, consistent with predicted performance. This provides a demonstration of the detector architecture, which is now ready for deployment in a dark matter search campaign to deliver SPLENDOR's first science results.
Finally, we present estimates of sensitivity reach in the parameter space of athermally produced relic dark matter under high- and low-background environments, and for various amplifier technology upgrades with increasing performance, including planned quantum sensing upgrades in order to achieve our ultimate goal of sub-electron resolution in optimized systems.
SPLENDOR provides a novel approach to dark matter direct detection, combining quantum sensing with material's design to open new avenues of exploration in the sub-MeV mass range of dark matter parameter space.
\end{abstract}

\preprint{LA-UR-25-26113}
\maketitle

\section{Introduction}
\label{sec:Intro}

The particle nature of dark matter is one of the most pressing outstanding questions in physics. Since the only direct evidence for dark matter is gravitational and on galactic scales or larger, the dark matter mass is largely unconstrained and can span 50 orders of magnitude or more~\cite{ParticleDataGroup:2024cfk, battaglieri2017cosmic}. This challenge calls for a diversity of search strategies, from multi-ton noble liquid detectors to quantum-limited sensors~\cite{Angloher_2017,Abdelhameed_2019,Alkhatib_2021,Ren_2021,Fink_2021, Amaral_2020, Barak_2020, Arnquist_2023, aguilararevalo2022oscura, TESSERACT, PhysRevD.100.092007, anthonypetersen2023applying, polar1, Griffin:2018bjn, Griffin_2021, Kurinsky_2019, Marshall_2021, Hochberg_2018, Geilhufe_2020, Hochberg_2016, osti_1872396, das2022dark, Hochberg_2019, Chiles_2022, PhysRevB.79.144511, 2018NatAs...2...90E, fink2023superconducting}. Focusing specifically on the scenario wherein dark matter deposits energy in a detector through non-relativistic scattering, a particularly difficult regime of parameter space is the mass range lighter than $\mathcal{O}(\text{MeV})$, where a typical dark matter particle carries kinetic energy of $\mathcal{O}(\text{eV})$ or less. Detecting such light dark matter candidates requires materials with sub-eV excitation energies; several ideas have been proposed, including phonons~\cite{Hochberg:2015fth,Hochberg_2016,polar1,Griffin:2018bjn}, magnons~\cite{Trickle:2019ovy}, and electron-hole pairs in Dirac materials~\cite{Geilhufe_2020, Hochberg_2018}.

\begin{figure}
     \centering
     \includegraphics[width=1.0\linewidth]{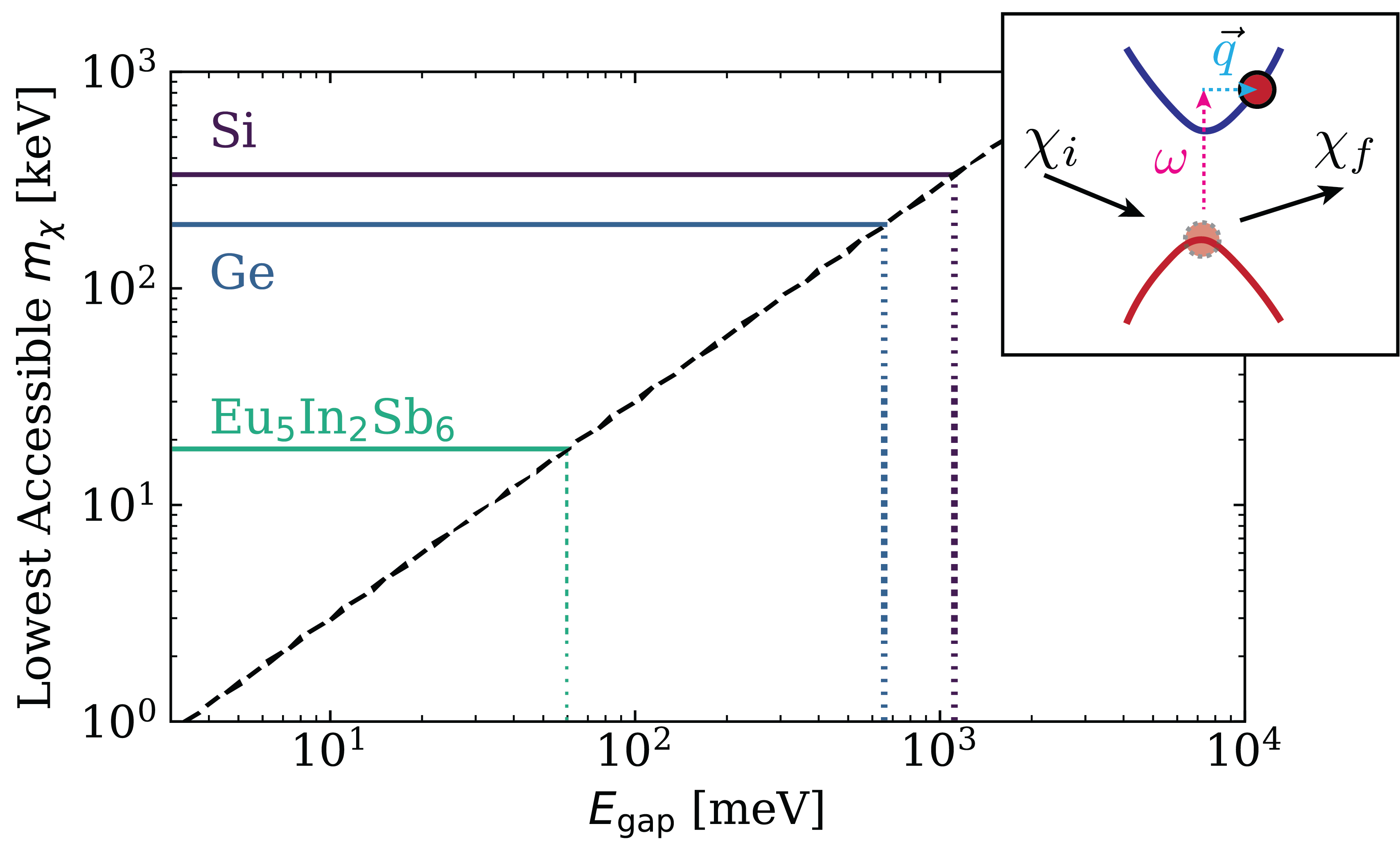}
     \caption{Lower bound on the range of dark matter masses that are kinematically accessible through scattering given a bandgap $E_\text{gap}$ (black line).
     }
     \label{fig:mMin_vs_Egap}
\end{figure}

In terms of technology-readiness, the detection of ionization signals in semiconductor targets is relatively mature compared to other more exotic ideas. Therefore, a promising direction in sub-MeV dark matter sensing is the use of semiconductors with narrow, sub-eV bandgaps, for which the energy cost to excite electrons to the conduction band is much lower than in Si or Ge. Fig.\,\ref{fig:mMin_vs_Egap} shows the lowest dark matter mass that is kinematically accessible through electron scattering given a target material's band gap $E_\text{gap}$.

In this paper, we introduce SPLENDOR\footnote{SPLENDOR is a loose acronym for “{\bf S}earch for {\bf P}articles of {\bf L}ight Dark Matt{\bf e}r with {\bf N}arrow-Gap Semicon{\bf d}uct{\bf or}s.”}, a dark matter detector platform whose core goal is the engineering and deployment of narrow-gap $f$-electron materials as dark matter detector targets.
These materials exhibit quantum phenomena via a combination of strong spin-orbit coupling and electronic correlations, and their crystal structures are remarkably versatile. Notably, stoichiometric $f$-electron materials can exhibit clean electronic bandgaps 1–2 orders of magnitude smaller than in conventional semiconductors, as well as anisotropic electronic structures that provide directional sensitivity of the target crystal to the incoming dark matter flux \cite{Hochberg_2018,Coskuner:2019odd,Geilhufe_2020,Boyd2023}. This latter feature results in a daily modulation of the dark matter scattering rate due to the Earth’s sidereal rotation relative to the dark matter wind \cite{PhysRevD.37.1353,Gondolo:2002np,Ahlen:2009ev}, enabling the subtraction of reducible and irreducible (isotropic) backgrounds even in environments with limited shielding, and is a key feature of the SPLENDOR detection strategy.

In addition to narrow-gap semiconductors, another enabling innovation of SPLENDOR is its cryogenic instrumentation to read out the dark matter-induced ionization signal with a resolution of a few electrons.
We deploy novel cryogenic charge amplifiers based on high electron mobility transistors (HEMTs) \cite{anczarski2023twostage}; similar HEMTs have been used by other collaborations (e.g. SuperCDMS \cite{PHIPPS2019181} and RICOCHET \cite{augier2023demonstration}). These amplifiers have the advantage of being easily adaptable to the target substrate, thereby allowing for rapid improvements as new target materials become available.
The amplifiers utilize a two-stage design to minimize parasitic capacitance when coupling to the detector material. The base temperature buffer stage is a charge integrating amplifier that produces a voltage signal which is less susceptible to noise. That voltage signal is amplified at the 4K gain stage before reaching a filtering board, commercial low-noise amplifier, and a data acquisition system at room temperature.

Together, these two detector components promise sensitivity in the infrared (10–100 meV) energy range, and with 1–2 orders-of-magnitude lower energy thresholds compared with existing semiconductor-based detector technology.

For the initial stages of the SPLENDOR experimental program, Eu$_5$In$_2$Sb$_6$, to be described in more detail in upcoming sections, was selected as the target material for the detector prototype due to several desirable properties, such as a narrow bandgap, anisotropic electronic structure, suppressed dark currents, and efficient charge collection. Eu$_5$In$_2$Sb$_6$ was only recently synthesized, for the first time, in single-crystal form by LANL members of the SPLENDOR collaboration~\cite{rosa2020colossal}.

Importantly, the SPLENDOR detectors are designed to be modular: should a new promising target material be identified, samples can be readily integrated with the charge amplifiers without the need for substrate-tailored device fabrication. In addition, upgrades based on quantum amplifier technologies are currently being explored by the collaboration
to achieve sub-electron resolution. The adaptive nature of the SPLENDOR detector platform allows us to modify the mK stage of the readout scheme to take advantage of these developments once they are demonstrated and ready for deployment.

This paper introduces the conceptual design and current status of SPLENDOR, and discusses its viability, scalability, and sensitivity to probe the low-mass frontier of particle dark matter. SPLENDOR’s ultimate science goal is to achieve sensitivity to relic-density cross sections of freeze-in dark matter \cite{Gopalakrishna:2006kr,Hall:2009bx,Bernal:2017kxu,Dvorkin:2019zdi} in the currently inaccessible sub-MeV mass range.

The remainder of this paper summarizes the framework of the SPLENDOR experimental program, including all aspects of its cross-disciplinary scope and components -- materials synthesis and characterization, detector development, and theoretical modeling -- and paves the way for a prototype detector that is expected to perform its first dark matter search in the near future, which will be described in an upcoming publication. Sec.~\ref{sec:Materials} describes the materials selection, synthesis, and characterization of our candidate detector material Eu$_5$In$_2$Sb$_6$. Sec.~\ref{sec:Detector} describes the detector concept, from idealized design to prototype performance. Sec.~\ref{sec:Sensitivity} presents the projected sensitivities of SPLENDOR detectors under different scenarios for backgrounds, exposures, and charge amplifier technologies. The supplemental materials provide further details on the synthesis and growth optimization for \Eu; our transport, photoresponse, and spectroscopy measurements; our radioassaying results and simulations of backgrounds resulting from radiological impurities; and our first-principles electronic structure calculations to estimate the dielectric response of \Eu, which was then used as input to determine the dark matter signal rate.
We conclude in Sec.~\ref{sec:Conclusion} with the outlook for a prototype detector and the broader SPLENDOR experimental program, underscoring the advantages of close collaboration among high energy theorists and experimentalists, condensed matter theorists and experimentalists, and materials scientists in probing unexplored parameter space for dark matter and paving a plausible path for discovery.

\section{SPLENDOR materials}
\label{sec:Materials}

\begin{figure}
    \centering
    \includegraphics[width=1.0\linewidth]{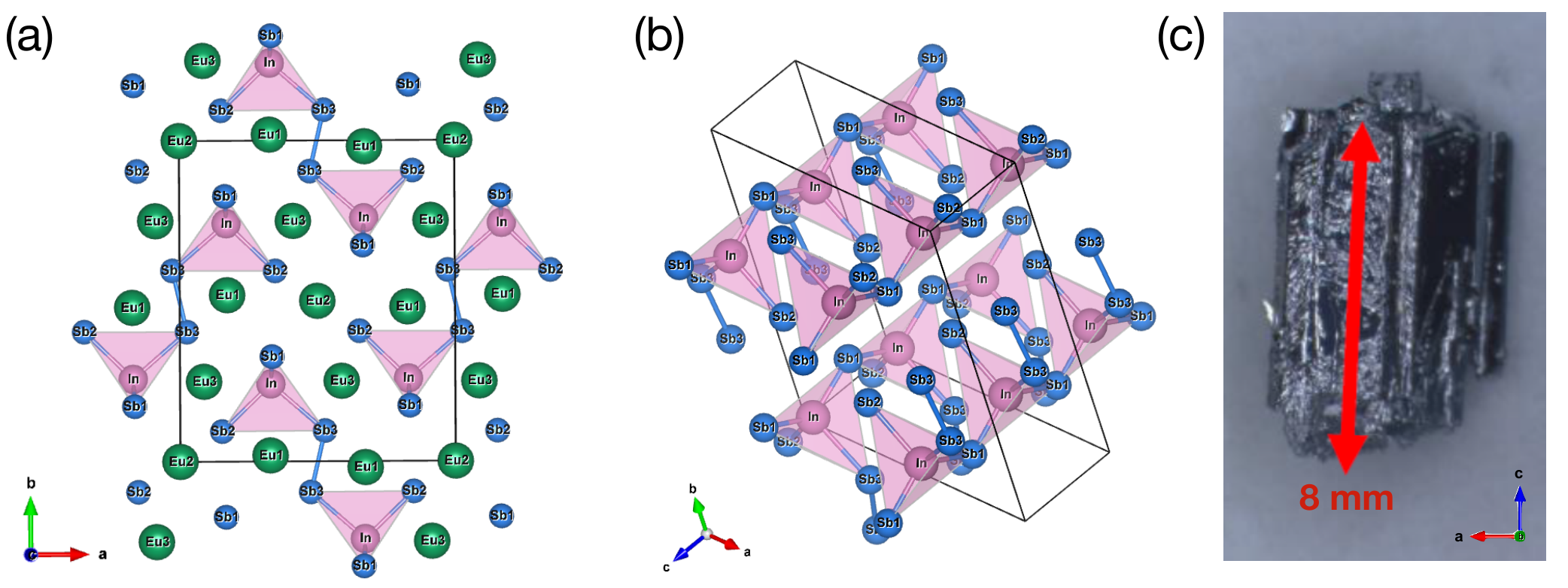}
    \caption{The orthorhombic crystal structure of Eu$_5$In$_2$Sb$_6$ with (a) InSb$_{4}$ tetrahedra connected by Sb-Sb dumbbells in the \textit{ab} plane and (b) InSb$_{4}$ chains along the \textit{c} axis. (c) Photo of an as-grown Eu$_5$In$_2$Sb$_6$ single crystal weighing $\sim$\,0.8\;g.}
    \label{fig:Eu526structure}
\end{figure}

In high energy physics, efforts to advance detector technology often focus primarily on sensor R\&D, while the detector target material is chosen based on commercial availability and by leveraging industrial developments. The materials R\&D component of the SPLENDOR program, on the other hand, aims to innovate in materials design in addition to sensor development. The collaboration has sought to identify and synthesize semiconductors for light dark matter detection with the following characteristics:
\begin{enumerate}[wide, labelwidth=!, itemindent=!, labelindent=0pt, itemsep=0pt, topsep=3pt]
\item a sub-eV band gap (ideally in the 1–100 meV range) to enable sensitivity to sub-eV energy deposits from light dark matter candidates with masses in the sub-MeV range;
\item an anisotropic electronic structure, which induces a daily modulation of the dark matter scattering rate, enabling the subtraction of reducible and irreducible (isotropic) backgrounds;
\item suppressed dark currents from in-gap states and surface states, which constitute one of the main backgrounds in semiconductor-based rare-event searches; 
\item large density of states near the band-edge to maximize the kinematical phase space for dark matter scattering;
\item low dielectric screening to prevent the dark matter rate from being suppressed;
\item  high-quality single crystals of large enough size (i.e., $\sim 0.1- 1$~g) to enable future scalability of the prototype, and with a low number of impurities to minimize dark rates and improve charge collection efficiency. 
\end{enumerate}

Well-known narrow-gap semiconductors exist, such as InSb and Hg$_{1-x}$Cd$_x$Te, which are used in infrared photodetection \cite{ROGALSKI1988,Rogalski2005HgCdTe}.
The Dirac semimetal ZrTe$_5$ has also been considered in theoretical studies for dark matter detection \cite{Hochberg_2018,Geilhufe_2020,Coskuner:2019odd}. Unfortunately, all of these possible candidates suffer from high dark rates, which makes them too noisy for use in rare-event, low-threshold detectors. InSb and ZrTe$_5$ suffer from impurity states caused by crystallographic defects such as dislocations and twinning \cite{Hollis1967,Chu2010,10.1063/1.4922977,PhysRevLett.121.187401}. In Hg$_{1-x}$Cd$_x$Te, which has been studied for decades \cite{Lawson1959HgCdTe,Rogalski_2005,10.1116/1.576215}, the band gap can be tuned via chemical substitution, but this process introduces disorder in the material.

Instead of tuning the band gap via substitution, our collaboration instead sought novel materials using the Zintl concept as the design principle \cite{Kauzlarich2024} and constituent elements with strong spin-orbit coupling. Zintl intermetallic phases are a combination of alkaline, alkaline-earth, or rare-earth cations and a covalently bonded polyanionic structure that achieves an octet in the valence electron shell. Zintl compounds are thus valence-precise, and electron transfer between cations and anions is expected to create a semiconducting state with low levels of disorder. Also, elements with strong spin-orbit coupling, such as $f$-electron elements, are prone to band inversion and tunable narrow gaps~\cite{PhysRevB.105.235128}. Notably, lanthanide materials offer a promising yet underexplored platform for the realization of these principles.

Candidate materials that were identified, synthesized, and studied by our collaboration included: \Eu~\cite{rosa2020colossal}, La$_3$Cd$_2$As$_6$ \cite{doi:10.1021/acs.chemmater.1c00797,kengle2025putative}, Ce$_3$Cd$_2$As$_6$ \cite{doi:10.1021/acs.chemmater.1c00797,PhysRevB.105.094443}, Eu$_7$Ga$_6$Sb$_8$~\cite{Cook2023}, \Eub~\cite{PhysRevB.106.054420,Krebber2023PRB,cook2025magneticpolaronformationeuzn2p2}, and Eu$_2$Cd$_2$As$_3$ \cite{WANG2013116}.
The majority of them were discarded due to either having low (longitudinal and Hall) resistivity or not yielding large enough high-quality single crystals after efforts to optimize synthesis and growth. The candidates best satisfying our preliminary criteria were \Eu~and \Eub. Upon further characterization, \Eu~was found to have higher carrier mobility than \Eub; additionally, the synthesis of \Eub~has not yet been fully optimized---further work is needed to avoid flux inclusions and to obtain single-crystal samples heavier than $\sim0.1$ g. As a result, \Eu~was selected as the best performing target material for the initial phase of SPLENDOR. A more detailed discussion of its properties now follows.

The stoichiometric Zintl compound Eu$_{5}$In$_{2}$Sb$_{6}$ was first synthesized in single-crystal form by LANL members of the collaboration \cite{rosa2020colossal,Rosa2019}. Eu$_{5}$In$_{2}$Sb$_{6}$ crystallizes in a low-symmetry orthorhombic structure (space group $Pbam$), which leads to a highly anisotropic crystal structure as shown in Figs.\;\ref{fig:Eu526structure}a–\ref{fig:Eu526structure}b. In addition, its structure is very stable, in contrast to alkaline-earth counterparts, which are known to quickly oxidize when exposed to air.
For the synthesis and growth optimization of \Eu, a self-flux technique with an In-Sb flux was used. The optimization of the growth conditions was performed through more than twenty different growth batches that varied parameters such as temperature steps, cooling profiles, and stoichiometry. After optimization, the best crystals were obtained using the molar ratio 5:35:17.5 of Eu:In:Sb. This led to an increase in the mass of a single crystal from less than 0.1\;g to a final size of approximately 0.8\;g (Fig.\;\ref{fig:Eu526structure}c).

\begin{figure}
    \centering
    \includegraphics[width=\columnwidth]{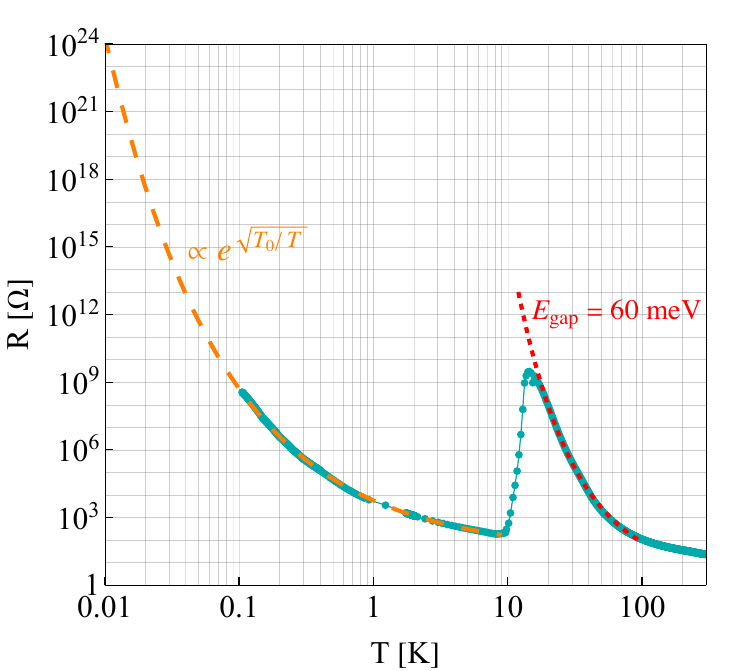}
    \caption{Resistance vs temperature of a high-quality single crystal of Eu$_5$In$_2$Sb$_6$. Fitting the resistance in the temperature range above $\sim$20 K to an activated behavior \mbox{$R\propto \text{exp}[E_\text{gap}/(2\,k_B T)]$}, a narrow transport gap of \mbox{$E_\text{gap}=60$\;meV} is obtained.
    In the range of 0.1–10\;K, the temperature dependence of the resistance is consistent with variable-range hopping (VRH) behavior. A VRH fit 
    extrapolated down to 10~mK suggests a bulk dark current in this sample of $\mathcal{O}(10^{-6})$\;attoA/V at the base-temperature of a dilution refrigerator, which would roughly translate to a dark rate of $\mathcal{O}(10^{-7})$\;Hz/g at 20 mV of applied voltage.}
    \label{fig:Eu526_Res_v_Temp}
\end{figure}

In addition to the increase in sample mass, electrical properties also improved. Compared to the previous generation of samples \cite{rosa2020colossal}, a larger rise in resistance
was observed upon cooling, indicating a smaller number of crystalline defects and free carriers resulting from them. Indeed, the low-temperature specific heat coefficient of our \Eu~samples extrapolates to zero within experimental accuracy, consistent with a clean semiconducting gap. Fig.\;\ref{fig:Eu526_Res_v_Temp} shows the electrical resistance versus temperature for one of our highest-quality single crystals. Although the resistance initially drops upon cooling below the magnetic ordering temperature of 14~K, it begins to rise sharply again upon further cooling. Below 10~K, the resistance displays the temperature dependence of Efros–Shklovskii variable-range hopping (VRH) behavior \cite{Efros_1975,PhysRevB.105.054206,10.1063/10.0034343,Shklovskii_book}, $R\propto \text{exp}[\sqrt{T_0/ T}]$, all the way down to our lowest measured temperature of 100~mK, when it reaches  $5\times10^8\;\Omega$. Sample heating, combined with limitations of the readout electronics of our setup, prevented accurate measurements below $\sim$\,100\;mK.
Extrapolating our VRH fit down to 10~mK (the expected operating temperature of our detector target) suggests an electrical resistance in this sample
of $10^{24}\;\Omega$,  which would translate to a bulk dark current of $\mathcal{O}(10^{-6})$\;attoA/V, or, equivalently, to a dark rate of $\mathcal{O}(10^{-7})$\;Hz/g at 20~mV of applied voltage.

\begin{figure}
\centering
\includegraphics[width=\columnwidth]{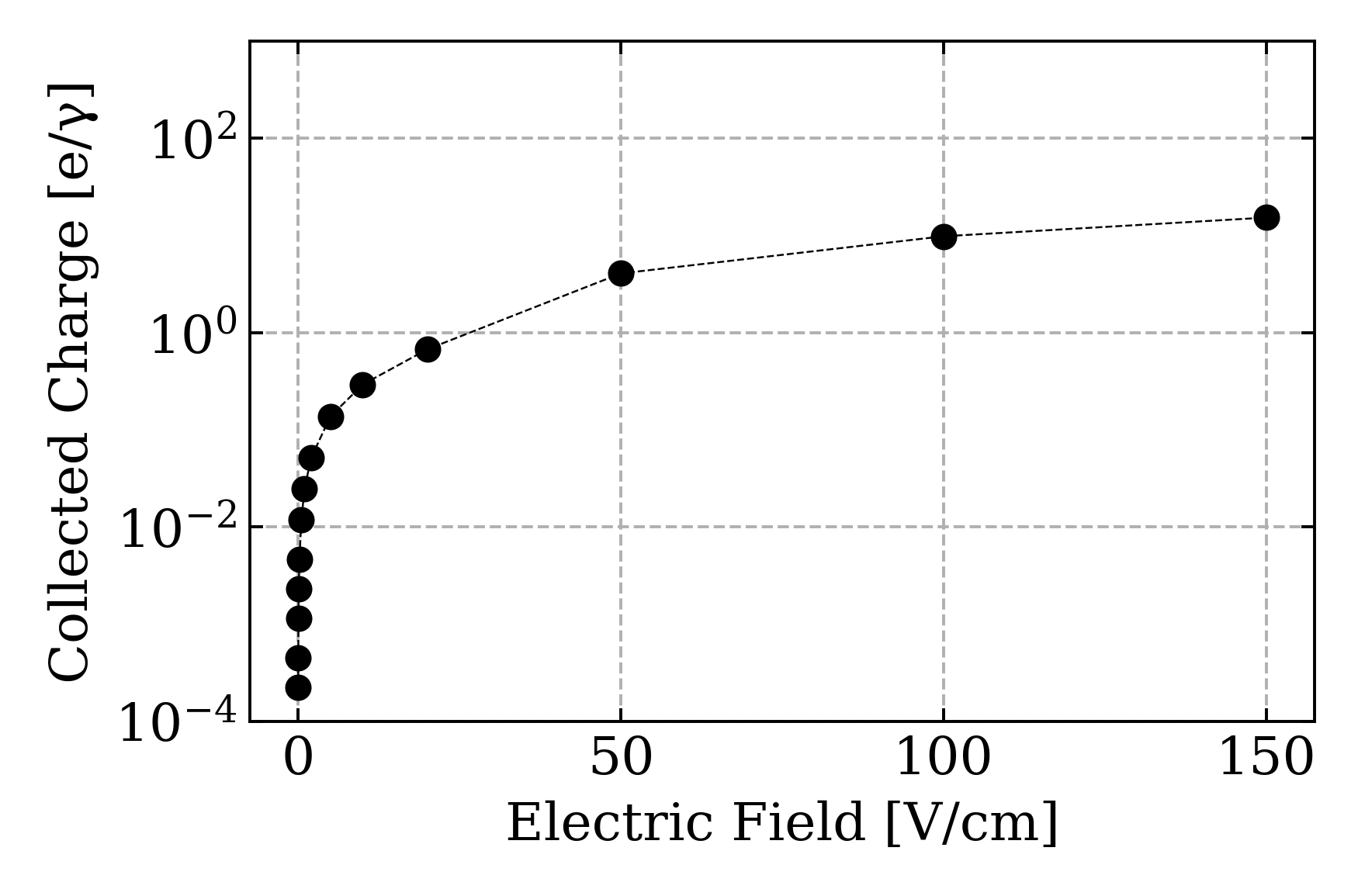}
\caption{The measured charge yield per $1300\,\mathrm{nm}$ photon for a Eu$_5$In$_2$Sb$_6$ sample as a function of applied electric field at 16\,K.}
\label{fig:QE_526}
\end{figure}

Measurements of AC Hall resistivity in our samples indicated a charge carrier mobility of 5\;cm$^2$/(V$\cdot_{\,}$s) at 16\;K, increasing to 50\;cm$^2$/(V$\cdot_{\,}$s) at 2\;K
(see discussion and Fig.\;\ref{fig:mobility} in Appx.\,\ref{Appx:ACHall}).
Although this is 1--2 orders of magnitude lower than in silicon, it is still high enough to enable (near) full charge collection at modest bias voltages, as shown by the plateauing of the photocurrent response curve in Fig.\,\ref{fig:QE_526}.

The presence of multiple heavy elements in \Eu~implies strong spin-orbit coupling, which likely contributes to its reduced band gap. The exact determination of the direct and indirect band gaps of \Eu{} is challenging, and it is critical to use as many probes as possible, as each probe has its own limitations. For instance, transport is a momentum- and real-space-averaged probe, whereas scanning tunneling microscopy (STM) is a real-space probe that is highly surface-sensitive, and momentum-resolved electron energy loss spectroscopy (M-EELS) is a surface-sensitive momentum-space probe limited by the quality of sample preparation and cleaving.

We have found a consistent band gap range of $\sim$\,(60-100)\;meV for \Eu~from several independent transport measurements.
In particular, the temperature dependence of the electrical resistance in our highest-quality samples reveals strongly insulating transport behavior with a transport gap of approximately 60\;meV in the temperature range above $\sim$\,20\;K; see Fig.\,\ref{fig:Eu526_Res_v_Temp} and Appx.\,\ref{appx:RvTdiscussion} for a more detailed discussion about inferring the bandgap from electrical resistance data.
Likewise, in our AC Hall measurements, the temperature dependence of the carrier concentration follows the expected exponential dependence of a narrow-gap semiconductor with an energy gap of 60\;meV, see Fig.\,\ref{fig:526_ACHALL} in Appx.\;\ref{Appx:ACHall}.
Finally, in our photoresponse measurements (described in more detail in Appx.\;\ref{Appx:photoresponse}), the charge yield per $1300\,\mathrm{nm}$ photon at $16\,\mathrm{K}$, shown in Fig.~\ref{fig:QE_526}, is consistent with a bandgap of $\sim$$10-100$\;meV, assuming fully saturated charge collection and ignoring the presence of thermally excited free carriers at 16 K.

Our theoretical determination of the band structure of \Eu, based on first-principles electronic structure calculations described in detail in Appx.\;\ref{DFTtheory}, also yielded values for the direct and indirect bandgaps consistent with transport measurements, of about 40 meV and 20 meV, respectively.

We also explored spectroscopy techniques to determine the band gap and dielectric response of \Eu.  Scanning tunneling microscopy gave a gap similar to transport measurements \cite{Crivillero2022PRB}. Unfortunately, additional spectroscopic measurements, including M-EELS and Fourier Transform Infrared Spectroscopy (FTIR), gave inconclusive results due to experimental systematics and difficulties in sample preparation and cleaving. To perform an EELS measurement with momentum resolution (M-EELS), the sample must usually be cleaved {\it in situ} in the high vacuum environment of the M-EELS chamber. After several unsuccessful cleaving attempts of \Eu~samples, we laser-etched the perimeter of a sample in order to cleave it, but the roughness of the resulting cleave meant that the EELS measurement could not be performed with any momentum resolution. Instead, it was only possible to obtain momentum-integrated data, which,
compounded by a broad elastic peak and additional measurement artifacts below 150~meV,
hindered the determination of the expected low gap values.  In our FTIR data, the most prominent feature was a large rise in optical conductivity for photon energies above 600 meV, with a low-energy tail extending down to $\mathcal{O}(10)$\;meV. Unfortunately, prominent phonon peaks below 25 meV prevented the determination of the electronic gap associated with this lower energy tail. A detailed discussion of our EELS and FTIR measurements can be found in Appxs.\,\ref{Appx:EELS},\,\ref{Appx:FTIR}.

More precise determinations of the bandgap of \Eu~are planned with a full detector calibration in the dilution refrigerator with the working charge amplifier using a longer-wavelength LED. For the projected sensitivities discussed in Sec.\,\ref{sec:Sensitivity}, we will adopt, for concreteness, the value of 60 meV for the bandgap of \Eu~inferred from resistivity measurements of our highest-quality single crystals.


\begin{figure}[b]
    \centering
    \includegraphics[width=1\linewidth]{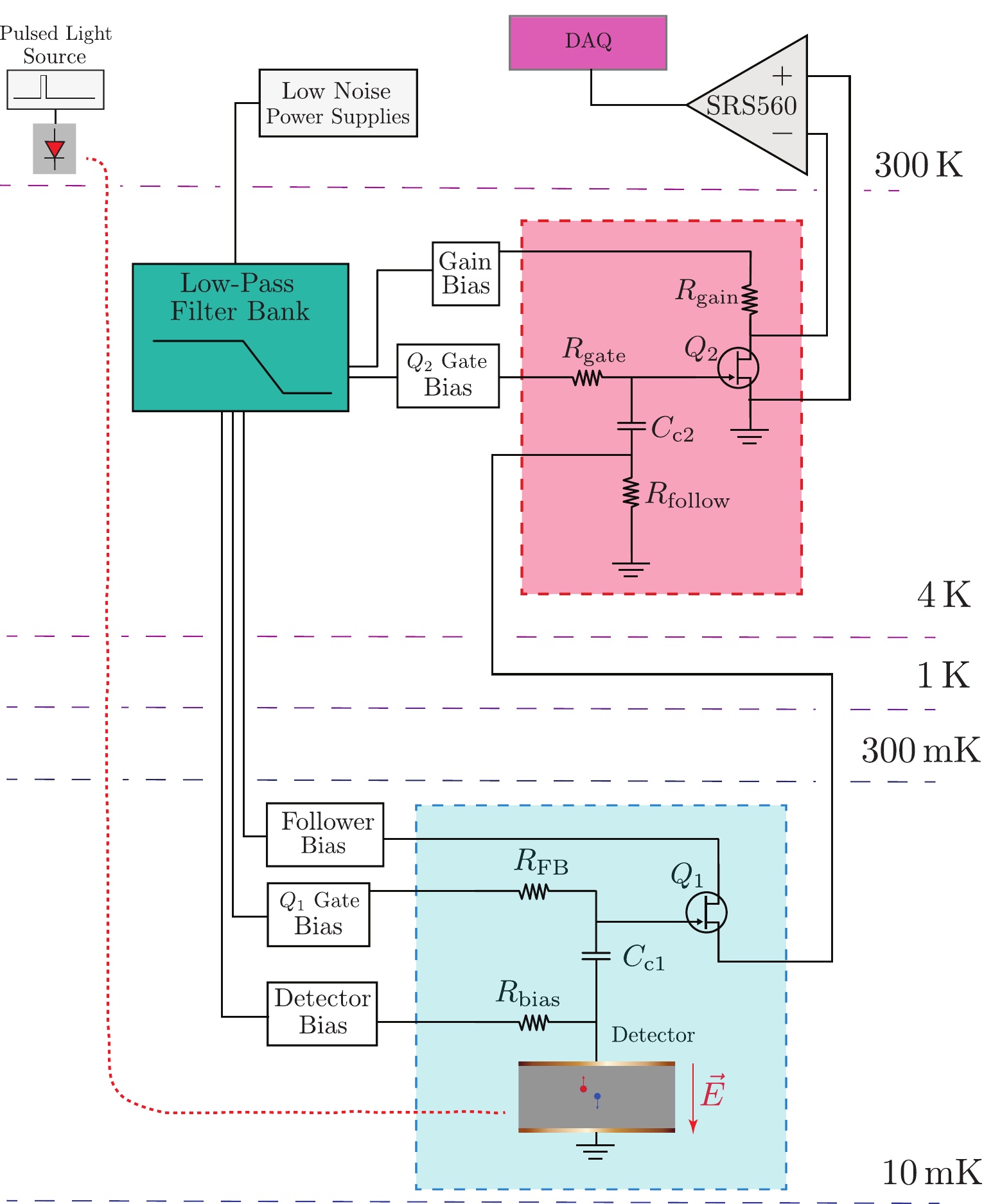}
    \caption{Schematic diagram of the simplified amplifier topology and physical location of various components in the cryostat. The detector and front-end follower amplifier are connected in the same housing on the mixing chamber of the dilution refrigerator. The charge from the detector target is integrated onto $Q_1$, which is then amplified by $Q_2$ mounted to the 4K plate.
    This signal is then amplified by a room temperature amplifier (SRS SR560; 1~Hz – 1~MHz) and digitized. The bias lines are filtered at room temperature as well as at 4K, where low-pass filters are mounted.}
    \label{fig:schematic}
\end{figure}

\section{Detector Concept and Calibration of Prototype Amplifier}
\label{sec:Detector}

To enable sensitivity to sub-eV energy deposits in our narrow-bandgap semiconductor targets, SPLENDOR employs a cryogenic ionization-based detection scheme using a custom-designed charge amplifier capable of near-single-electron resolution. The signal of interest arises from charge carriers generated by dark matter scattering or absorption. These charges are collected in a point-contact geometry and capacitively coupled to a low-noise, HEMT amplifier system.

\begin{figure*}
\centering
\includegraphics[width=\linewidth]{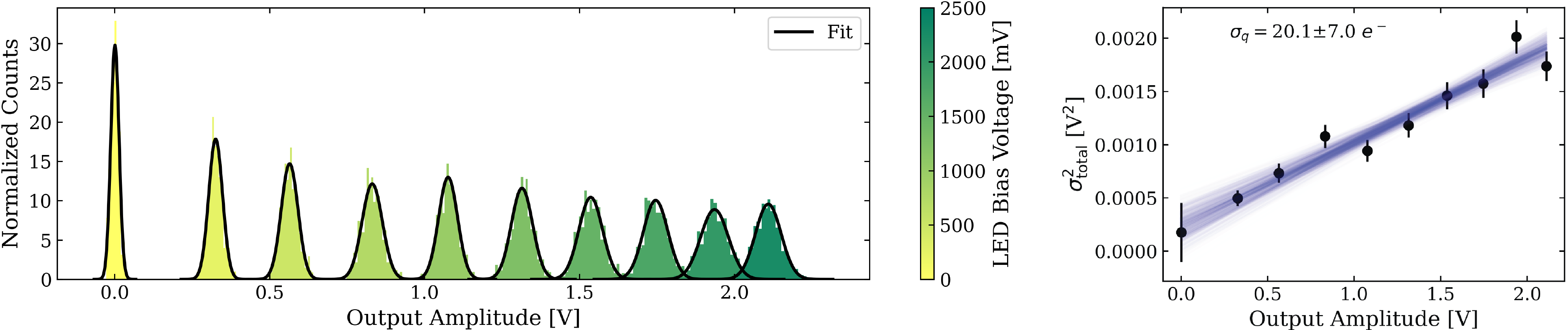}
\caption{\emph{Left:} Pulse amplitude distribution from a calibration dataset in which a 635\;nm pulsed-light source shined on a Si test sample. Different colors correspond to different LED powers. \emph{Right:} Linear fit relating the variance and mean of the pulse amplitude distributions shown in the left panel, from which the charge resolution was extracted. The shaded band represents the variance in the fit.}
\label{fig:calibrationANDfit}
\end{figure*}

\begin{figure}[b]
    \centering
    \includegraphics[width=\linewidth]{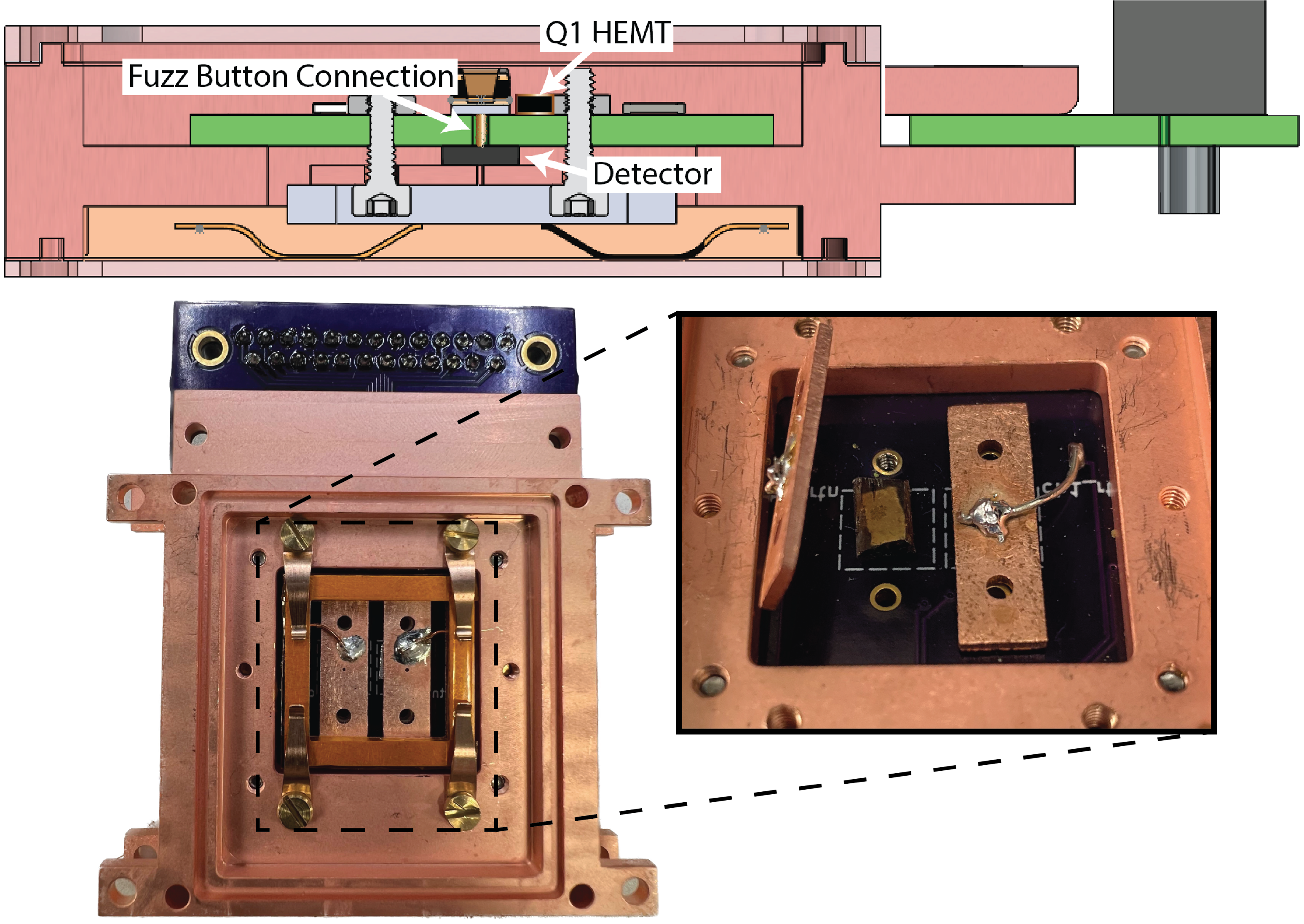}
    \caption{\textit{Top:} Cutaway view of the CAD model for the detector housing showing
    the low-capacitance connection between the detector target and the front-end HEMT of the two stage amplifier.  \textit{Bottom~left:} Picture of the backside of the prototype detector housing with Si and \Eu{} samples installed. \textit{Bottom~right:} \Eu{} shown before installation of backing plate and detector clamp. Adapted from Ref.~\cite{anczarski2023twostage}.}
    \label{fig:housing}
\end{figure}

Our charge readout system is based on a two-stage cryogenic HEMT amplifier previously described in~\cite{anczarski2023twostage}. The amplifier’s base stage consists of a low-capacitance ($C_\text{gs} \approx 1.6\,\mathrm{pF}$) HEMT operating at 10 mK, configured as a voltage buffer in a common-drain topology (Fig.~\ref{fig:schematic}). This stage is integrated directly into the detector housing, with the gate coupled to the detector target via a fuzz button contact (Fig.~\ref{fig:housing}). This compact geometry minimizes parasitic capacitance ($C_\text{par} \approx 1\,\mathrm{pF}$), which is critical for achieving low noise. The buffer stage output is sent to a second-stage 200 pF HEMT amplifier mounted at 4 K, followed by further room-temperature amplification and digitization.

The amplifier’s charge resolution $\sigma_q$ scales approximately as:
\begin{equation}
    \sigma_q \propto \frac{N_V}{\varepsilon_\text{CCE}\,\tau_\text{BW}}\left[C_\text{det}+C_\text{in}+C_\text{par}\right]
\end{equation}
where $N_V$ is the voltage noise density, $\varepsilon_\text{CCE}$ is the charge collection efficiency, $\tau_\text{BW}$ is the integration bandwidth, and $C_\text{det}$, $C_\text{in}$, and $C_\text{par}$ are the detector, input transistor, and parasitic capacitances, respectively. In our prior benchmarking of the voltage amplifier, we have measured an integrated noise equivalent to approximately 7 electrons under an assumed capacitive load of $\lesssim 5$ pF~\cite{anczarski2023twostage}. This measured noise was slightly elevated relative to the expectation based on the HEMT model used in~\cite{Juillard_2019}, and was attributed to excess vibrations from the pulse tube cryocooler, which we have since been able to mitigate. This reduced vibrational background improves our expected charge resolution to 5 electrons.

We are in the process of modifying our prototype amplifier to further improve charge resolution.
We plan to replace the feedback resistor with a separate HEMT as an active-reset circuit, allowing us to remove the Johnson noise from the large feedback resistor and increase our signal bandwidth, which would improve the expected resolution to 3 electrons. 
An additional simple upgrade we plan on implementing is parallel amplification, in which the signal is measured on both sides of the crystal, each with its own HEMT amplifier; since the signal is correlated on both HEMT channels while the noise on each channel is random, parallel amplification improves the signal-to-noise ratio by a factor of $\sqrt{2}$, effectively improving the expected resolution from 3 to 2 electrons.

To validate amplifier performance and establish an empirical calibration of the signal chain, we conducted a photon-counting calibration using a Si test sample in one channel of the amplifier. Optical pulses from a 635 nm fiber-coupled LED were used to shine photons on the silicon sample. By varying the LED drive voltage and using a synchronized TTL signal, we collected over 600 single-shot waveforms per LED setting. The resulting pulse amplitude distribution is shown in Fig.~\ref{fig:calibrationANDfit}\,(left).

\begin{figure*}
\centering
\includegraphics[width=\textwidth]{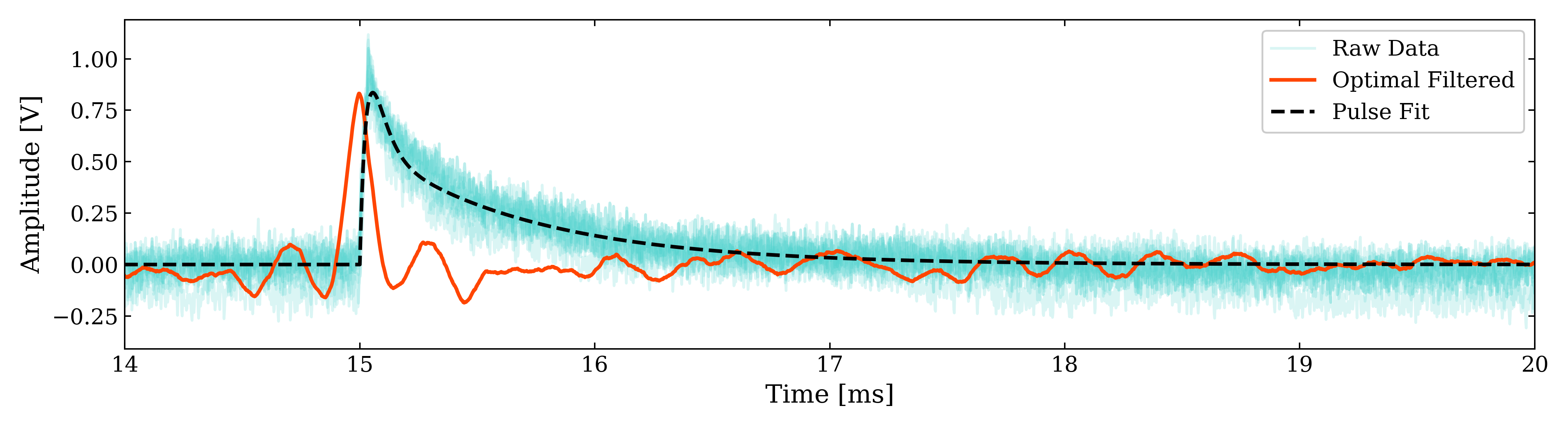}
\caption{Example of a pulse from a charge signal in a silicon chip as measured by our current cryoHEMT-based charge amplifier. A subset of the raw data is shown in turquoise, the optimal filtered data (i.e., the data filtered by the measured noise power spectral density and pulse shape~\cite{Gatti:1986cw, golwala}) is shown in red, and the corresponding pulse fit to the data using the optimal filter formalism in SPLENDAQ~\cite{watkins2023splendaq} is shown in black.}
\label{fig:pulse_si}
\end{figure*}

The mean $\left<V\right>$ and variance $\sigma^2_V$ of the pulse amplitudes were computed as a function of LED power. The Poissonian nature of photon number fluctuations results in a linear relationship between them,
\begin{equation}
    \sigma^2_V = m\left<V\right> + b,
\end{equation}
where the slope $m$ and intercept $b$ encode the system gain and the baseline amplifier noise floor, respectively. Fitting this relation across the LED power range yields a calibrated charge resolution of $20 \pm 7$ electrons; see Fig.~\ref{fig:calibrationANDfit}\,(right). The $\mathcal{O}(1)$ degradation in resolution relative to the 5-electron prediction from the amplifier noise model can be attributed to non-ideal charge collection efficiency in the silicon sample, likely limited by surface effects from the interface contact. We are actively working on optimizing the electrode parameters.

Finally, the calibrated gain allows us to convert pulse amplitude to generated charge and thus to deposited energy. This energy calibration provides a robust performance benchmark for upcoming measurements on Eu$_5$In$_2$Sb$_6$, for which the photon-generated charge yield is expected to exceed that of silicon, given its much smaller bandgap.

For data acquisition (DAQ), the collaboration has developed an open-source software package called \mbox{SPLENDAQ}~\cite{watkins2023splendaq}. The primary functionality of this software is to acquire continuous time streams of data using Moku:Pro~\cite{liquidinstruments}, a commercial off-the-shelf system, and implement offline triggering based on a matched (optimal) filter triggering algorithm~\cite{Gatti:1986cw, golwala}. In Fig.~\ref{fig:pulse_si}, we show an example of the matched filter applied to raw data. By working with continuous time streams of data, we are able to easily quantify our triggering efficiency and live times for a variety of analysis strategies.

This validated readout platform enables high-precision measurements on SPLENDOR target materials and can be used to investigate both ionization efficiency and event discrimination in low-background environments. As our initial silicon-based calibration is now complete, we are in the process of calibrating the \Eu-based prototype to begin science-driven detector runs.

Looking forward towards scalability, our detector housing design is also being modified to enable multi-channel readout with several target crystals.

Additionally, we are exploring upgrades to the cryo-HEMT amplifier that will enable sub-electron charge resolution in this same modular architecture. We plan to leverage the sensitivity of charge amplifiers based on cavity-embedded Cooper-pair transistors (cCPTs)~\cite{cCPT}. This work will be the subject of a separate publication.

Dark matter sensitivity projections based on these future improvements are discussed in Sec.\,\ref{sec:Sensitivity}.

\section{Sensitivity Projections}
\label{sec:Sensitivity}

\subsection{Event Rate Predictions}

The SPLENDOR experimental program aims to probe models in which dark matter interactions induce perturbations in the electron number density of the target material. This can occur when dark matter particles scatter off of valence electrons in the target, or when they are absorbed by it (if bosonic), transferring an energy $\omega$ and momentum $\vec{q}$ to excite charge carriers in the semiconducting medium. For these models, the key physical property of the target material that controls its response to dark matter interactions is the dielectric loss function, $\mathcal{W}(\omega,\vec{q\,})$ \cite{Kahn_2022,Knapen:2021run,Geilhufe_2020}. In the kinematic regime relevant for absorption of (vector) dark matter, where $|\vec{q}\,|\ll\omega$, the loss function is given by the average over polarizations, $\mathcal{W}(\omega)=-\frac{1}{3}\sum_{i=1}^3\text{Im}[1/\epsilon_{ii}(\omega,\vec{q}\approx 0)]$, where $\epsilon_{\mu\nu}$ is the dielectric tensor \cite{Geilhufe_2020}. Conversely, in the kinematic regime relevant for dark matter scattering, where $|\vec{q}\,|\gg\omega$, the loss function is given by $\mathcal{W}(\omega,\vec{q\,})=-\text{Im}[1/\epsilon_{L}(\omega,\vec{q\,})]$, where the longitudinal component of the dielectric tensor, $\epsilon_{L}\equiv\epsilon_{00}$, is related to the electric susceptibility, $\chi$, by $\epsilon_{L}(\omega,\vec{q\,})=1-\frac{4\pi}{|\vec{q}|^2}\chi(\omega,\vec{q\,})$ \cite{Knapen:2021run}.

\begin{figure}
    \centering
    \includegraphics[width=1\linewidth]{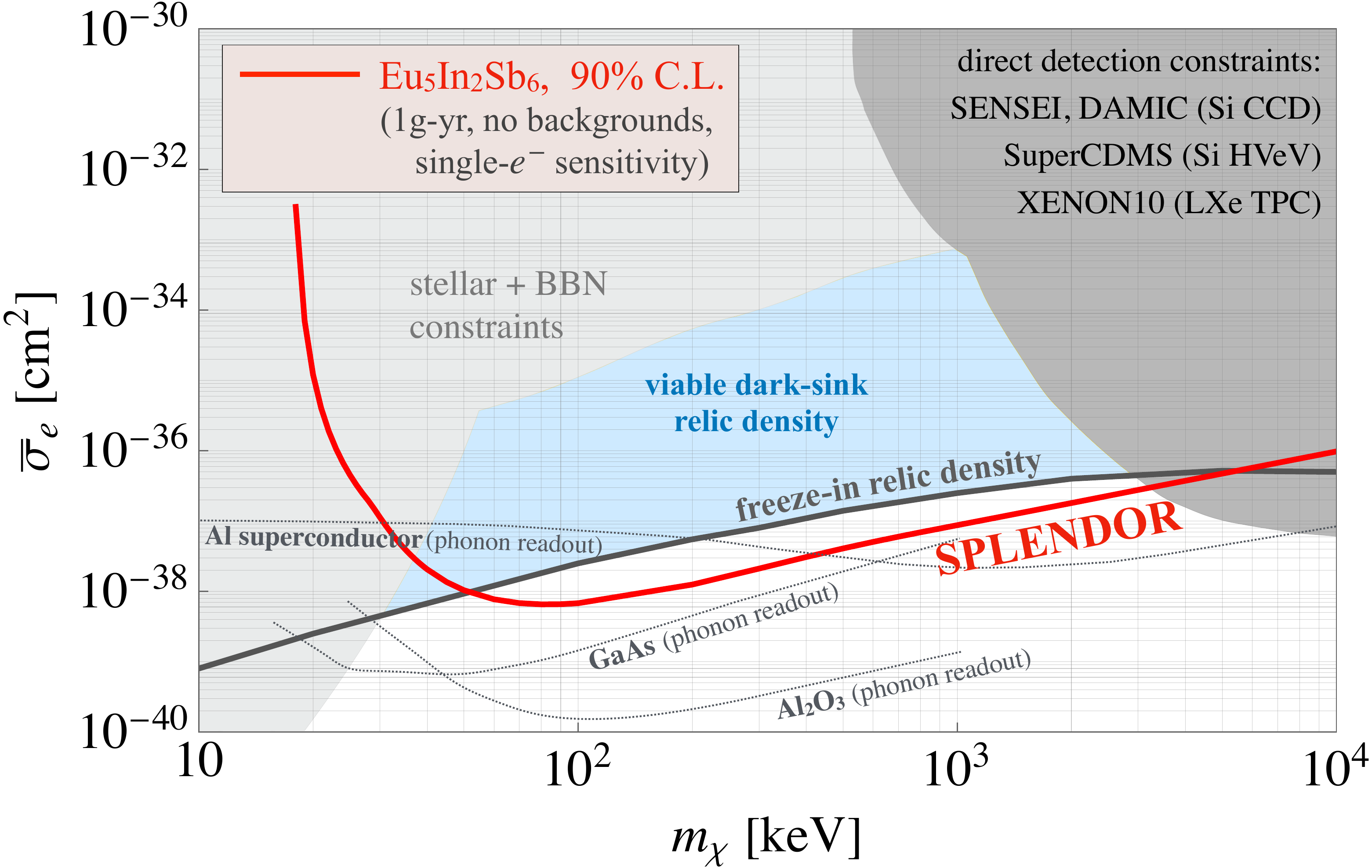}
    \caption{Red curve: the 90\% C.L.~sensitivity reach of \Eu~to dark matter-electron scattering in a model of freeze-in dark matter interacting through an ultralight mediator, where $\bar{\sigma}_e$ is the conventional dark matter-electron scattering cross-section (see Appx.\,\ref{appx:DMrate} for details), and $m_\chi$ is the dark matter mass. The solid gray curve indicates the parameter space for which the observed dark matter relic density is predicted in the standard freeze-in scenario \cite{Dvorkin:2019zdi}. The blue-shaded region above this curve can also be consistent with the observed dark matter relic abundance if a dark-sink mechanism to siphon off energy from dark matter to dark radiation is present in the early Universe \cite{Bhattiprolu:2024dmh}. The projected reach assumes 1\;g-yr of exposure, no backgrounds, and optimal detector performance. For comparison, the dashed gray curves indicate the projected sensitivity of low-energy phonon-detection schemes proposed in the literature \cite{Hochberg:2015fth,Hochberg_2016,polar1,Griffin:2018bjn}, under the same idealized assumptions for exposure, backgrounds, and detector performance. Gray-shaded regions indicate excluded parameter space from direct detection experiments \cite{Essig:2017kqs,EDELWEISS2020,DAMIC2019,Arnquist_2023,SENSEI:2023zdf,SuperCDMS:2024yiv,DAMIC-M:2025luv} (dark gray) and from combined bounds from BBN and stellar cooling \cite{Davidson:2000hf,Feng:2009hw,Vogel:2013raa,Essig:2015cda,An:2020bxd} (light gray).}
    \label{fig:dm_scattering}
\end{figure}

The dielectric loss function can be calculated from the material's electronic structure, which can in turn be evaluated from {\it ab initio} condensed matter theory methods given a material's crystal structure and constituent elements.
To obtain the electronic structure of \Eu, we used the all-electron full-potential linear-augmented plane wave (FP-LAPW) method \cite{blaha1990full}.
Fully relativistic effects were included to capture the strong orbital and spin anisotropies present in compounds with heavy elements, and exchange-correlation effects were treated within the Perdew, Burke, and Ernzerhof generalized gradient approximation (PBE-GGA) \cite{perdew1996generalized}.
For computing the electronic susceptibility, the collaboration developed a custom Green's function-based code that uses a tight-binding local orbital basis set. A scissor-shift correction
was applied to our results to match the indirect bandgap of \Eu~to the value inferred from our resistivity measurements, namely, $\Delta_\text{indirect}\approx60$\;meV. Once these corrections were applied, our calculated electronic band dispersions then predicted a direct gap for \Eu~of $\Delta_\text{direct}\approx80$\;meV. Assuming that at $\mathcal{O}(\text{mK})$ temperatures phonon-assisted electronic transitions are suppressed, the direct gap sets a lower bound on the dark matter mass range accessible through absorption. A more detailed discussion of our first-principles electronic structure calculations for \Eu~can be found in Appx.\,\ref{DFTtheory}.

Our numerical results for the dielectric loss function were then used as input in the calculation of the dark matter scattering and absorption rates in \Eu, including their energy spectrum and daily modulation. Details of our dark matter event rate calculations are provided in Appx.\,\ref{appx:DMrate}.

\subsection{Idealized Reach Projections}

\begin{figure}
    \centering
    \includegraphics[width=1\linewidth]{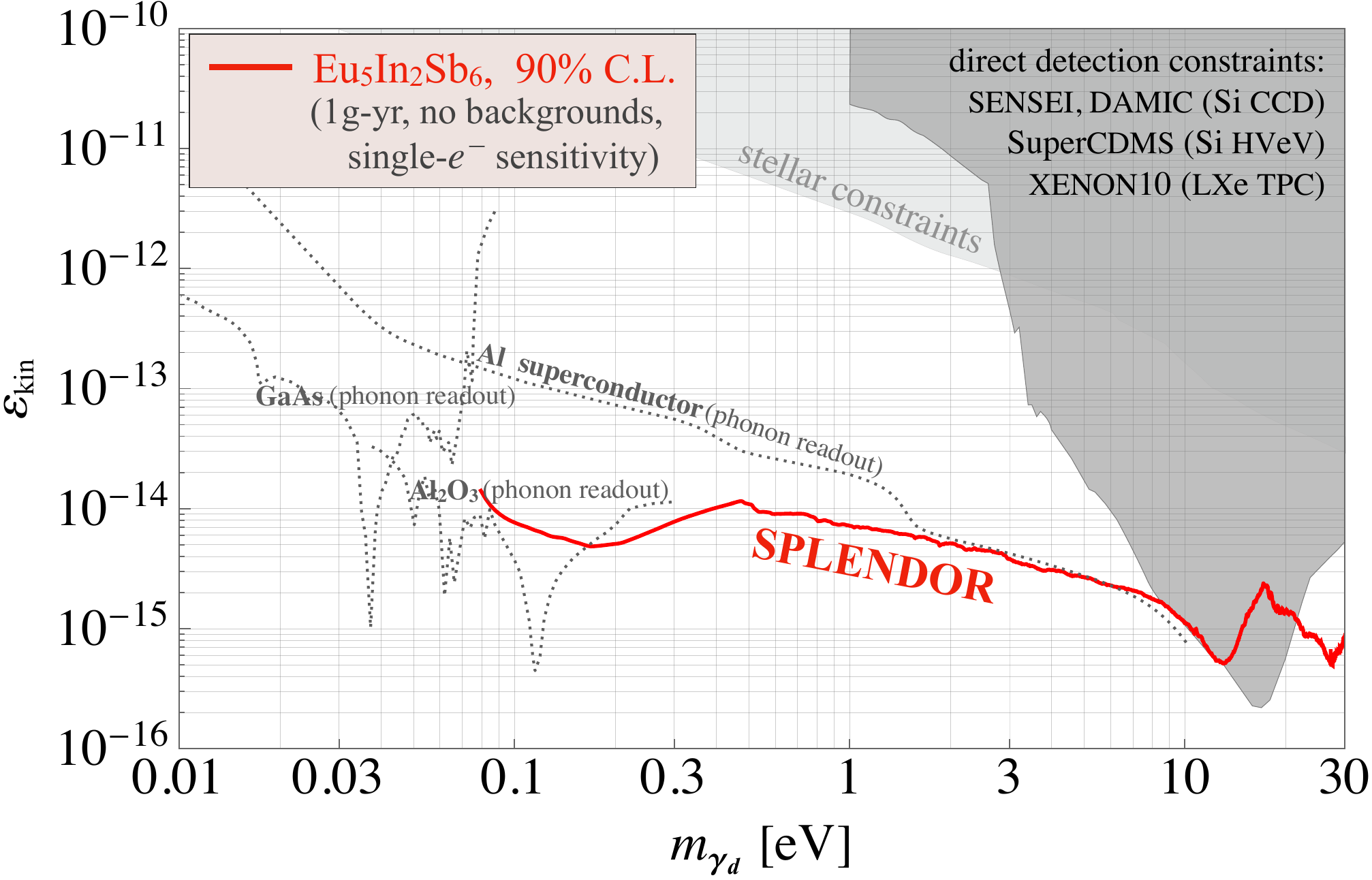}
    \caption{Red curve: the 90\% C.L.~sensitivity reach of \Eu~to dark-photon dark matter absorption in the parameter space of dark photon mass, $m_{\gamma_{\text{d}}}$, vs kinetic mixing parameter, $\varepsilon_\text{kin}$. The projected reach assumes 1\;g-yr of exposure, no backgrounds, and optimal detector performance. For comparison, the dashed gray curves indicate the projected sensitivity of low-energy-phonon detection schemes proposed in the literature \cite{Hochberg:2015fth,Hochberg_2016,polar1,Griffin:2018bjn}, under the same idealized assumptions for exposure, backgrounds, and detector performance. Gray-shaded regions indicate excluded parameter space from direct detection experiments \cite{An:2014twa,Bloch:2016sjj,EDELWEISS2020,DAMIC2019,Arnquist_2023,SENSEI:2023zdf,SuperCDMS:2024yiv,DAMIC-M:2025luv} (dark gray) and from bounds on stellar cooling \cite{Davidson:2000hf,Feng:2009hw,Vogel:2013raa,Essig:2015cda,An:2020bxd} (light gray).}
    \label{fig:DM_abs}
\end{figure}

In order to compare the sensitivity of SPLENDOR to alternative detector technologies that have been proposed based on low-energy phonon readout \cite{Hochberg:2015fth,Hochberg_2016,polar1,Griffin:2018bjn}, we first consider the dark matter reach of \Eu~under idealized assumptions for detector performance (i.e., 100\% detection efficiency down to single-electron events) in a background-free scenario. For this exercise, we considered two benchmark dark matter models commonly adopted in the literature:
\begin{enumerate}[wide, labelwidth=!, itemindent=!, labelindent=0pt, noitemsep]
\item \textbf{\textit{dark-photon dark matter}} ($\gamma_\text{d}$)
\cite{Nelson:2011sf,Fabbrichesi:2020wbt}, which can be absorbed by electrons in the target material, leading to the excitation of an electron-hole pair; and
\item \textbf{\textit{freeze-in dark matter}}
\cite{Gopalakrishna:2006kr,Hall:2009bx,Bernal:2017kxu,Dvorkin:2019zdi}, wherein dark matter particles ($\chi$) can excite electron-hole pairs in the target material through electronic scattering mediated by an ultralight dark boson.
\end{enumerate}

\begin{table*}
\includegraphics[width=1\textwidth]{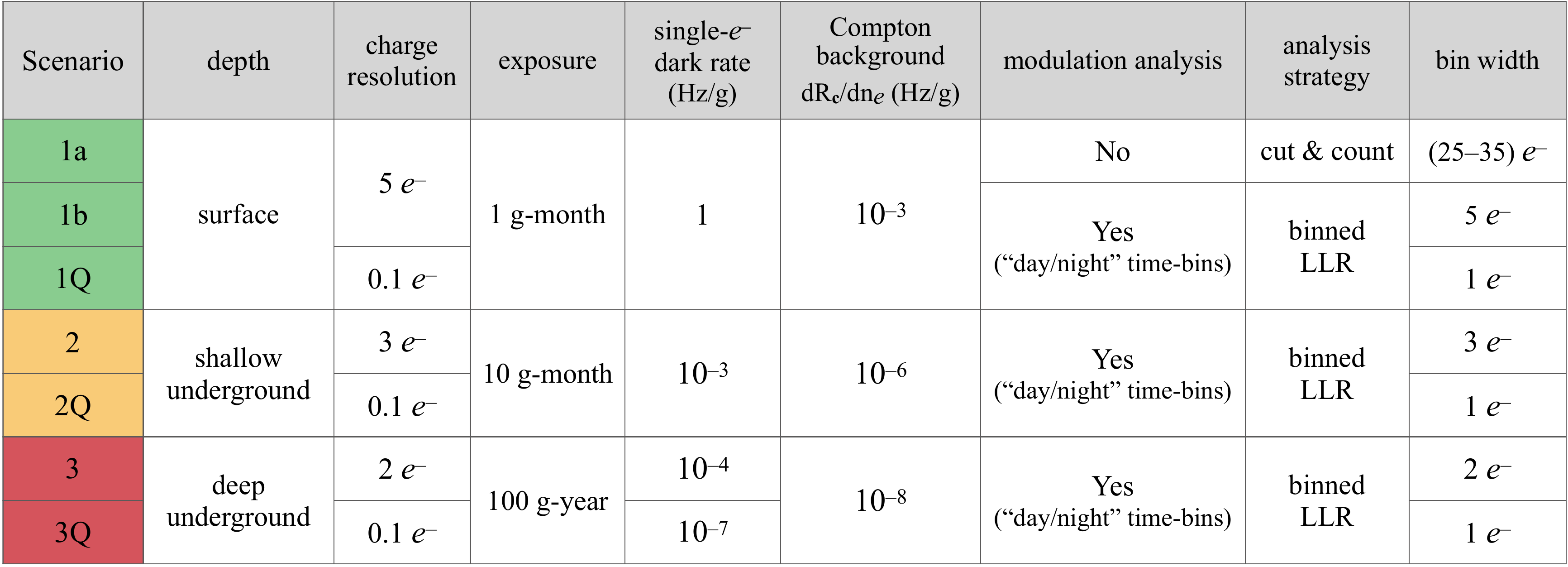}
\caption{Parameters characterizing the progression of SPLENDOR modular upgrades through three site-overburden scenarios. The baseline scenarios\;1a/b,\;2,\;3 are based on cryoHEMT readout, whereas the quantum-enabled scenarios\;1Q,\;2Q,\;3Q assume quantum cCPT readout with sub-electron charge resolution.}
\label{Scenarios_Table}
\end{table*}

\noindent Additionally, we also consider the freeze-in regime in the presence of a \textbf{\textit{dark sink}} \cite{Bhattiprolu:2023akk,Bhattiprolu:2024dmh}, which depletes the dark matter abundance in the early Universe through dark matter annihilation into dark radiation. In these models, stronger dark matter interactions with the Standard Model are needed to reproduce the correct dark matter relic abundance, effectively expanding the viable relic density parameter space to a range of cross sections above the standard freeze-in target.

The 90\% confidence level (C.L.) projected sensitivities of a \Eu-based SPLENDOR detector to models 1 and 2 above are shown in Figs.\;\ref{fig:dm_scattering}\;and\;\ref{fig:DM_abs},
assuming a fiducial exposure of \mbox{1~g–year}.
While these sensitivity projections are unrealistic, their comparison with the idealized reach of proposed low-energy phonon-readout technologies (also shown in Figs.\;\ref{fig:dm_scattering}\;and\;\ref{fig:DM_abs}) indicate the promise of SPLENDOR materials in the exploration of dark matter models in challenging low-mass regimes.

\subsection{\texorpdfstring{Sensitivity Studies:\\Current Prototype and Future Upgrades}{Sensitivity Studies: Current Prototype and Future Upgrades}}

We end this paper with sensitivity studies to map out the technology upgrades, scalability, and background mitigation that will be necessary in order to achieve sensitivity to the relic density target of sub-MeV freeze-in dark matter. In addition to the baseline \mbox{cryoHEMT} readout technology discussed in this paper, we also considered scaling up the detector fiducial volume from single-crystal prototype to multi-crystal arrays and multi-channel cryogenic readout, as well as integration with quantum sensing technology based on cCPT readout, which will enable significant sensitivity gains through improved charge resolution. We defined three benchmark scenarios for the depth of the experiment (surface, shallow underground, and deep underground), along with representative estimates for expected backgrounds based on existing runs by SENSEI~\cite{Barak_2020,SENSEI:2023zdf}, SuperCDMS HVeV~\cite{Amaral_2020}, and DAMIC-M~\cite{DAMIC-M:2025luv}. Table\;\ref{Scenarios_Table} summarizes the chosen parameters for the different scenarios we considered.

For the purposes of obtaining reach estimates, the parameter we used to distinguish between the baseline and quantum readout schemes was their charge resolution, which we assumed to be 2--5 electrons for cryoHEMTs vs 0.1 electrons for cCPTs.
Specifically, a resolution of $5e^-$ was chosen for the cryoHEMT readout scheme in scenario\;1 based on the current performance we were able to demonstrate with a passive reset (see discussion in Sec.\;\ref{sec:Detector}). Active reset will improve this resolution to approximately $3e^-$, which was the value we adopted for scenario\;2. Finally, parallel amplification will provide a further $\sqrt{2}$-improvement, from $3 e^{-}$ to $2 e^{-}$, which was the value we adopted for scenario\;3.

Concurrently, our collaboration is working on developing cCPT-based readout technology which is expected to achieve sub-electron resolution. Our choice of $0.1 e^{-}$ for the sensitivity projections below was a fiducial choice for illustration only; different numerical choices should have little effect on the expected sensitivity, as long as sub-electron resolution is achieved.

In addition to the overburden increase and associated background reduction in each scenario, we assumed increased exposures at each stage through heavier target masses and/or longer running times. The increase in payload, in particular, will be enabled by R\&D on cryogenic multi-channel readout.\footnote{Continued efforts in crystal synthesis and growth optimization might deliver single crystals heavier than $\sim$1\;g, which is the current level demonstrated by the collaboration. For the purposes of obtaining reach estimates, we conservatively assumed a mass of 1\;g for each crystal.}
For simplicity, for multi-crystal arrays we assumed that several single-channel amplifiers would run in parallel, with one channel per 1-g crystal.

\begin{figure*}
\includegraphics[width=1\linewidth]{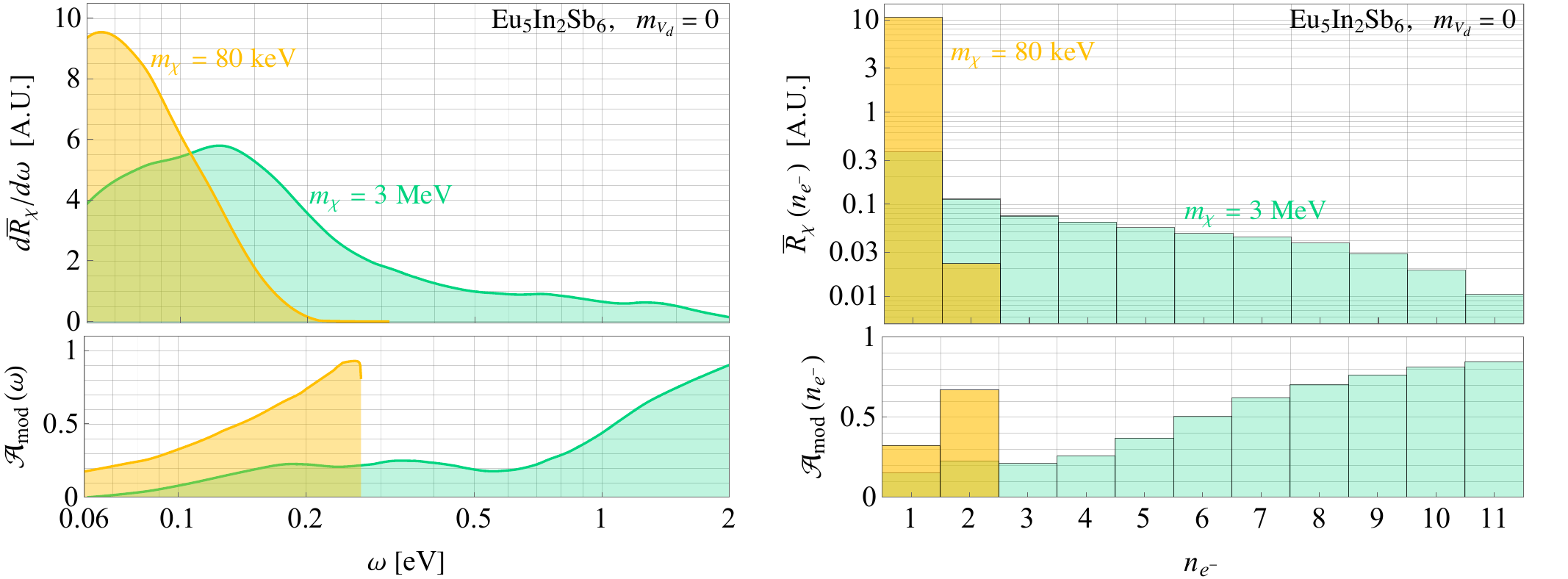}
\caption{\emph{Left, top panel:} The time-averaged differential energy spectrum of the signal rate, $d\overline{R}_\chi/d\omega$, for two dark matter benchmark examples at $m_\chi=80$\;keV and $m_\chi=3$\;MeV. \emph{Left, bottom panel:} the fractional modulation amplitude of the signal energy spectrum, $\mathcal{A}_\text{mod}(\omega)$, defined as the amplitude of the daily variation in $d{R}_\chi/d\omega$ divided by the daily-averaged rate, $d\overline{R}_\chi/d\omega$.
\emph{Right, top panel:} The time-averaged $n_{e^-}$ distribution of the signal rate, $\overline{R}_\chi(n_{e^-})$, for the same benchmarked $m_\chi$ examples. \emph{Right, bottom panel:} the fractional modulation amplitude of the signal $n_{e^-}$ distribution, $\mathcal{A}_\text{mod}(n_{e^-})$, defined in an analogous manner as $\mathcal{A}_\text{mod}(\omega)$.
}
\label{BenchmarkSpectra}
\end{figure*}

Finally, for the purposes of background modeling, we considered three general types of backgrounds: dark currents, Compton scattering, and amplifier noise.

\vspace{4pt}
\noindent\textit{\textbf{Dark currents}} were benchmarked by four parameters: the single-electron dark rate $R^\text{d}_{1e^-}$, the total exposure time $t_\text{exp}$, the amplifier's pulse integration time $t_\text{int}$ (which, for concreteness, was taken to be 100\;ms for all scenarios), and the number of amplifier channels $n_\text{channels}$.
These parameters determine the number of timing bins per channel, $n_\text{bins}=t_\text{exp}/t_\text{int}$, and the expected dark count in each timing bin, $R^\text{d}_{1e^-\!\!}\times t_\text{int}$. From Poissonian statistics it then follows that the total number of timing bins with an expected pileup dark count of $n_{e^-}$ electrons is given by:
\begin{equation}
N_\text{dark}(n_{e^-})=P_\text{Poisson\!}\big[n_{e^-}\big|\,R^\text{d}_{1e^-\,}t_\text{int\,}\big] \times n_\text{bins} \times n_\text{channels},\nonumber
\end{equation}
where $P_\text{Poisson}\big[\lambda\,|\,\mu\big]=\mu^\lambda\,e^{-\mu}/\lambda!$~~is the Poisson probability for $\lambda$ occurrences given a mean expectation of $\mu$. Based on the performance of Si detectors, we benchmarked our scenarios on single-electron dark rates of $R^\text{d}_{1e^-}=1$\;Hz/g at the surface \cite{Amaral_2020,Du:2020ldo}, $R^\text{d}_{1e^-}=10^{-3}$\;Hz/g at a shallow underground site with some shielding \cite{Barak_2020}, $R^\text{d}_{1e^-}=10^{-4}$\;Hz/g with a shielded detector deep underground \cite{SENSEI:2023zdf,SENSEI:2024yyt,DAMIC-M:2025luv}, and $R^\text{d}_{1e^-}=10^{-7}$\;Hz/g as the ultimate target needed in order to achieve sensitivity to relic-density cross sections of our freeze-in dark matter benchmark model.

\vspace{4pt}
\noindent\textit{\textbf{Compton scattering}} of environmental photons off of electrons in the target crystal can produce
ionization events that constitute the dominant background in the tail of the $n_{e^-}$ distribution at high-exposures and ultra-low background environments \cite{Essig:2023wrl}. We modeled this background as a flat distribution in $n_{e^-}$ with rates of $dR_\text{c}/dn_{e^-}=10^{-3}$~Hz/g at the surface, $dR_\text{c}/dn_{e^-}=10^{-6}$~Hz/g at a shallow underground site \cite{Barak_2020}, and $dR_\text{c}/dn_{e^-}=10^{-8}$~Hz/g at a deep underground site \cite{SENSEI:2023zdf}.

\begin{figure*}
\includegraphics[width=1\linewidth]{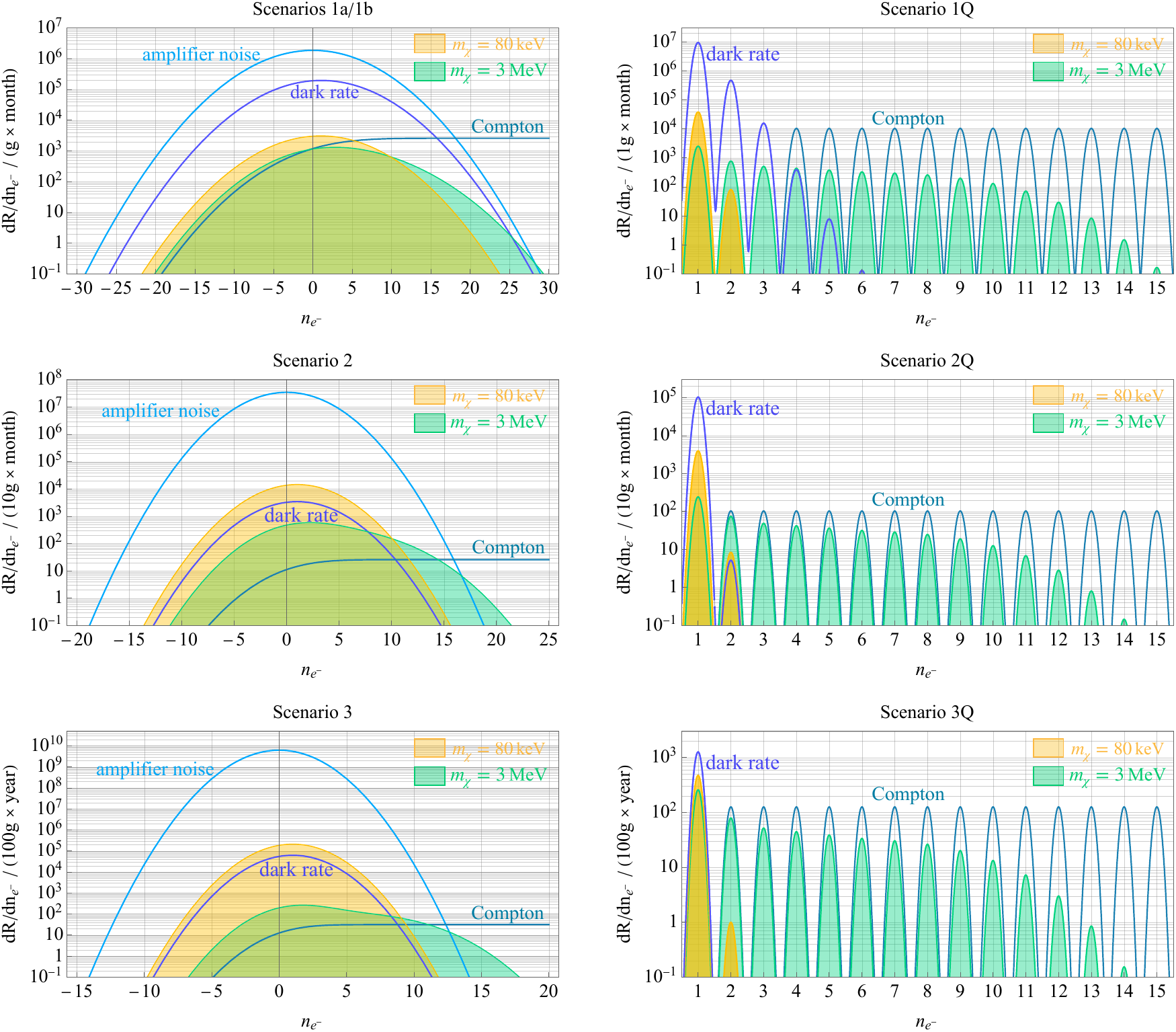}
\caption{Differential $n_{e^-}$ distributions for backgrounds (solid blue curves) and benchmark signal examples (shaded areas) for each of the scenarios summarized in Table~\ref{Scenarios_Table}. The cross section chosen for each signal example coincides with the 90\% C.L. expected reach in the respective scenario. Scenarios in the left column assume cryoHEMT readout with charge resolution of $5e^-$ (top), $3e^-$ (middle), and $2e^-$ (bottom). Given the significant amplifier noise in these scenarios, daily modulation analyses with low-$n_{e^-}$ thresholds are critical to tease out signals from very light dark matter ($m_\chi\lesssim 300$ keV). Scenarios in the right column assume quantum cCPT readout with sub-$e^-$ resolution, and are dominated by dark rates in the low-$n_{e^-\!\!}$ bins and Compton scattering backgrounds in the high-$n_{e^-\!\!}$ bins. The effects of these backgrounds are also mitigated by daily modulation analyses.}
\label{DiffSpectra}
\end{figure*}

\vspace{4pt}
\noindent\textit{\textbf{Amplifier noise}} was modeled by taking the ``truth-level'' $n_{e^-}$ distributions described above and smearing them with a Gaussian distribution with standard deviation $\sigma$ given by the amplifier's charge resolution. In particular, in the $n_{e^-}$ distributions shown in Fig.~\ref{DiffSpectra} (which we shall discuss shortly), the curves labeled ``amplifier noise'' refer to the noise-only timing bins (i.e., the timing bins with zero ``truth-level'' dark counts smeared by the charge resolution).

For all readout schemes, we assumed 100\% detection efficiency in the signal region. (In Si-based CCDs and HVeV detectors, detection efficiencies are indeed within the $\sim 90-100\%$ range.) Strictly speaking, the quoted exposures in our reach projections should be interpreted as being efficiency-corrected. 

For the signal modeling, we took as input the prediction for the differential signal rate, $dR_\chi/d\omega$, as a function of the energy transfer to the detector target material (\Eua). These predictions were based on our numerical modeling of \Eua's dielectric response to dark matter energy-momentum transfers described in Appx.\;\ref{DFTtheory}.

A proper determination of how a given energy transfer $\omega$ translates into a number of conduction-band electrons $n_{e^-}$ created in the target material requires the development of data-driven Monte Carlo simulations with input from charge transport measurements and detector calibration. For the purposes of obtaining approximate reach estimates, we based our modeling on a simplified, albeit roughly universal heuristic description of the ionization yield in semiconductors. Specifically, for energy transfers much higher than the material's band gap, $\omega\gg E_\text{gap}$, we assume that the average number of electron-hole pairs created in the material is given by $\langle n_{eh}\rangle = \frac{1}{3}\,\omega/E_\text{gap}$, where the factor of 1/3 can be roughly understood as accounting for the proportion of the total deposited energy that ends up in ionization rather than phonons. (Unless the detector material has exceptionally weak or strong electron-phonon coupling, this factor is generically expected to be of $\mathcal{O}(1)$.)

With this assumption, we can adopt a simplified
probability $p_{n_e\!}(\omega)$ for the production of $n_{e^-}$ electrons given an energy transfer $\omega$~\cite{Ramanathan:2020fwm},
\begin{equation}
    p_{n_e\!}(\omega) = \left\{ \begin{array}{cl}
        0 & \mathrm{for} \ \omega<E_\text{gap} \vspace{7pt} \\
        ~~~~\,\delta_{1,n_e} & \mathrm{for} \ E_\text{gap}\leq\omega<2E_\text{gap}\vspace{5pt} \\
        \frac{1}{N}\,\text{exp}{\left[{-\frac{\left(n_e-\langle n_{eh}\rangle\right)^2}{2F\langle n_{eh}\rangle}}\right]} & \mathrm{for} \ \omega\geq2E_\text{gap} \\
    \end{array}
    \right.\nonumber
\end{equation}
where, above, $N={\sqrt{2\pi F\langle n_{eh}\rangle}}$, and $F$ is the Fano factor, a material-specific parameter for which we adopt a value similar to Si of $F=0.15$.

The energy spectrum of the dark matter rate, $dR_\chi/d\omega$, can then be converted to a  $n_{e^-}$ distribution through the following convolution:
\begin{align}
R_\chi(n_{e^-})=\int d\omega\,\frac{dR_\chi}{d\omega}~p_{n_e\!}(\omega).\nonumber
\end{align}
Finally, this ``truth-level prediction'' for the dark matter-generated charge yield in the material was smeared by the amplifier's charge resolution in the same way as was done for the backgrounds.

The top panels in Fig.\;\ref{BenchmarkSpectra} show, for two signal examples at low ($m_\chi=80$~keV) and high ($m_\chi=3$~MeV) dark matter masses, the differential energy spectrum $dR_\chi/d\omega$ and resulting truth-level $n_{e^-}$ distribution obtained by the procedure just outlined. Fig.\;\ref{DiffSpectra} shows, for each of the scenarios summarized in Table~\ref{Scenarios_Table}, the resolution-smeared $n_{e^-}$ distributions for backgrounds as well as for the two signal examples at $m_\chi=80$~keV and $m_\chi=3$~MeV. The cross sections for these signal examples were chosen to coincide with the projected 90\% C.L. exclusion reach under the assumptions of each scenario (to be discussed shortly).

It is clear from these distributions that the amplifier noise is the dominant low-energy background for the scenarios based on cryoHEMT readout, and that any high-$n_{e^-}$ threshold set to eliminate this background would also eliminate any potential signal at low dark matter masses. The purpose of scenario\;1a was to illustrate how high-$n_{e^-}$ thresholds would affect the reach in dark matter parameter space: its signal region was chosen to be $n_{e^-}\geq25e^-$~\cite{Mancuso:2017ffg}, and a straightforward cut-and-count analysis (i.e., ignoring any spectral information) was performed to determine the sensitivity reach.

Such high-threshold strategies, however, are suboptimal, since they ignore a key property of \Eu, namely, its anisotropic response to the incoming flux of dark matter particles and resulting modulation of the signal rate over a sidereal day.
The bottom panels in Fig.\;\ref{BenchmarkSpectra} provide an illustration of the magnitude of this daily modulation for the two benchmark examples at $m_\chi=80$~keV and $m_\chi=3$~MeV. Specifically, these panels display the fractional modulation amplitudes $\mathcal{A}_\text{mod}$ of the signal distributions in both $\omega$ and $n_{e^-}$,
where $\mathcal{A}_\text{mod}(n_{e^-})$ is defined as
\begin{equation}
\mathcal{A}_\text{mod}(n_{e^-})~\equiv~\frac{1}{2}\,\frac{R_\chi(n_{e^-},\,t_\text{max})-R_\chi(n_{e^-},\,t_\text{min})}{\overline{R}_\chi(n_{e^-})},\nonumber
\end{equation}
with $t_\text{max}$ ($t_\text{min}$) being the time of day at which the rate reaches its maximum (minimum), and $\overline{R}_\chi(n_{e^-})$
being the sidereal daily average of $R_\chi(n_{e^-},\,t)$. $\mathcal{A}_\text{mod}(\omega)$ is defined in an analogous manner.

The data analysis can be designed to harness the expected signal modulation in order to discriminate the dark matter signal from (non-modulating) backgrounds\footnote{Terrestrial backgrounds are not expected to exhibit a daily modulation, with the exception of the cosmic ray muon background
 \cite{COSINE:2020jml}. The solar diurnal modulation of the muon rate, with an amplitude of 0.1-0.6\%, is correlated with atmospheric temperature, and is therefore out of phase with the dark matter signal. The sidereal modulation of the muon rate for energies below 500~GeV (relevant for shallow underground sites) is due to solar wind effects. For energies in the TeV range or above (relevant for deep underground sites), a modulation amplitude of $\sim$$10^{-3}$ with a phase uncorrelated with the direction of the Cygnus constellation was reported by the MACRO collaboration \cite{MACRO:2002qsl}.}. This, in particular, enables the setting of much lower $n_{e^-}$ thresholds in order to increase the signal acceptance at lower dark matter masses. Although the amplifier noise increases exponentially as the $n_{e^-}$ threshold is lowered, it can be subtracted via sidereal daily averaging of the observed rate, along with all other background contributions that are constant over time, without any precise determination of their shape or overall normalization, hence eliminating significant sources of systematic uncertainties. The residual, background-subtracted data, can then be analyzed in order to identify (or exclude) a modulating dark matter signal with a known expected phase.

\begin{figure*}
\centering
\includegraphics[width=0.75\linewidth]{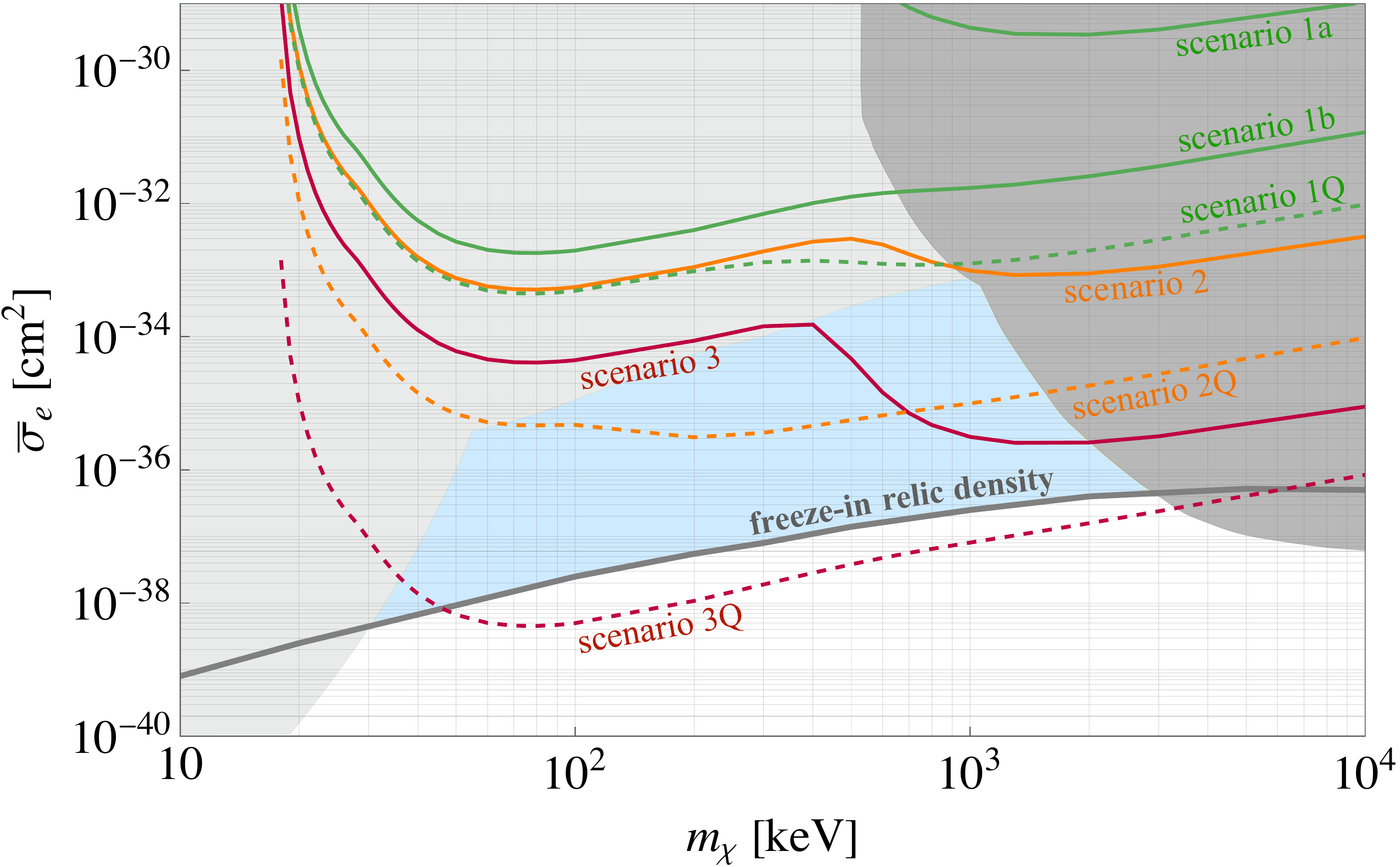}
\caption{The 90\% C.L. sensitivity reach estimates for SPLENDOR detector upgrades under different scenarios summarized in Table~\ref{Scenarios_Table}. The baseline scenarios\;1a/b,\;2,\;3 assume cryoHEMT charge readout, whereas the quantum-enabled scenarios\;1Q,\;2Q,\;3Q assume quantum cCPT readout with sub-electron charge resolution.
The blue-shaded region denotes viable parameter space in dark-sink models \cite{Bhattiprolu:2024dmh}, and the gray-shaded regions are excluded by indirect astrophysical and cosmological constraints (light gray) \cite{Essig:2015cda,Davidson:2000hf,Feng:2009hw}, and by direct detection experiments (dark gray, upper right) \cite{EDELWEISS2020,DAMIC2019,Arnquist_2023,SENSEI:2023zdf,SuperCDMS:2024yiv,DAMIC-M:2025luv}.
Scenarios\;1b, \;1Q,\;and\;2, which are realistic within a $\sim$\,5-year timescale, can set the first terrestrial limits on dark matter-electron scattering at masses below 0.5 MeV. Scenario\;2Q can probe unexplored dark matter parameter space and set world-leading limits in the sub-MeV mass range. The end goal of the SPLENDOR experimental program is to probe the freeze-in relic density target denoted by the dark gray curve \cite{Dvorkin:2019zdi}, and will require the technology, exposure, background levels outlined in scenario\;3Q.}
\label{ReachPlot}
\end{figure*}

With this in mind, we adopted a simplified daily modulation analysis strategy to take advantage of the target material anisotropy. Specifically, after setting
the $n_{e^-}$ threshold to $0.5 e^-$ and binning the $n_{e^-}$ spectra in each scenario according to the charge resolution (see Table\;\ref{Scenarios_Table} for details), we further split the count in each charge-bin into two time-bins, each integrating the observed rate over a fixed period within the sidereal day. We label these time-bins ``dark-matter day'' and ``dark-matter night.'' The sidereal time marking the end of ``dark-matter day'' and beginning of ``dark-matter night'' was chosen such that the difference in the expected signal count between day and night was maximized. Given the expected signal across these $(n_{e^-},\,\Delta t)$ bins, we then performed a binned Log-Likelihood Ratio (LLR) analysis \cite{Cowan:1998ji} to obtain the value of the signal cross section that could be excluded at 90\% confidence level (C.L.). 

Fig.\,\ref{ReachPlot} shows our results for the sensitivity reach of each of the seven scenarios we considered in the parameter space of dark matter–electron scattering cross section, $\bar\sigma_e$, versus dark matter mass, $m_\chi$. They allow us to make three important observations:

\begin{enumerate}[wide, labelwidth=!, itemindent=!, labelindent=0pt,itemsep=0pt, topsep=3pt]
\item Comparing the resulting reach in scenario 1 between the high-threshold analysis (1a) and the low-threshold modulation strategy (1b), the advantage of the latter is clear: the reach is not only improved towards lower cross sections, but it also extends into the $(0.01-0.5)$\;MeV dark matter mass range, which is beyond any detector technology that has been demonstrated to date. The critical innovation enabling SPLENDOR's potential to probe this low-mass range is the use of a highly anisotropic, narrow-gap target material.
\item As the resolution of the cryoHEMT amplifier improves, the tail of the amplifier noise retreats to lower $n_{e^-}$ values, improving the experimental sensitivity in the heavier dark matter mass range where the signal spectrum is harder.
This explains the peculiar shape of the reach curves for scenarios 2 and 3, for which the analysis strategy included spectral information.
\item Comparing the sensitivities of scenarios 3 and 3Q, the need for upgrading the baseline amplifier technology to quantum readout is clear: only sub-electron resolution will enable SPLENDOR to achieve sensitivity to the standard freeze-in relic density target in the sub-MeV mass range.
\end{enumerate}

\section{Conclusion and Outlook}
\label{sec:Conclusion}

We have presented the conceptual design and current status of the SPLENDOR experimental program, including the selection and characterization of \Eu~as the target material for the initial SPLENDOR detector prototype, as well as the design and calibration results for the cryoHEMT-based charge amplifier architecture in Si test samples. 

Our results provide a demonstration of the modular SPLENDOR detector platform and position it to be the first experiment capable of advancing the light dark matter frontier with narrow-gap quantum materials.

Ongoing work includes calibration of the final amplifier prototype with a \Eu~sample, background characterization under varied operating conditions, and refinement of analysis pipelines for low-threshold and modulated signal searches. Looking ahead, the SPLENDOR architecture allows for straightforward payload scalability through multi-crystal arrays and multi-channel cryogenic readout.
Planned upgrades include implementation of RF multiplexing and integration of cavity-coupled Cooper pair transistor amplifiers with sub-electron resolution. 
Our sensitivity studies indicate that these future upgrades offer a realistic pathway towards excluding the relic density target of athermally produced relic dark matter in the sub-MeV mass range.

SPLENDOR's ultra-low energy ionization thresholds are enabled by low-noise, cryogenic charge readout combined with the narrow bandgap and anisotropic electronic structure of the target material, leading to a daily modulation of the dark matter event rate and enabling the subtraction of unmodulated backgrounds. With these combined innovations, SPLENDOR offers a flexible and scalable path forward for rare-event detection, capable of probing unexplored parameter space of dark matter-electron interaction strengths in the sub-MeV mass regime.

\section{Acknowledgments}
We wish to thank Dan Baxter, Tongyan Lin, Michael McGuire, David Meyerhofer, and Boris Shklovskii for helpful feedback.
This work was supported by the Laboratory Directed Research and Development program of Los Alamos National Laboratory under project numbers 20220135DR, 20220252ER, 20230777PRD1, and 20230782PRD1.
Los Alamos National Laboratory is operated by Triad National Security, LLC, for the National Nuclear Security Administration of the U.S. Department of Energy (DOE) under contract number 89233218CNA000001.
NAK was supported in part by the DOE Early Career Research Program (ECRP) under FWP\;100872. JA was supported in part by a Kavli Institute for Particle Astrophysics and Cosmology Chabollah Fellowship.
YK was supported in part by DOE grant DE-SC0015655.
DSMA and MLG were supported in part by the DOE Office of Science High Energy Physics under contract number DE-AC52-06NA25396.
FTIR measurements were supported by the DOE Office of Basic Energy Sciences, Division of Materials Science and Engineering.
Laser etching of \Eu{} crystals was performed using a femtosecond laser direct writing system at the Center for Integrated Nanotechnologies, a DOE Office of Science User Facility \&  Nanoscale Science Research Center.
The computational work was supported by the Center for Integrated Nanotechnologies in partnership with the LANL Institutional Computing Program for computational resources. Additional computations were performed at the National Energy Research Scientific Computing Center (NERSC), supported by NERSC award number ERCAP0028014. NERSC is a DOE Office of Science User Facility located at Lawrence Berkeley National Laboratory, operated under contract number DE-AC02-05CH11231.

\bibliography{refs}

\appendix 
\section{\texorpdfstring{\\Synthesis and Characterization of \Eu}{Synthesis and Characterization of Eu5In2Sb6}}

\subsection{Synthesis and Growth Optimization}
\label{Appx:Synthesis}

To reveal the intrinsic properties of semiconductors and certify their readiness for applications, the synthesis of high-quality single crystals is indispensable. For instance, the precise synthesis of single-crystalline silicon achieved decades ago underpinned recent progress in the semiconducting industry, ranging from dense computer chips to solar cells. Importantly, electrical transport properties in polycrystals may be drastically altered by grain boundaries, hindering the determination of a semiconducting gap, affecting charge collection, and averaging-out any directional dependence of the material to the incoming dark matter flux. In fact, the temperature-dependent electrical resistivity of \Eu~polycrystals was reported in 2016 to show a weak temperature dependence, in agreement with the behavior of a poor metal or a heavily-doped semiconductor \cite{Subbarao2016}. In 2020, members of our collaboration reported the synthesis of \Eu~single crystals grown through a combined Indium/Antimony flux technique \cite{Rosa2019,rosa2020colossal}.

More recently, our collaboration has optimized the synthesis protocol for \Eu~using the self-flux technique with an In-Sb eutectic flux.
The raw starting materials were Eu ingot (Ames, 99.95$\%$), In ingot (Alfa Aesar, 99.9995$\%$) and Sb shot (Alfa Aesar, 1-3 mm, 99.9999$\%$). Numerous attempts were made to maximize the size and quality of the single crystals by systematically altering the relative concentrations of In and Sb. The current optimized molar ratio was found to be 5:35:17.5~=~Eu:In:Sb. This ratio produced the largest crystals with the desired transport properties of a clean semiconducting ground state. 

The Eu ingots are air sensitive and thus were stored, cut, and weighed within an argon-filled glovebox with oxygen and water levels kept below 1 ppm. All the reactants were loaded into an alumina crucible that was then placed within a large fused quartz ampoule with a 19 mm inner diameter. The ampoule was evacuated to approximately 30 mTorr and backfilled with argon gas before being flame-sealed under vacuum with a hydrogen-oxygen torch. The sealed ampoule was heated in a box furnace at a rate of 70${\degree}$C/hr to 1050${\degree}$C and held there for 10 hours to allow the Eu to dissolve into the melt. The furnace was then allowed to cool at a rate of 2${\degree}$C/hr to 350${\degree}$C. The growth was finally brought back up to 500${\degree}$C to allow the flux to completely melt and was then spun in a centrifuge to decant any excess flux. The resulting single crystals were rod-like with well-formed facets and lengths of up to 1 cm (see Fig.\,\ref{fig:Eu526structure}c in Sec.\;\ref{sec:Materials}). The crystals were air-stable for several months, but degradation was observed if crystals were left in air for longer periods of time. Therefore, crystals were stored under vacuum or in an inert atmosphere.

\subsection{Bandgap Extraction from Electrical Resistance Measurements}
\label{appx:RvTdiscussion}

At temperatures of $\sim$\,20\;K and above, the electrical resistance of our \Eu{} samples is dominated by thermal excitation of electrons across the bandgap, and, therefore, a reliable inference of a thermally activated transport gap can be made. At much lower temperatures, especially in the sub-K range, other competing effects become important in determining the resistance, such as residual impurities and disorder in the sample \cite{10.1063/10.0034343}, as well as ionization induced by external backgrounds. Hence, the sub-K range of our electrical resistance data cannot be used to reliably infer a transport gap from thermally activated behavior \cite{Shklovskii_book}. 

In this paper, we work under the assumption that the bandgap of \Eu{} in the sub-K antiferromagnetic phase is not significantly different from its counterpart in the paramagnetic phase above 14\;K. While, in general, the appearance of different broken symmetry phases as the temperature is lowered can alter the electronic band structure in many ways, in \Eu{} we expect that the low-temperature bandgap should be relatively insensitive to the magnetic orderings that occur below $T_{N1}=14$\;K and $T_{N2}=7$\;K. Firstly, the heat capacity goes smoothly to zero below $T_{N2}$, indicating the presence of a finite gap in the system. Secondly, the Eu $f$-states are not expected to overlap significantly with the electronic states close to the Fermi level, since they tend to be localized and minimally hybridized. Indeed, theoretical band structure calculations in \cite{rosa2020colossal} confirm this intuition by showing that the gap forms in the $s$-, $p$-, and $d$-states of the system, while $f$-states are centered around $\sim$\,1.5\;eV below the Fermi level. Supplementary Fig.\;S9 of \cite{rosa2020colossal} also indicates that the low-energy bands near the Fermi level in the antiferromagnetic phases remain virtually identical to those in the paramagnetic phase.

\subsection{AC Hall Measurements}
\label{Appx:ACHall}

\begin{figure}
    \centering
    \includegraphics[width=0.95\linewidth]{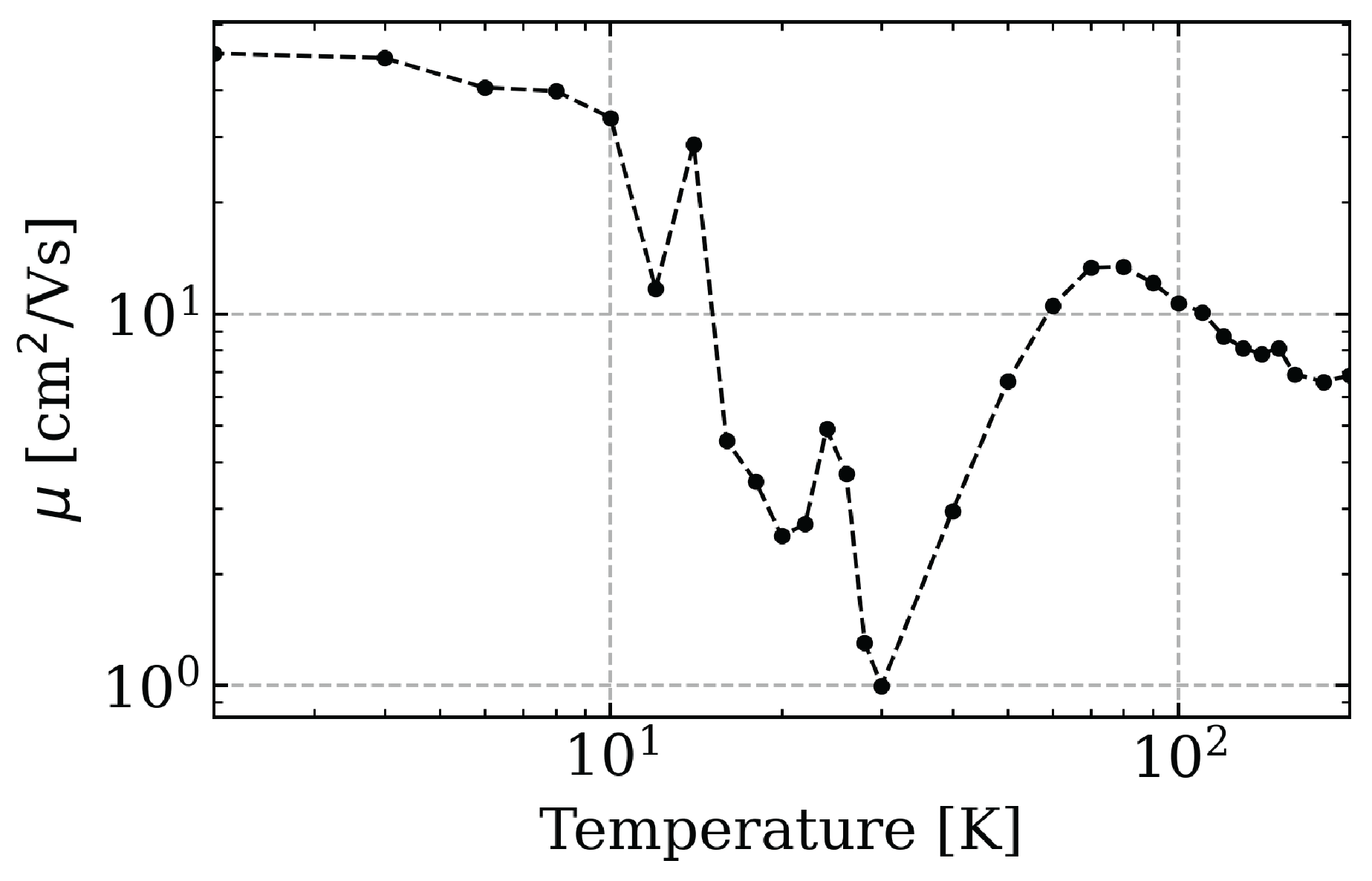}
    \caption{\label{fig:mobility} Temperature dependence of the carrier mobility $\mu$ in Eu$_5$In$_2$Sb$_6$ inferred from measurements of the AC Hall resistivity.}
\end{figure}

\begin{figure}
    \centering
    \includegraphics[width=\columnwidth]{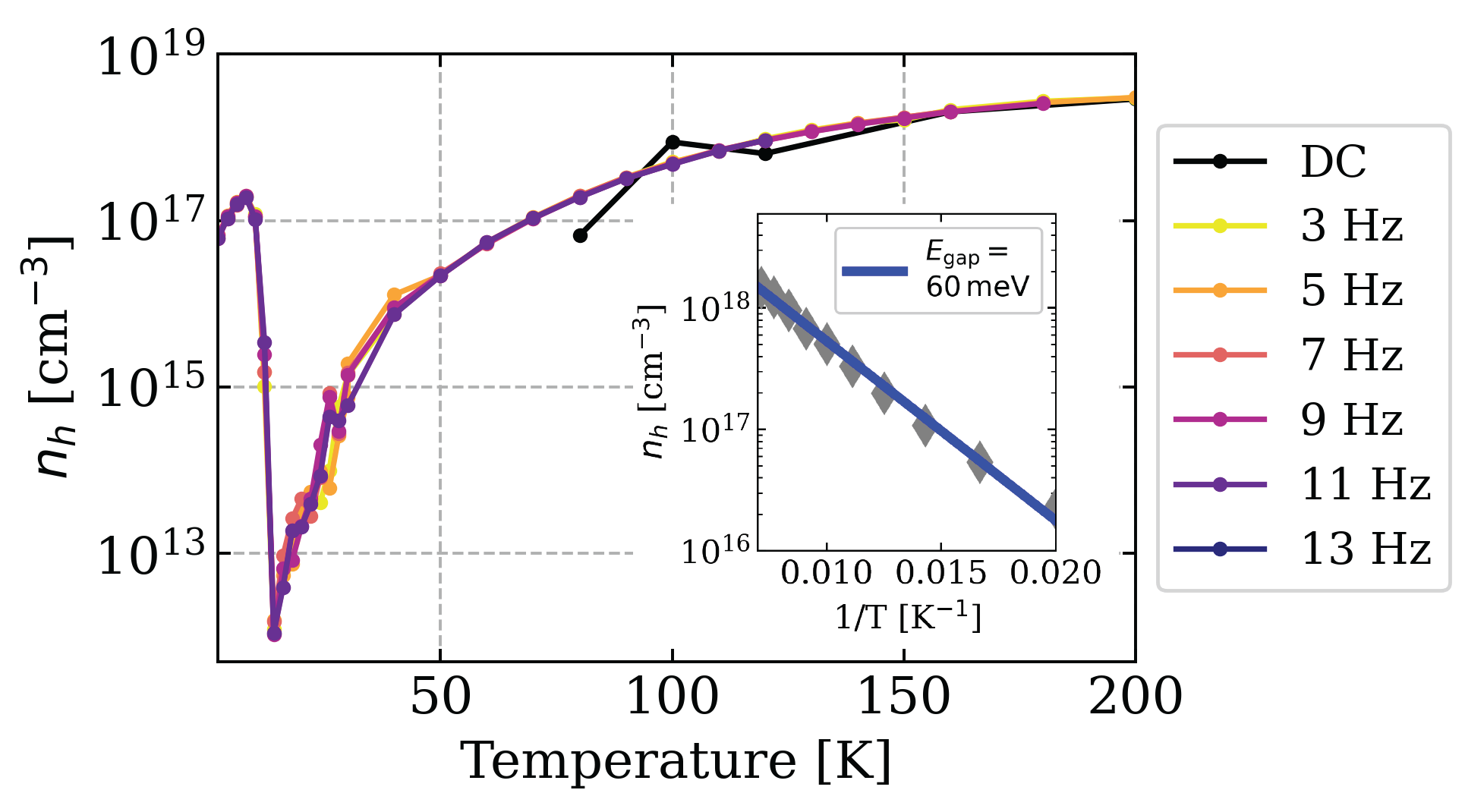}
    \caption{\label{fig:526_ACHALL} Temperature dependence of the carrier concentration $n_h$ in Eu$_5$In$_2$Sb$_6$ determined at various frequencies using the AC Hall method. The carrier concentration determined from traditional DC Hall measurements is shown in black. The inset shows the expected exponential dependence $n_h\propto\text{Exp}[-E_\text{gap}/(2\,k_BT)]$ of a narrow-gap semiconductor with an energy gap $E_\text{gap}=60$\;meV.}
\end{figure}

To characterize the narrow-gap transport properties of Eu$_5$In$_2$Sb$_6$, we measured the temperature dependence of the carrier mobility and concentration using the AC Hall effect. This method was chosen over the traditional DC Hall effect as the offset voltage from imperfect contact geometry becomes prohibitively large at low temperatures, when the longitudinal resistivity, $\rho_{xx}$, becomes much larger than the transverse resistivity, $\rho_{yx}$. This measurement technique is able to separate the spurious contributions to the Hall voltage in frequency space, such that the Hall coefficient $R_H$ can be reliably measured using a low-frequency AC magnetic field and lock-in techniques.

Our results for the carrier mobility $\mu$ of \Eu~are shown in Fig.\,\ref{fig:mobility}, where $\mu$ was inferred by assuming a single band and using the relationship $\mu=R_H/\rho_{xx}$.

The measured carrier concentration $n_h$ of Eu$_5$In$_2$Sb$_6$ at different magnetic field frequencies is shown in Fig.\,\ref{fig:526_ACHALL}.
Capacitive contributions to the AC voltage can be ruled out as the measurements of carrier concentration at different frequencies lay on top of each other. The carrier concentration as determined from DC Hall measurements at higher temperatures was quantitatively consistent with the AC field results.

\subsection{Photoresponse}
\label{Appx:photoresponse}

We investigated the photoresponsiveness of \Eu{} to ensure its viability as a detector target material. The response to light for multiple Eu$_5$In$_2$Sb$_6$ samples was measured with the following setup. A $1300\,\mathrm{nm}$ laser was coupled into a super guide SFS1500 multi-mode fiber. The fiber was placed in a Quantum Design PPMS photo probe, shining uncollimated  onto a sample stage at a distance of roughly 1\;cm. A FDG03-CAL calibrated Ge photodiode with a known responsivity curve was used to measure the irradiance on the sample stage in our system. A Eu$_5$In$_2$Sb$_6$ sample was then placed on the sample stage at the same height as the calibrated photodiode. The sample was sputtered with gold on each end and leads were silver painted on. A Keithley 2400 source meter was used to both apply a bias voltage and measure the corresponding current through the sample. The current through the sample was measured as a function of applied bias voltage and temperature with both the laser on and off. For the laser off data, the laser was disconnected and the fiber capped at the top of the cryostat to minimize the amount of ambient light. The photocurrent was quantified by subtracting the light and dark current data for each voltage and temperature. The charge yield per $1300\,\mathrm{nm}$ photon at a temperature of about $16\,\mathrm{K}$ was shown in Fig.~\ref{fig:QE_526} of Sec.\,\ref{sec:Materials}. Although at this temperature there are still thermally excited free carriers, the charge yield already showed an asymptotic trend toward fully saturated charge collection.

\subsection{Electron Energy Loss Spectroscopy}
\label{Appx:EELS}

\begin{figure}[b]
    \centering
    \includegraphics[width=1.0\linewidth]{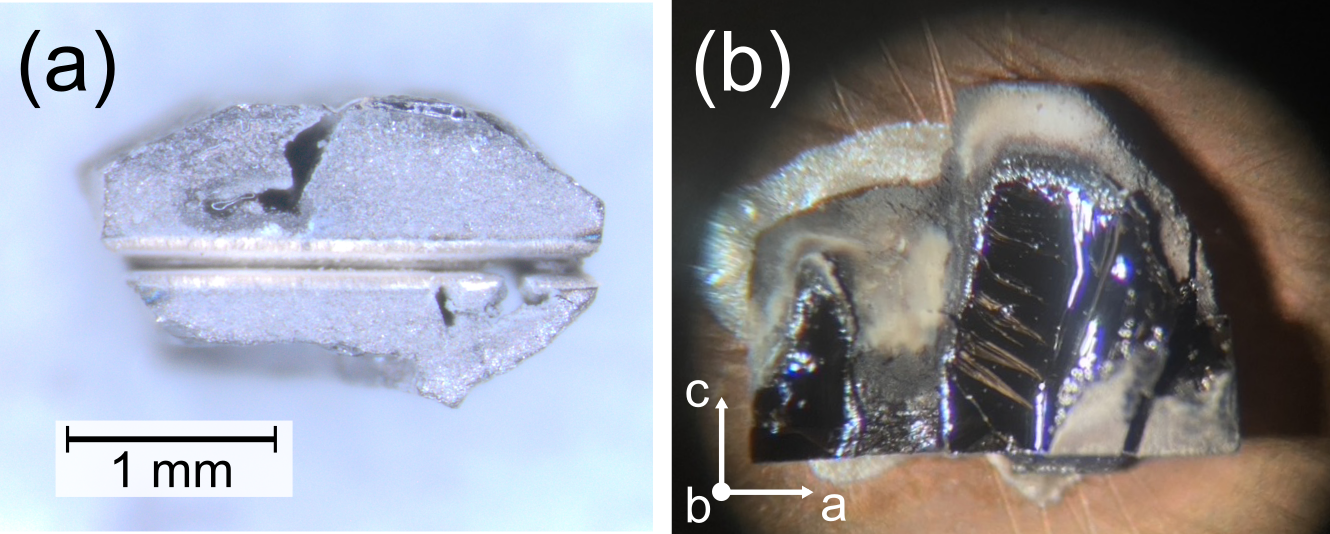}
    \caption{Femtoscribed Eu$_5$In$_2$Sb$_6$ sample. (a) Sample of Eu$_5$In$_2$Sb$_6$ with a $\sim$ 70 $\mu$m wide trough cut normal to the b axis. (b) The resulting cleaved surface. The scale is the same for both pannels.}
    \label{fig:Eu526scribed}
\end{figure}

The interaction of dark matter with solid-state targets is hypothesized to occur via conventional particle scattering, which has a rich history in condensed matter physics.
In the simplest case in which dark matter interactions with electrons or nuclei induce a scalar density-density response in the target,
Electron Energy Loss Spectroscopy (EELS) \cite{Egerton_2011,Fink_2014} provides a direct probe of the inelastic scattering cross section \cite{Hochberg:2021pkt, Boyd2023}.  Despite the $\mathcal O(10$\,meV$)$ resolution of modern EELS implementations, the sizable elastic peak dwarfs inelastic losses (i.e., energy transfers of electrons to the sample) up until around 0.5–1 eV \cite{Schuster_2015,Roest_2021}.  In a reflection geometry, {\it High-Resolution} EELS (HREELS) \cite{Ibach_1977,Ibach_1982} can be utilized to achieve low-energy sensitivity, where energy transfers from the meV \cite{Ibach_1996,Veal_2004} to the eV scale \cite{Pfau_1994,Bronner_2013} are regularly studied.  Further, finer control over the momentum resolution at a fixed energy transfer is possible through {\it Momentum-resolved} EELS (M-EELS) \cite{MEELS}. 

\begin{figure}
    \centering
    \includegraphics[width=0.9\linewidth]{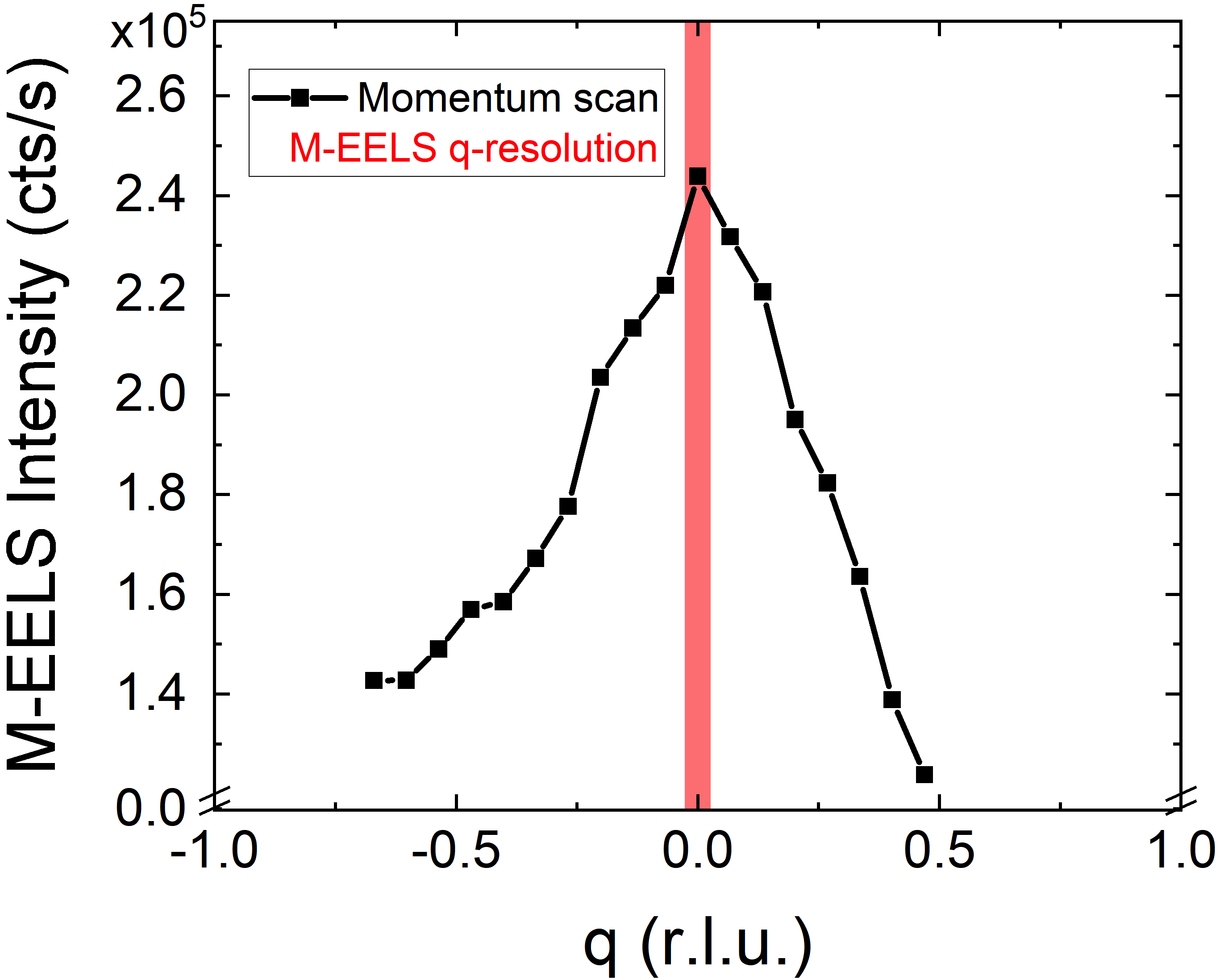}
    \caption{Zero-loss EELS momentum scan of Eu$_5$In$_2$Sb$_6$ measured at T = 300 K.}
    \label{fig:Eu526MEELS_scan}
\end{figure}

Material properties can be characterized at energy and momentum scales relevant to dark matter scattering through M-EELS; however, the reflection geometry probes the density response in a different, more surface-sensitive manner than EELS \cite{MEELS,Mills_1975}.  Nevertheless, M-EELS directly probes the (possibly anisotropic) dielectric response in the long wavelength limit \cite{Lucas_1984}. Further, M-EELS can resolve both direct \cite{Bronner_2013} and indirect \cite{Chen_1999} band gaps in conventional semiconductors.

M-EELS measurements were performed on samples of Eu$_5$In$_2$Sb$_6$. The technique requires clean, crystallographically-pristine surfaces which are typically achieved via cleaving {\it in situ}. Since \Eu~is not a layered material, it requires additional sample preparation in order to be cleaved. 
A notch was cut on three sides of each sample using the FemtoScribe at the Center for Integrated Nanotechnologies at Los Alamos National Laboratory. The sample was cut parallel to the $ac$ plane, as shown in Fig.~\ref{fig:Eu526scribed}. During the notching process, the material surrounding the trench turned a light, whitish color. This is likely due to oxidation of the sample caused by the high temperature of the laser. However, the bulk of the sample was unaffected.

A sample cleave pin was then affixed to the top. Samples were fractured in ultra-high vacuum (pressure $10^{-10}$ Torr), revealing clean scattering surfaces. An example of such a surface is shown in Fig.\;\ref{fig:Eu526scribed}(b). Data were taken with the electron beam incident on the flat, shiny portion of the sample.

Fig.\;\ref{fig:Eu526MEELS_scan} shows a scan of a Bragg peak of a Eu$_5$In$_2$Sb$_6$ sample. The red bar indicates the typical momentum resolution of the M-EELS instrument. The Bragg peak is significantly broader than the resolution limit, due mostly to the quality of the cleaved sample surface. Despite the efforts described above, the as-measured surface did not provide highly momentum-resolved data. Our energy loss scans of this sample, therefore, could not be considered momentum-resolved, but rather momentum-\textit{integrated}. The scattering plane of this data was the $ac$-plane, so that the data in Fig.\;\ref{fig:Eu526MEELS_scan} should be considered as being integrated over the $\Gamma - X$ and $\Gamma-Z$ directions. Improvements to the momentum-resolution may be achievable by cleaving the samples at low temperature \cite{Crivillero2022PRB}.

\begin{figure}
    \centering
    \includegraphics[width=0.9\linewidth]{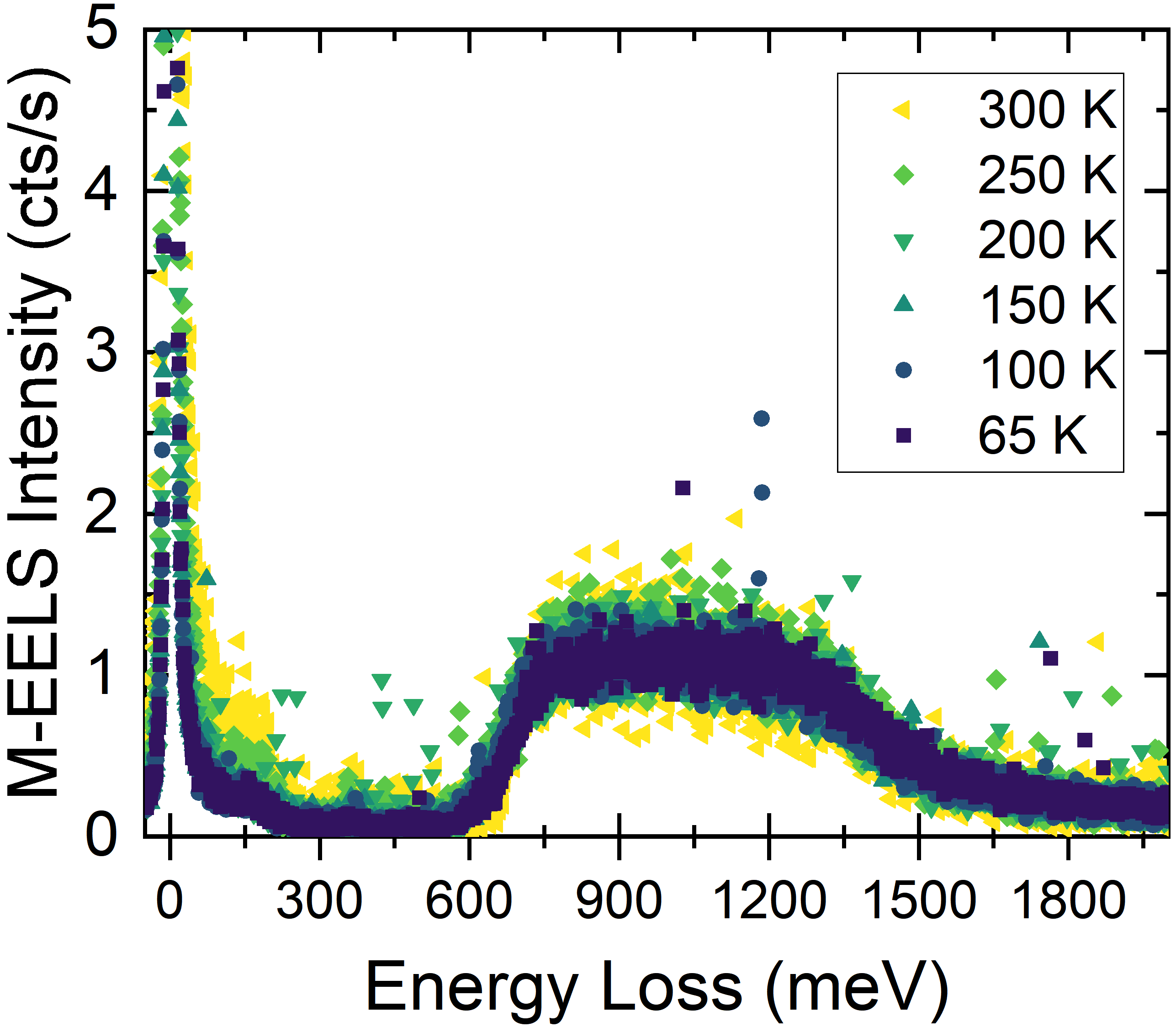}
    \caption{M-EELS spectra on Eu$_5$In$_2$Sb$_6$ at various temperatures. A significant onset of intensity is observed near 600~meV. The shoulder seen at $\sim$\,150~meV is a measurement artifact.}
    \label{fig:Eu526MEELS_spectra}
\end{figure}

Energy loss scans at various temperatures are shown in Fig.\;\ref{fig:Eu526MEELS_spectra}. At $\sim$\,150 meV, a measurement artifact from the copper puck on which the \Eu{} sample was mounted is visible as a low-intensity shoulder on the elastic line. The spectra remain nearly featureless from $\sim$250\;meV until $\sim$600\;meV, at which point a flat, continuum-like feature onsets. The continuum in energy is indicative of an interband transition that, under additional assumptions, can be used to infer the value of the band gap. Such inference would suggest a gap in Eu$_5$In$_2$Sb$_6$ closer to $600$~meV instead of the $\sim$$20-100$~meV values obtained from our transport measurements and DFT calculations. This discrepancy is likely due to the fact that, because of the roughness of the cleave, the entire $ac$ plane was being sampled.

\subsection{Fourier Transform Infrared Spectroscopy}
\label{Appx:FTIR}

\begin{figure}[t]
    \centering
    \includegraphics[width=1.0\linewidth]{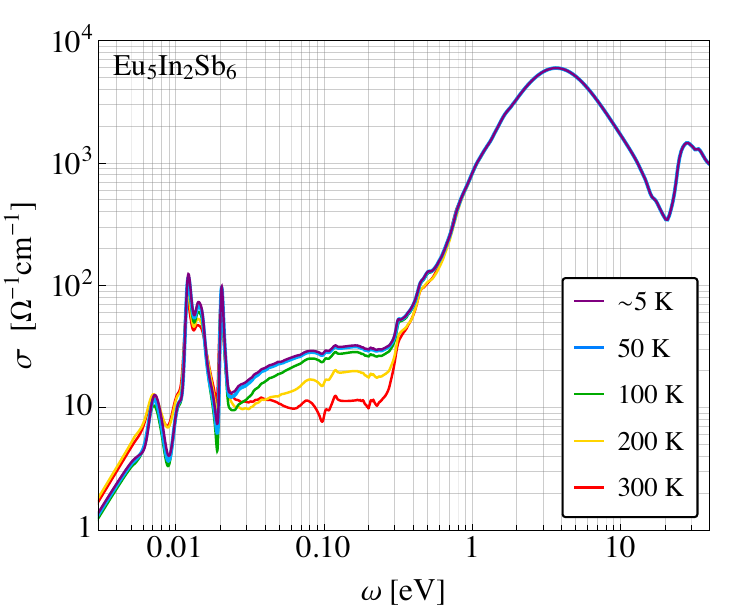}
    \caption{Optical conductivity spectrum of Eu$_5$In$_2$Sb$_6$ at various temperatures obtained from a smooth interpolation of FTIR and x-ray reflectivity data via a Kramers-Kronig transform.}
    \label{optical Conductivity}
\end{figure}

Optical conductivity measurements are a complementary probe of the electronic structure and can therefore help elucidate the results found from the M-EELS measurement. We performed Fourier transform infrared (FTIR) spectroscopy on a polished $ac$-plane of a single crystal of Eu$_5$In$_2$Sb$_6$ in the energy range from 5~meV to 2.4~eV for temperatures in the range of $\sim$\,5 – 300~K. The reflectivity data was normalized by measuring the reflectivity of the sample before and after evaporating gold (silver) on the sample for energies below (above) 1.4~eV to account for sample imperfections \cite{Homes1993optics}. The resulting reflectivity $R(\omega,T)$ was converted into optical conductivity $\sigma(\omega,T)$ by means of a Kramers-Kronig transform. To this end, the reflectivity was extended by a constant function towards $\omega \rightarrow 0$, and at high energies, the FTIR data was smoothly interpolated to the reflectivity inferred from x-ray data above 10 eV. The resulting optical conductivity as a function of photon frequency is shown in Fig.\,\ref{optical Conductivity} for measurements performed at various temperatures.

Due to the small momentum of photons at these frequencies, the optical conductivity predominantly measures direct interband transitions, although phonon-assisted indirect transitions can also contribute. Fig.\,\ref{optical Conductivity} shows high energy interband transitions peaking at $\omega\simeq4$\;eV. The onset of these transitions at approximately $\omega\sim600$\;meV and their lack of temperature dependence are consistent with our M-EELS measurements.
Transitions between additional low-energy states that were not resolved in the M-EELS data are observed in the mid-energy range of $\omega\sim25-300$\;meV, with an increase in optical conductivity with decreasing temperature.
Unfortunately, the low-energy tail of the $\mathcal{O}$(eV) interband transitions, combined with pronounced phonon peaks appearing below $\omega\sim25$\;meV, precluded the determination of a direct bandgap in this region.

\subsection{Radiological Impurities}
\label{Appx:Radioassay}

Radiological impurities are a significant concern of experiments searching for dark matter and other rare events~\cite{DAMIC:2021crr}. Natural elements like uranium and thorium are common sources of radioactivity in most materials~\cite{Hoppe:2014nva}. These radioactive backgrounds can interfere with the detection of dark matter signals, potentially suppressing or mimicking the signals from dark matter particles.

To establish a baseline for the radioactive background level, radioactive assays~\cite{MAJORANA:2016lsk} were conducted to estimate the radioactive content of the \Eu~samples used as detector targets. Moreover, the radioactive isotope $^{115}$In is naturally present in indium, a constituent element of the crystal samples. \texttt{GEANT4}~\cite{Allison:2016lfl} simulations were employed to further investigate its impact.

\subsubsection{Radioactive Assays}

The assays were conducted at the tunnel facility at Los Alamos National Laboratory (LANL). The tunnel is situated at the base of Los Alamos canyon, approximately 300~m below the ground level of Los Alamos, NM, at an altitude of approximately 2300~m.

A low-background p-type high-purity germanium (HPGe) detector, specifically the ORTEC GWL Series Coaxial HPGe Detector, was housed within a carbon fiber cryostat endcap, surrounded by lead shields with a thickness of at least 8 inches. Oxygen-free high thermal conductivity (OFHC) copper was utilized for nearby components. The detector's efficiency, adjusted for geometry- and energy-dependent effects, was determined through simulations.

\begin{figure}[t]
    \centering
    \includegraphics[width=0.5\textwidth]{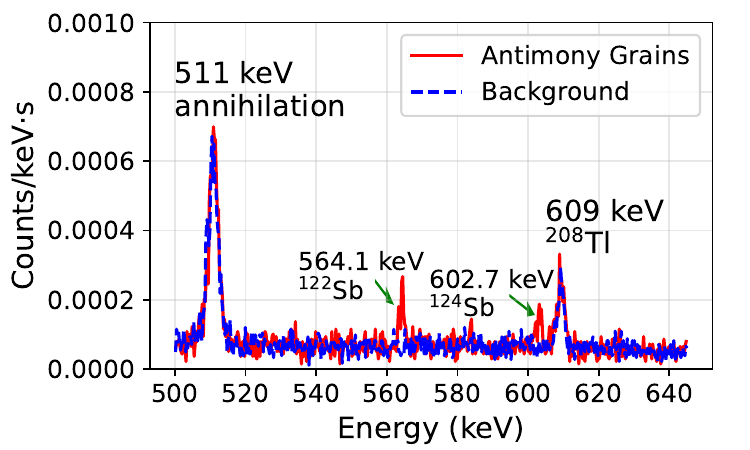}
    \caption{The radioassay spectrum of the antimony sample reveals prominent peaks at 564.1 keV from $^{122}$Sb and at 602.7 keV from $^{124}$Sb, becoming apparent after subtracting the background spectrum.}
    \label{fig:assayspectrum}
\end{figure}

The natural abundance of antimony (Sb) is comprised of two stable isotopes, namely,  $^{121}$Sb at 57$\%$, and $^{123}$Sb at 43$\%$. The sample utilized in this study consisted of 17.015 grams of antimony beads contained within a polyethylene test tube. The total background rate above 50 keV was measured to be 0.1676 counts/s, while the rate for the antimony sample was 0.1684 counts/s. This suggests relatively low activity, with a difference of around 69.12 counts/day. As shown in Fig.~\ref{fig:assayspectrum}, several peaks were detected in the gamma-spectrum of the antimony sample, notably at 564.1 keV from $^{122}$Sb and 602.7 keV from $^{124}$Sb. These suggest that natural antimony undergoes cosmogenic activation, resulting in the production of $^{122}$Sb and $^{124}$Sb through neutron capture. Both isotopes have relatively short half-lives of 2.7 and 60 days, respectively, and will rapidly decay away once isolated underground. Table~\ref{tab:radioactive_sb} indicates that the overall activity is consistent with negligible levels of radioactivity, especially if $^{122}$Sb and $^{124}$Sb are ignored.

\begin{table}[b]
  \centering
  \begin{tabular}{@{}lrr@{}}
    \toprule
    \textbf{Isotope} & \textbf{Sb (Bq/kg)} & \textbf{In (Bq/kg)}\\ \midrule
    Total U-chain & $0.05 \pm 0.05$ & $0.09 \pm 0.06$ \\
    Total Th-chain & $-0.015 \pm 0.045$ & $-0.07 \pm 0.05$\\
    $^{228}$Th chain & $-0.05 \pm 0.05$ & $-0.08 \pm 0.06$\\
    $^{60}$Co & $0.007 \pm 0.035$ & $0.017 \pm 0.040$\\
    $^{40}$K & $-0.56 \pm 0.39$ & $-0.15 \pm 0.45$\\
    $^{137}$Cs & $0.019 \pm 0.032$ & $-0.009 \pm 0.037$\\
    $^{122}$Sb & $0.135 \pm 0.036$ & $-0.012 \pm 0.040$\\
    $^{124}$Sb & $0.076 \pm 0.026$ & $0.008 \pm 0.029$\\ \bottomrule
  \end{tabular}
  \caption{Radioactive isotope concentrations. Antimony is not anticipated to be present in the indium sample. These values are provided solely for comparison purposes.}\label{tab:radioactive_sb}
\end{table}

Indium is composed of two natural isotopes: 4.3$\%$ $^{113}$In and 95.7$\%$ $^{115}$In. The isotope $^{115}$In undergoes beta decay with a half-life of $4.4\times 10^{14}$ years at a $Q$-value of 497.5~keV~\cite{PhysRevC.19.1035}, emitting minimal gamma radiation~\cite{Cattadori:2004vi}. This beta decay will contribute to unavoidable background in the experiment. The sample utilized in this study consisted of 50.674~grams of indium cast as an ingot. The total background rate above 50 keV was measured to be 0.1676~counts/s, while the rate for the indium sample was 0.1691~counts/s. This suggests relatively low activity, with a difference of around 129.6~counts/day. Table\;\ref{tab:radioactive_sb} indicates that the overall activity is low but not zero. However, the radioassay is only sensitive to gammas, not alpha or beta radiation. The radioactive background from In was therefore estimated using \texttt{GEANT4} simulations, to be described shortly.

Europium has one stable isotope, $^{153}$Eu (52.19$\%$), and one unstable isotope, $^{151}$Eu (47.81$\%$), which can undergo alpha-decay with a half-life of $4.6\times 10^{18}$ years at a $Q$-value of 1949~keV. However, the radioassay is not sensitive to alpha particles. Additionally, its contribution to low-energy backgrounds, particularly in the $\mathcal{O}(1\!-\!100)$~meV region of interest for SPLENDOR, is insignificant. Therefore, no radioassay or \texttt{GEANT4} simulation was conducted for Eu.

\begin{figure}[b]
    \centering
    \includegraphics[width=0.46\textwidth]{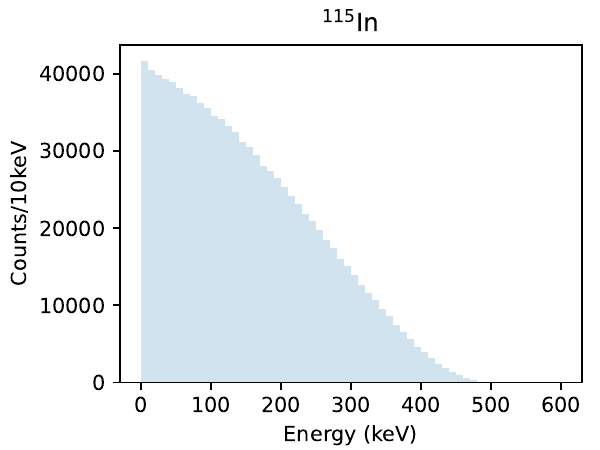}\\
    \includegraphics[width=0.46\textwidth]{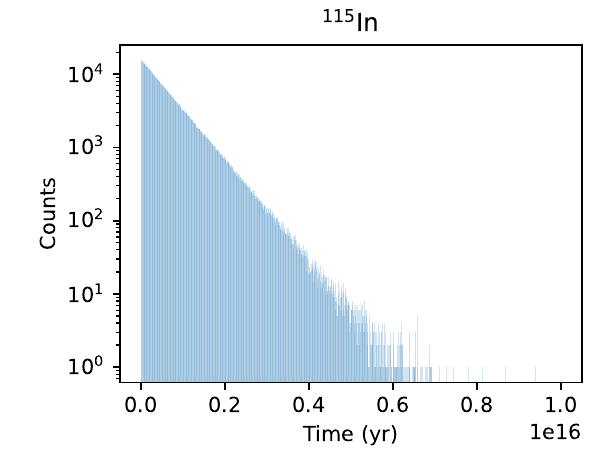}
    \caption{Simulated $^{115}$In $\beta$-decays, with one million events generated. \textit{Top:} $\beta$-decay energy spectrum with $Q$-value of 497.489~keV. \textit{Bottom:} Time distribution of $\beta$-decay events with a half-life of $4.4\times 10^{14}$ years.}
    \label{fig:indiumsimulation}
\end{figure}

\subsubsection{GEANT4 simulations}

The beta decay of $^{115}$In can contribute to low-energy backgrounds in dark matter searches due to the continuous nature of the beta energy spectrum. For simulating $^{115}$In decay in $\text{Eu}_{5}\text{In}_{2}\text{Sb}_{6}$, we utilized the physics list \texttt{FTFP\_BERT\_HP} and \texttt{Livermore} in \texttt{GEANT4.10.07.p04}. This physics list employs the Bertini-style cascade for hadrons at energies below 5~GeV, along with all standard and low-energy electromagnetic processes.

The detector target was modeled as a cylindrical $\text{Eu}_{5}\text{In}_{2}\text{Sb}_{6}$ sample with a length of 1~cm and a radius of 0.1~cm, resulting in a sample mass of 0.213~grams given a density of 6.77~g/cm$^{3}$ for \Eu~\cite{PhysRevB.105.035135}. The $1.425\times 10^{20}$ $^{115}$In isotopes were randomly distributed within the sample, and the beta particles were emitted isotropically in the simulation. We generated one million events and recorded their energy spectrum and time distribution, which are shown in Fig.\,\ref{fig:indiumsimulation}. Given the $Q$-value of 497.489~keV, the resulting energy spectrum shows that 5$\%$ of events have energies below 10~keV.
Given the half-life of $^{115}$In of $4.4\times 10^{14}$ years, we estimate approximately 221,532  $^{115}$In decays in the target sample in one year. Normalizing the corresponding energy counts in the simulation to the decay counts and total target mass over one year, we arrive at a count rate of $(1.4-1.6)\times 10^{-3}$~Hz/g at energies below 10 keV.

\section{\texorpdfstring{\\First-Principles Electronic Structure Calculations}{First-Principles Electronic Structure Calculations}}
\label{DFTtheory}

To theoretically determine both scattering and absorption rates of light dark matter in real materials, an accurate description of the electronic structure is required. Over the past five decades, electronic structure theory of solids has undergone significant leaps in advancement where today both the single- and two-particle properties of materials composed of elements from across the periodic table may be determined in good accord with experimental values. Despite these successes, the many-body problem is far from solved, therefore careful analysis and the judicious use of  approximations is necessary to yield reliable results in correlated electron systems. To elucidate the electronic states of Eu$_{5}$In$_{2}$Sb$_{6}$, we took a systematic approach: (i) the electronic band dispersions were obtained within standard electronic structure methodologies, then (ii) the response of the system to an impinging particle (scattering and absorption processes) was calculated using the single-particle states within the many-body Green's function formalism. The following sections present the details of each step of the calculations.

\subsection{Band Structure}
To obtain the electronic band structure of Eu$_{5}$In$_{2}$Sb$_{6}$, DFT calculations were performed using the full potential linearized augmented plane-wave+local orbitals (L/APW+lo) method~\cite{blaha1990full,  blaha2001wien2k} as implemented in the WIEN2k code~\cite{blaha2020wien2k}. Exchange-correlation effects were treated within the Perdew-Burke-Ernzerhof generalized gradient approximation (PBE-GGA)~\cite{perdew1996generalized}. The Brillouin zone was sampled on a $\Gamma$-centered grid using 1000 $k$-points. An energy cutoff of $-6.0$ Ry was used to delineate the core-valence separation. The value of $R_{mt}K_{max}$ was chosen to be 7. Furthermore, the $f$-electrons were treated within the core via the open-core method and relativistic effects were included self-consistently using a finite spin-orbit coupling.

\begin{figure}
\centering
\includegraphics[width=\columnwidth]{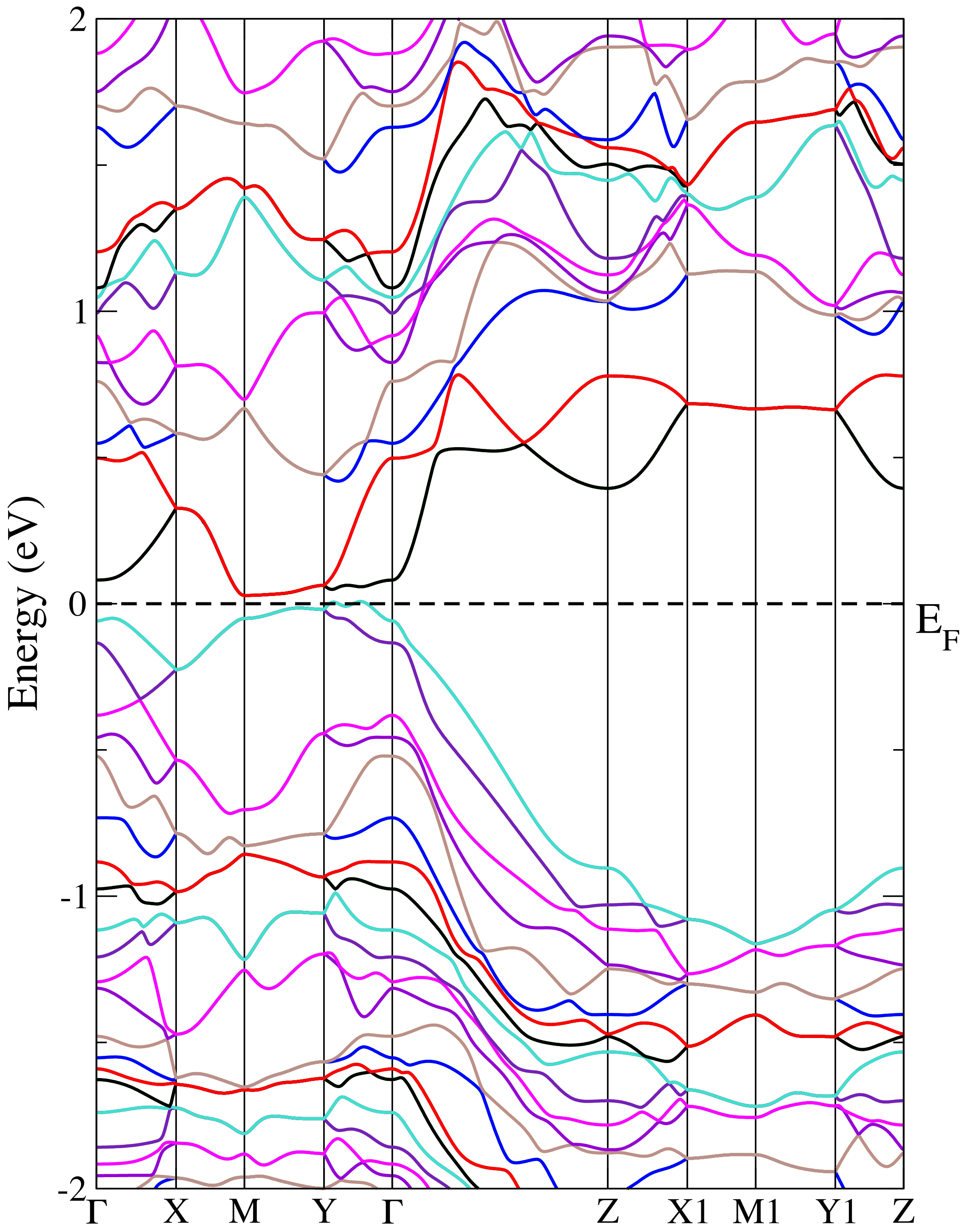}
\caption{Electronic band structure for bulk Eu$_{5}$In$_{2}$Sb$_{6}$ along various high-symmetry lines in the Brillouin zone.}
\label{fig:Eu526Bands}
\end{figure}

Fig.\;\ref{fig:Eu526Bands} presents the band structure of Eu$_{5}$In$_{2}$Sb$_{6}$ about the Fermi level. A clear 20 meV indirect (40 meV direct) band gap is observed separating valence and conduction states. Moreover, the bands surrounding the Fermi level are relatively flat, thereby suggesting a strong dielectric response. Our results are in good agreement with previously published results \cite{rosa2020colossal,PhysRevB.105.235128}.

\vspace{5pt}
\noindent{\it Band gap determination in DFT:} Due to the sensitive role the band gap plays in determining the electronic response of the system, we must be careful when interpreting and using the band gap directly from DFT calculations. Rigorously, the Kohn-Sham eigenvalues are Lagrange multipliers used to ensure orthogonality of the Kohn-Sham wavefunctions during the variational minimization of the total energy of the full many-body problem of interacting particles with respect to the single-particle density~\cite{dreizler2012density}. Although the eigenvalues and eigenvectors of the Kohn-Sham equation do not correspond to excitations of the solid, if electrons are weakly to intermediately correlated, the energies often provide a very good starting point for computing the true excitation spectra via Hubbard U corrections~\cite{himmetoglu2014hubbard}, hybrid~\cite{garrick2020exact} and meta-GGA~\cite{pokharel2022sensitivity} functionals, or many-body perturbation theory (e.g., the GW approximation)~\cite{onida2002electronic}. However, if these higher levels of theory are prohibitively computationally expensive, e.g., owing to strong relativistic effects and a large number of atoms in the unit cell, a ``scissor correction'' may be applied~\cite{di2014first}. This procedure consists of rigidly and uniformly shifting the conduction bands relative to the Fermi level so that the bandgap matches a given value, typically the experimental value. As the curvature of the band structures produced by DFT are frequently qualitatively correct, the scissor correction scheme is a reasonable approach for highly complex materials. A shift in the conduction bands amounts to a simple rigid shift of the excitation spectrum in the various response functions, such as the loss function. Specifically, energy differences across the gap $\varepsilon_{n'\mathbf{k}+\mathbf{q}} - \varepsilon_{n\mathbf{k}}$ are mapped to $\varepsilon_{n'\mathbf{k}+\mathbf{q}} - \varepsilon_{n\mathbf{k}}+\Delta$, where $\varepsilon_{n\mathbf{k}}$ is the Kohn-Sham eigenvalue for the state with band index $n$ and crystal momentum $\mathbf{k}$, and $\Delta$ is the induced correction in the band gap from the scissor shift.
The resulting effect in the loss function is easily seen, for example, in the random phase approximation, where the integrand in the sum over Kohn-Sham eigenvalues is proportional to a Dirac delta function, $\delta(\varepsilon_{n'\mathbf{k}+\mathbf{q}} - \varepsilon_{n\mathbf{k}} - \omega)$, due to the Cauchy identity.
Hence, a scissor correction effectively amounts to a shift in the energy transfer argument of the loss function, i.e.,
\begin{equation}
\mathrm{Im}\bigg[-\frac{1}{\epsilon(\mathbf{q},\omega)}\bigg]~\rightarrow~\mathrm{Im}\bigg[-\frac{1}{\epsilon(\mathbf{q},\omega-\Delta)}\bigg].
\label{equation:scissor_shift}    
\end{equation}
For our present purposes, a scissor correction of $\Delta=40$\;meV was applied to our numerical results to match the transport gap of 60 meV observed in our \Eu~samples, as mentioned in Sec.\;\ref{sec:Sensitivity}.

\subsection{Response Functions}
 To facilitate the evaluation of the response functions over a dense $k$-point grid in the Brillouin zone, we employed a real-space tight-binding model Hamiltonian (obtained via the Wien2Wannier interface~\cite{kunevs2010wien2wannier}) to interpolate the {\it ab initio} electronic band structure. For Eu$_5$In$_2$Sb$_6$, the full manifold of Eu-$5d$, In-$4p$, and Sb-$4p$ states was included in generating the orbital projections.
 
 As a matter of convention, we adopted the Cartesian axes' labels $\bm{x}$,\,$\bm{y}$,\,$\bm{z}$ for the spatial components of the dielectric tensor and momentum transfer. The Cartesian \mbox{$\bm{x}$-$\bm{y}$-$\bm{z}$} frame coincides with the crystal lattice frame \mbox{\textbf{\textit{a}}-\textbf{\textit{b}}-\textbf{\textit{c}}} defined by the lattice vectors displayed in Figs.~\ref{fig:Eu526structure}a,\,\ref{fig:Eu526structure}b of Sec.~\ref{sec:Materials}.

\vspace{5pt}
\noindent{\it Dark Matter Absorption:} The central quantity governing the absorption rate of dark matter is the dynamical dielectric tensor $\epsilon_{\alpha \beta}(\omega,\mathbf{q})$. This quantity is related to the dynamical conductivity tensor~\cite{dressel2002electrodynamics} by 
\begin{align}
\epsilon_{\alpha \beta}(\omega,\mathbf{q})&=\delta_{\alpha \beta}+i\frac{1}{\epsilon_0}\frac{\sigma_{\alpha \beta}(\omega,\mathbf{q})}{\omega},
\label{eq:EpsilonSigma}
\end{align}
where $\epsilon_0$ is the vacuum permittivity.
For the given Hamiltonian the conductivity can be calculated directly via the current-current correlator and the kinetic energy flow~\cite{zhu2016bogoliubov}
\begin{align}
\sigma_{\alpha \beta}(\omega,\mathbf{q})&=\frac{1}{i\omega}
\left[ -\Pi_{\alpha \beta}(\omega,\mathbf{q})+K_{\alpha}\delta_{\alpha\beta}  \right],
\end{align}
with the current-current correlation function given by
\begin{align}
&\Pi^{\alpha\beta}(\omega,\mathbf{q}) = 
\frac{1}{N_{\mathbf{k}}} \!\!\sum_{\substack{\mathbf{k},i,j,s\\s',\,s'',\,s'''}} 
\!\!v^{\alpha}_{\mathbf{k}+\mathbf{q},s,s^{\prime}}\,
v^{\beta}_{\mathbf{k},s^{\prime\prime},s^{\prime\prime\prime}}~~~~~\\
&~~~~~~~~~~~\times~
V^{\mathbf{k}}_{s^{\prime\prime\prime},i}\,
V^{*\mathbf{k}}_{s,i}\,
V^{\mathbf{k+q}}_{s^{\prime},j}\,
V^{*\mathbf{k+q}}_{s^{\prime\prime},j}\,\frac{f(\varepsilon^i_{{\mathbf{k}}}) -f(\varepsilon^j_{{\mathbf{k+q}}})   }{\omega+\varepsilon^i_{{\mathbf{k}}}-\varepsilon^j_{{\mathbf{k+q}}}+i\delta},\nonumber
\end{align}
and the kinetic energy flow by
\begin{align}
K_\alpha = \frac{1}{N_{L\,} a^3}\sum_{\mathbf{k},l,s,s^\prime} 
\kappa^\alpha_{\mathbf{k},s,s^\prime} V^{*\mathbf{k}\,}_{ls}V^{\mathbf{k}\,}_{ls^\prime}f(\varepsilon^l_\mathbf{k}),
\end{align}
where, for brevity, $s$ denotes a composite of spin and orbital indices, $v^{\alpha}_{\mathbf{k},s,s^{\prime}}$ are the electronic band velocities, $\kappa^\alpha_{\mathbf{k},s,s^\prime}$ are the directional kinetic energy factors, $V_{s,i}=\braket{s |i}$  are the matrix elements connecting the orbital-spin and band spaces found by diagonalizing the Hamiltonian, and $f(\varepsilon^i_{{\mathbf{k}}})$ are the Fermi functions. In terms of computational scalability, the calculation of $\Pi^{\alpha\beta}$ follows the same treatment adopted for the electric susceptibility, to be described shortly.
We note that our developed formalism contains the contribution from the entire momentum space in the Brillouin zone, which goes beyond earlier studies~\cite{Hochberg_2018}.

\begin{figure}
    \centering
    \includegraphics[width=1\linewidth]{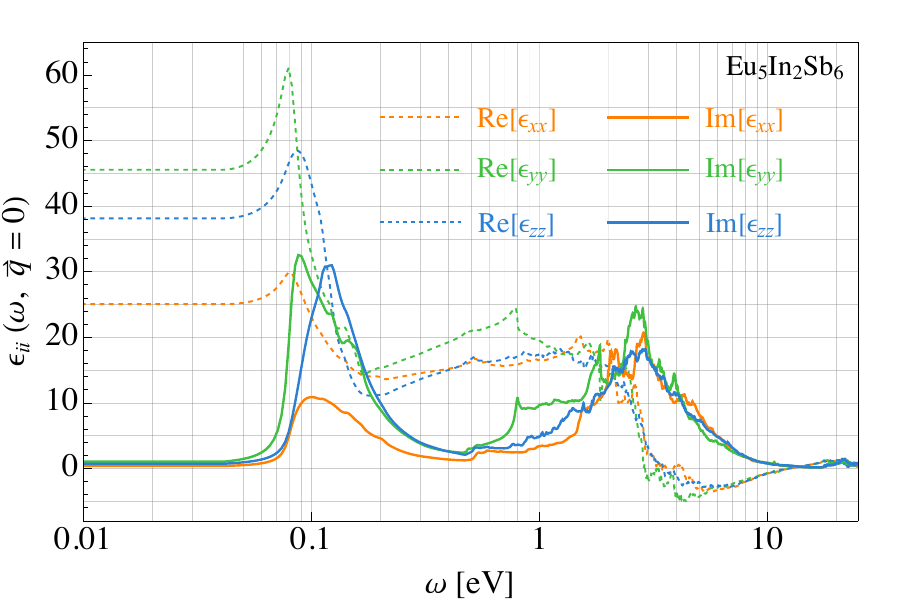}
    \caption{The frequency dependence of the real (dashed) and imaginary (solid) parts of the diagonal components of the  dielectric tensor for bulk Eu$_5$In$_2$Sb$_6$ at $\vec{q}=0$. A scissor correction of $\Delta=40$\;meV has been applied.}
    \label{fig:EpsAbs}
\end{figure}

\begin{figure}[t]
    \centering
    \includegraphics[width=1\linewidth]{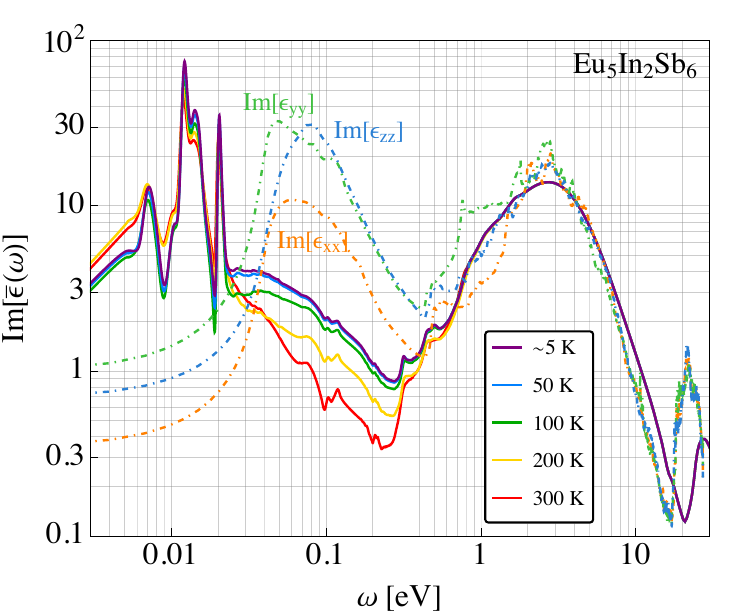}
    \caption{Comparison between our DFT prediction for the imaginary part of the dielectric tensor components $\text{Im}[\epsilon_{ii}(\omega)]$ at $\vec{q}=0$ (dot-dashed curves) and the polarization-averaged dielectric function $\text{Im}[\bar{\epsilon}(\omega)]$ (solid curves) inferred from our measurements of the optical conductivity of \Eu~at different temperatures (cf.\;Fig.\,\ref{optical Conductivity}). No scissor correction has been applied to the DFT predictions.}
    \label{fig:EpsFTIR}
\end{figure}

Fig.\,\ref{fig:EpsAbs} shows the frequency dependence of the dielectric tensor components for bulk Eu$_5$In$_2$Sb$_6$ along the three symmetry axes of the crystal evaluated at $\mathbf{q}=0$. The onset of prominent peaks in Im$[\epsilon_{yy}]$ and Im$[\epsilon_{zz}]$ is seen at $\omega=80$~meV and $\omega=90$~meV, respectively. A less pronounced peak in Im$[\epsilon_{xx}]$, which is weaker by a factor of three, is also seen in the same energy range. These peaks appear to originate from interband transitions
close to the Fermi level (see Fig.~\ref{fig:Eu526Bands}). At high frequencies, the dielectric tensor exhibits much more structure, with each component exhibiting similar line shapes and amplitudes. These transitions originate from the narrow bands along the X$-$M$-$Y (X1$-$M1$-$Y1) path in the Brillouin zone of Fig.~\ref{fig:Eu526Bands}.

We can compare our theoretical results for the dielectric tensor $\epsilon_{ii}(\omega,\,\mathbf{q}=0)$ with our measurements of the optical conductivity $\sigma(\omega)$ (cf.\;Fig.\;\ref{optical Conductivity} of Appx.\,\ref{Appx:FTIR}) through an adaptation of \eqref{eq:EpsilonSigma}:
\begin{equation}
\text{Im}[\bar{\epsilon}(\omega)]~=~\frac{1}{\epsilon_0}\,\frac{\sigma(\omega)}{\omega},
\end{equation}
where $\bar{\epsilon}(\omega)$ is a weighted average of the components of the dielectric tensor reflecting the unpolarized light used in our measurements of $\sigma(\omega)$.

Fig.\;\ref{fig:EpsFTIR} overlays our results for $\text{Im}[\overline{\epsilon}(\omega)]$ with the imaginary part of the diagonal components of the dielectric tensor predicted by DFT\footnote{Unlike the scissor-corrected curves for $\epsilon_{ii}(\omega)$ shown in Fig.\,\ref{fig:EpsAbs}, no scissor correction has been applied to $\text{Im}[\epsilon_{ii}(\omega)]$ in Fig.\;\ref{fig:EpsFTIR}.}. We note that there is good agreement between theory and experiment in the high-energy range above $\omega\sim400$\;meV.
In the very low-energy range of $\omega\lesssim20$\;meV, on the other hand, this comparison is precluded by the presence of pronounced phonon peaks. In the intermediate range of $\omega\approx(20-400)$\;eV, there is good alignment between the peak-like feature predicted by DFT from low-energy interband transitions and the observed rise in conductivity as the temperature is decreased. Indeed, for low-energy photons, a suppression in optical conductivity is expected at high temperatures. However,
it is not clear whether this suppression can fully account for the discrepancy in intensity between theory and experiment in the $\omega\approx(20-400)$\;eV range indicated in Fig.\;\ref{fig:EpsFTIR}. Further investigation is needed of additional effects that could potentially contribute to this discrepancy, such as magnetic-polaron dynamics, which is absent in our DFT modeling, as well as possible effects of impurity-generated Coulomb potentials on the charge-carrier recombination rate \cite{Ruzin_Shklovskii_1989}. Additional conductivity data at lower temperatures would also be valuable. Nonetheless, the comparison in Fig.\;\ref{fig:EpsFTIR} indicates that our DFT modeling of the electronic structure of \Eu{} is in reasonable qualitative agreement with the experimental data.

\begin{figure*}
    \centering
    \begin{subfigure}
        \centering
        \includegraphics[width=0.49\linewidth]{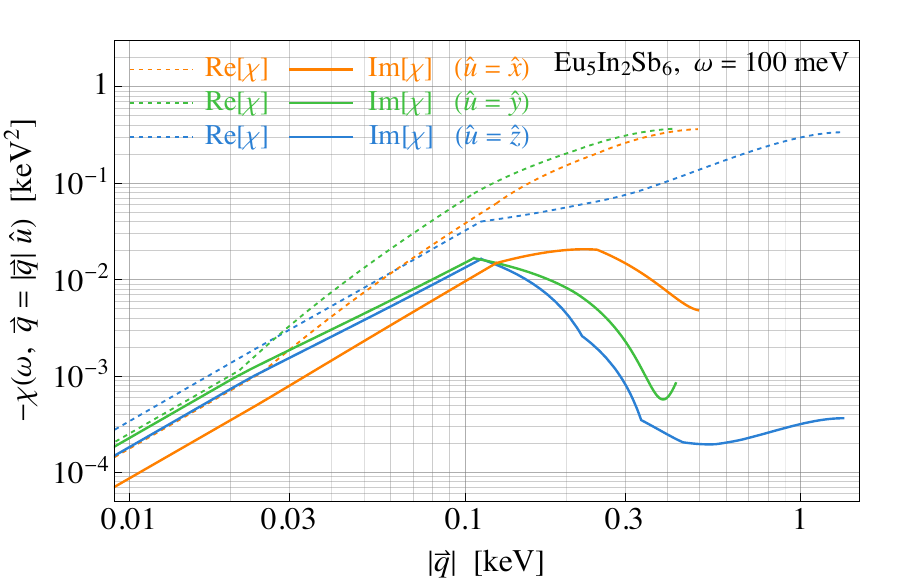}
    \end{subfigure}
    \begin{subfigure}
        \centering
        \includegraphics[width=0.49\linewidth]{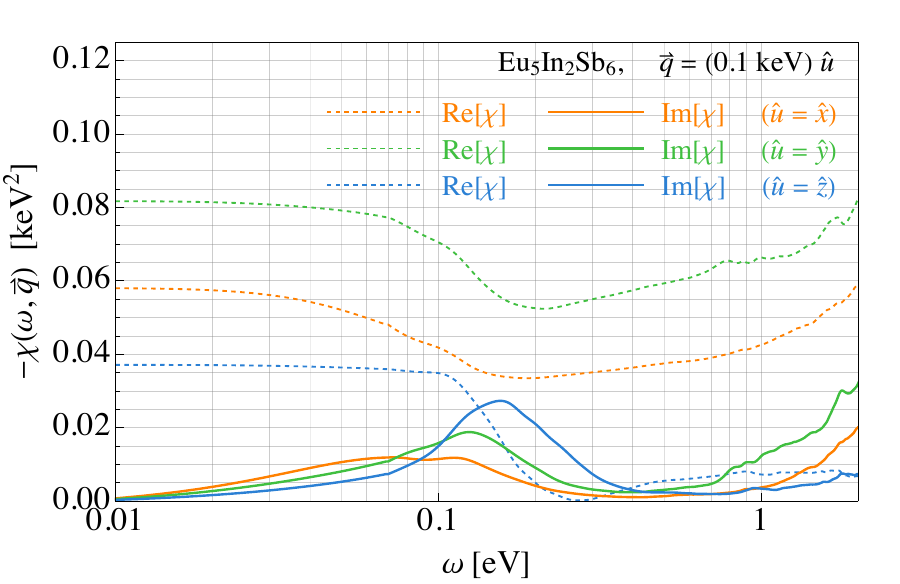}
        \label{fig}
    \end{subfigure}
    \caption{The real (dashed) and imaginary (solid) parts of the electric susceptibility of bulk Eu$_5$In$_2$Sb$_6$ along various high-symmetry lines in the Brillouin zone for a fixed energy of $\omega=60$\;meV (left panel) and a fixed momentum-transfer magnitude of $|\vec{q}|=0.1$\,keV (right panel). A scissor correction of $\Delta=40$\;meV has been applied.}
    \label{fig:Chi_q_omega}
\end{figure*}

\vspace{5pt}
\noindent {\it Dark Matter Scattering:} A key ingredient in determining the dark matter scattering rate is the electric susceptibility of the medium.
In the random phase approximation, the electric susceptibility can be written as: 
\begin{align}\label{eq:LindhardSF}
\chi(\omega,\mathbf{q})=~&\frac{1}{N_\mathbf{k}}\!\!\!\!\!\!\!\sum_{\substack{\mathbf{k},i,j \\\nu^{(\prime)} \mu^{(\prime)} a^{(\prime)}b^{(\prime)}}}\!\!\!\!\!\!\!\!
V^{\mathbf{k}+\mathbf{q}}_{(\nu^{\prime}b^{\prime})i}\,
 V^{\dagger~\mathbf{k}+\mathbf{q}}_{i(\mu a)}\,
 V^{\mathbf{k}}_{(\nu b)j}\,
 V^{\dagger~\mathbf{k}}_{j(\mu^{\prime} a^{\prime})}\nonumber\\
&~~~~~~~~~~~~~~\times\,\frac{f(\varepsilon^{j}_{\mathbf{k}})-f(\varepsilon^{i}_{\mathbf{k}+\mathbf{q}})}{\omega+\varepsilon^{j}_{\mathbf{k}}-\varepsilon^{i}_{\mathbf{k}+\mathbf{q}}+i\delta},
\end{align}
where Greek and Latin letters denote the orbital and spin degrees of freedom, respectively.

Computationally, the multiorbital, spin-dependent Lindhard susceptibility in standard form shown in \eqref{eq:LindhardSF}
scales linearly with the discretization of $\mathbf{q}$, $\mathbf{k}$, and $\omega$, and therefore calculations over dense meshes quickly become impractical. This presents a challenge, since, to obtain the frequency-dependence of the dielectric response for real materials with many bands, dense meshes are required to sufficiently capture the small energy and momentum scales. To reduce the computational complexity, we follow Ref.~\cite{lane2023competing} where the Lindhard susceptibility is written in the spectral representation as:
\begin{align}
&\chi^{\nu^{\prime}b^{\prime}\nu b}_{\mu^{\prime} a^{\prime}\mu a}(\omega,\mathbf{q})=\nonumber\\
&~\frac{1}{N_\mathbf{k}}\sum_{\mathbf{k},i,j}
 V^{\mathbf{k}+\mathbf{q}}_{(\nu^{\prime} b^{\prime})i}\,
 V^{\dagger~\mathbf{k}+\mathbf{q}}_{i(\mu a)}\,
 V^{\mathbf{k}}_{(\nu b)j}\,
 V^{\dagger ~ \mathbf{k} }_{ j(\mu^{\prime}a^{\prime}) } \\
&~~~~~~~~~~\times\int_{-\infty}^{+\infty}\!\!\!dX\,
\frac{f(\varepsilon^{j}_{\mathbf{k}})-f(\varepsilon^{i}_{\mathbf{k}+\mathbf{q}})}{\omega-X+i\delta}\,\delta(X-[\varepsilon^{i}_{\mathbf{k}+\mathbf{q}}-\varepsilon^{j}_{\mathbf{k}}]).\nonumber
\end{align}
The integral over $X$ and the sum over $\mathbf{k},\,i,\,j$ may be reordered, yielding
\begin{align}\label{eq:chi}
\chi^{\nu^{\prime}b^{\prime}\nu b}_{\mu^{\prime} a^{\prime}\mu a}(\omega,\mathbf{q})&=
\int_{-\infty}^{+\infty}\!\!\!dX~
\frac{ \hat{\chi}^{\,\nu^{\prime}b^{\prime}\nu b}_{\mu^{\prime} a^{\prime}\mu a}(X,\mathbf{q})}{\omega-X+i\delta},
\end{align}
where
\begin{align}
&\hat{\chi}^{\,\nu^{\prime}b^{\prime}\nu b}_{\mu^{\prime} a^{\prime}\mu a}(X,\mathbf{q})=\nonumber\\
&~~~~~~~~\frac{1}{N_\mathbf{k}}\sum_{\mathbf{k},i,j}
 V^{\mathbf{k}+\mathbf{q}}_{(\nu^{\prime} b^{\prime})i}\,
 V^{\dagger~\mathbf{k}+\mathbf{q}}_{i(\mu a)}\,
 V^{\mathbf{k}}_{(\nu b)j}\,
 V^{\dagger ~ \mathbf{k} }_{ j(\mu^{\prime}a^{\prime}) } \\
&~~~~~~~~~~~~~~~~~~\times(f(\varepsilon^{j}_{\mathbf{k}})-f(\varepsilon^{i}_{\mathbf{k}+\mathbf{q}}))\,\delta(X-[\varepsilon^{i}_{\mathbf{k}+\mathbf{q}}-\varepsilon^{j}_{\mathbf{k}}]).\nonumber
\end{align}

\noindent In this way, the computation of $\chi$ is split into two parts: (i)~a binning process over the excitation energies $\varepsilon^{i}_{\mathbf{k}+\mathbf{q}}-\varepsilon^{j}_{\mathbf{k}}$ for a given $\mathbf{q}$, and (ii)~a Hilbert transform of $\hat{\chi}$ to recover $\chi$. The pole in the denominator of the Hilbert transform in \eqref{eq:chi} is treated via a principal-value integral aided by the modified Gaussian rule. This splitting reduces the computational complexity from $N_{\mathbf{q}} \times N_{\mathbf{k}} \times N_{\omega}$ to $N_\mathbf{q} (N_\omega N_X + N_\mathbf{k})$, and yields a significant computational advantage when $N_X <N_{\mathbf{k}}$, which is usually the case. This scheme enables us to undertake calculations on dense $\mathbf{q}$, $\mathbf{k}$, and $\omega$ meshes. 

Finally, the total electric susceptibility is determined by summing over all the orbital and spin degrees of freedom,
\begin{align}
\chi(\omega,\mathbf{q})&=\sum_{\substack{\nu \nu^{\prime} \mu \mu^{\prime} \\ a a^{\prime} b b^{\prime} } }\chi^{\nu^{\prime}b^{\prime}\nu b}_{\mu^{\prime} a^{\prime}\mu a}(\omega,\mathbf{q}).
\end{align}

The left panel in Fig.\,\ref{fig:Chi_q_omega} shows the electric susceptibility $\chi(\omega,\mathbf{q})$ as a function of momentum transfer along the cardinal directions in the Brillouin zone for a fixed frequency of $\omega=100$~meV. For small momentum transfers, the imaginary part of the electric susceptibility is quite weak owing to the dispersive nature of the electronic bands within an energy window of $\Delta E=100$~meV about the Fermi level. As the momentum transfer increases, the magnitude of the imaginary part of the electric susceptibility behaves nonmonotonically, reaching a maximum of Im$[-\chi]\sim 0.02$~keV$^2$ for $|\mathbf{q}|\simeq(0.1-0.3)$~keV. Above $|\mathbf{q}|\sim0.1$~keV the anisotropy between the various directions in the Brillouin zone becomes more pronounced. The magnitude of the real part of the electric susceptibility increases monotonically with $|\mathbf{q}|$ over the range of momenta in the Brillouin zone. 

The right panel in Fig.\,\ref{fig:Chi_q_omega} shows the electric susceptibility $\chi(\omega,\mathbf{q})$ as a function of $\omega$ for momentum transfers along the cardinal directions of the Brillouin zone with a fixed magnitude of
$|\mathbf{q}|=0.1$~keV. The spectrum exhibits peaks at $\omega\sim (130-160)$ meV for momentum transfers along the $\bm{y}$- and $\bm{z}$-axes, and a double-peak structure at $\omega\sim70$~meV and $\omega\sim120$~meV along the $\bm{x}$-axis. Further transitions appear for $\omega\gtrsim 1$~eV due to the relatively narrow bands at $(E_F-E)$\,$\simeq1$~eV in the electronic band structure. As a consequence of the Kramers-Kronig relations, the real part of the electric susceptibility resembles the derivative of the imaginary part.

\section{Dark Matter Interaction Rate}
\label{appx:DMrate}

In this appendix, we provide further details on the computation of the dark matter signal rate for the benchmark dark matter models discussed in Sec.\;\ref{sec:Sensitivity}. This content is based on existing literature (see, e.g., \cite{Geilhufe_2020,Kahn_2022,Knapen:2021run,Mayet2016Direction,Blanco:2021hlm,Boyd2023}).

For relic dark photons, $\gamma_\text{d}$, the event rate per target mass for $\gamma_\text{d}$-absorption by electrons in the material, assuming $\gamma_\text{d}$ is polarized in the $i$-direction, is given by \cite{Geilhufe_2020}:
\begin{equation}\label{absorption_ii}
R_{\gamma_{\text{d}},\,i}^\text{abs}~=~\varepsilon_\text{kin}^2 \frac{\rho_{\gamma_{\text{d}}}}{\rho_T}\, \text{Im}\bigg[ \frac{-1}{\epsilon_{ii}(\omega=m_{\gamma_{\text{d}}}, \vec{q}\approx0)}\bigg],
\end{equation}
where $\varepsilon_\text{kin}$ is the kinetic mixing parameter between the Standard Model photon and the dark photon; $\rho_{\gamma_{\text{d}}}=0.4$~GeV/cm$^3$ is the local dark matter relic density; $\rho_T$ is the density of the target material, which, for \Eu, corresponds to $\rho_T=6.77$ ~g/cm$^3$; and $\epsilon_{ii}$ is the $ii$-component of the dielectric tensor of \Eu~(see Fig.\;\ref{fig:EpsAbs}).

Since relic dark photons are expected to be unpolarized, the total rate per target detector mass can be approximated by the average over the three polarizations:
\begin{equation}
R_{\gamma_{\text{d}}}^\text{abs}~=~\frac{1}{3}\sum_{i=x,y,z}R_{\gamma_{\text{d}},\,i}^\text{abs}.\vspace{5pt}
\label{absorptionRate}
\end{equation}
Note that $\omega\gg|\vec{q\,}|$ in the kinematic regime for absorption, and therefore the $|\vec{q\,}|$-dependence of the dielectric tensor in \eqref{absorption_ii} and~\eqref{absorptionRate} is neglected. Hence, the dark photon absorption rate does not exhibit any significant directional modulation, despite the anisotropy of the electronic structure of \Eu.

In the case of relic dark matter $\chi$ scattering off of electrons in the material, the instantaneous event rate per target mass is given by~\cite{Geilhufe_2020,Knapen:2021run}:
\begin{widetext}
\begin{equation}\label{scatteringRate}
R_{\chi}(t)=\frac{\bar\sigma_e}{\rho_T}\,\frac{\rho_\chi}{m_\chi}\,\frac{\pi}{\mu_{e\chi}^2}\int d\omega\,\frac{d^3q}{(2\pi)^3} ~|\mathcal{F}_\chi(\vec{q\,})|^2\,\frac{~|\vec{q\,}|^2}{2\pi\alpha}\,\text{Im}\bigg[ \frac{-1~~}{\epsilon_L(\omega, \vec{q\,})}\bigg]\int d^3v~f_\chi\big(\vec{v}-\vec{v}_T(t)\big)\; \delta\bigg(\omega+\frac{|\vec{q\,}|^2}{2m_\chi}-\vec{q}\cdot\vec{v}\bigg),
\end{equation}
\end{widetext}
where, above, $\bar\sigma_e$ is the conventional dark matter-electron scattering cross section\footnote{In terms of the mass and couplings of the dark vector boson which mediates the interaction between dark matter and electrons, the conventional dark matter-electron scattering cross section is defined as
$\bar\sigma_e\equiv (1/\pi)\,g^2_e\,g^2_\text{d}\,\mu^2_{e\chi}/(\alpha^2m_e^2+m^2_{V_\text{d}})^2$, where $m_{V_d}$ is the mediator mass, and $g_\text{d}$ and $g_e$ are its couplings to dark matter and electrons, respectively.}; $m_\chi$ is the dark matter mass; $\mu_{e\chi}\equiv m_e m_\chi/{(m_e+m_\chi)}$ is the reduced mass of the dark matter-electron system; $\rho_T$ is the density of the target material; $\rho_\chi=0.4$ GeV/cm$^3$ is the local dark matter relic density;  $\alpha=1/137$ is the fine-structure constant; $\mathcal{F}_\chi(\vec{q\,})\equiv(\alpha^2m_e^2+m^2_{V_\text{d}})/(|\vec{q\,}|^2+m^2_{V_\text{d}})$ is the dark matter form factor in the scattering regime $|\vec{q}\,|\gg\omega$; $\epsilon_L(\omega,\vec{q\,})$ is the longitudinal component of the dielectric tensor; $f_\chi(\vec{v})$ describes the local dark matter velocity distribution in the Milky Way halo frame; and $\vec{v}_T(t)=v_\oplus\,\hat{v}_T(t)$ is the velocity of the detector target in this frame, with $v_\oplus\simeq235$\;km/s being the galactic Earth's speed.

\begin{figure}[t]
    \centering
    \includegraphics[width=1\linewidth]{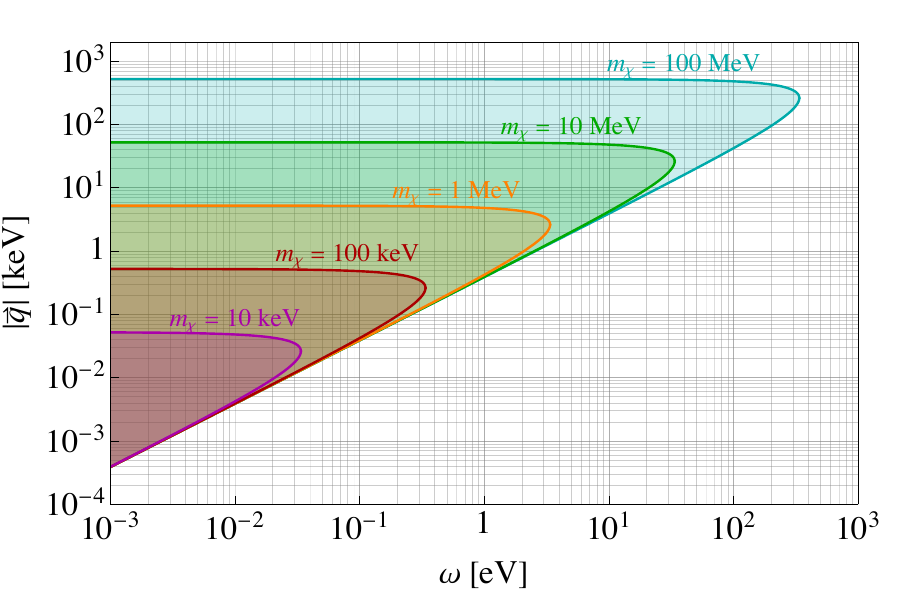}
    \caption{The kinematically allowed phase space for dark matter scattering (shaded regions) for a few illustrative choices of dark matter mass. These regions are determined by the escape velocity in the solar neighborhood through the constraint $v_\text{esc}+v_\oplus>v_\text{min}(\omega,|\vec{q\,}|)$, where $v_\text{min}(\omega,|\vec{q\,}|)$ has been defined in \eqref{vmin}.}
    \label{fig:PhaseSpace}
\end{figure}

Assuming that the Milky Way halo is well-described by the Standard Halo Model, the integration over the dark matter velocity distribution can be performed analytically, greatly simplifying the evaluation of \eqref{scatteringRate} \cite{Boyd2023}:
\begin{eqnarray}\label{DMvelocityIntegral}
&&\int d\vec{v}~f_\chi\big(\vec{v}-\vec{v}_T(t)\big)\;\delta\bigg(\omega+\frac{|\vec{q\,}|^2}{2m_\chi}-\vec{q}\cdot\vec{v}\bigg)~~~~~~~~~~~~~~~~~\nonumber\\
&&~~~~~~~~~~~~~~~~=~\frac{1}{|\vec{q\,}|}\frac{\pi v_0^2}{\mathcal{N_\chi}}\left(e^{-v_{-}^2/v_0^2}-e^{-v^2_\text{esc}/v_0^2}\right),
\end{eqnarray}
where $v_0\simeq220$\;km/s is the local dispersion velocity, $v_\text{esc}\simeq550$\;km/s is the local escape velocity,
\begin{equation}
\mathcal{N_\chi}=\big(\sqrt{\pi}\,v_0\big)^3\Bigg[\text{Erf}\bigg(\!\frac{v_\text{esc}}{v_0}\!\bigg)-\frac{2}{\sqrt{\pi}\,}\frac{v_\text{esc}}{v_0}\,\text{Exp}\bigg(\!\!\!-\frac{v^2_\text{esc}}{v^2_0}\!\bigg)\Bigg],\nonumber
\end{equation}
and
\begin{equation}
v^2_{-}~\equiv~\text{Min}\bigg[v^2_\text{esc},\;\Big(v_\text{min}(\omega,|\vec{q\,}|)-\hat{q}\cdot\vec{v}_T(t)\Big)^2\,\bigg],\nonumber
\end{equation}
where
\begin{equation}\label{vmin}
v_\text{min}(\omega,|\vec{q\,}|)~\equiv~\frac{\omega}{|\vec{q\,}|}+\frac{|\vec{q\,}|}{2m_\chi}.
\end{equation}
Note that \eqref{DMvelocityIntegral} vanishes when $v_\text{min}(\omega,|\vec{q\,}|)\geq v_\text{esc}+v_\oplus$. This sets a kinematic constraint on the phase space available for dark matter scattering, which is illustrated in Fig.\,\ref{fig:PhaseSpace} for a few choices of dark matter mass. The electronic structure of the target material will set additional constraints on the values of $\omega$ and $\vec{q}$ allowed in dark matter-induced electronic transitions.

\begin{figure}
    \centering
    \includegraphics[width=1\linewidth]{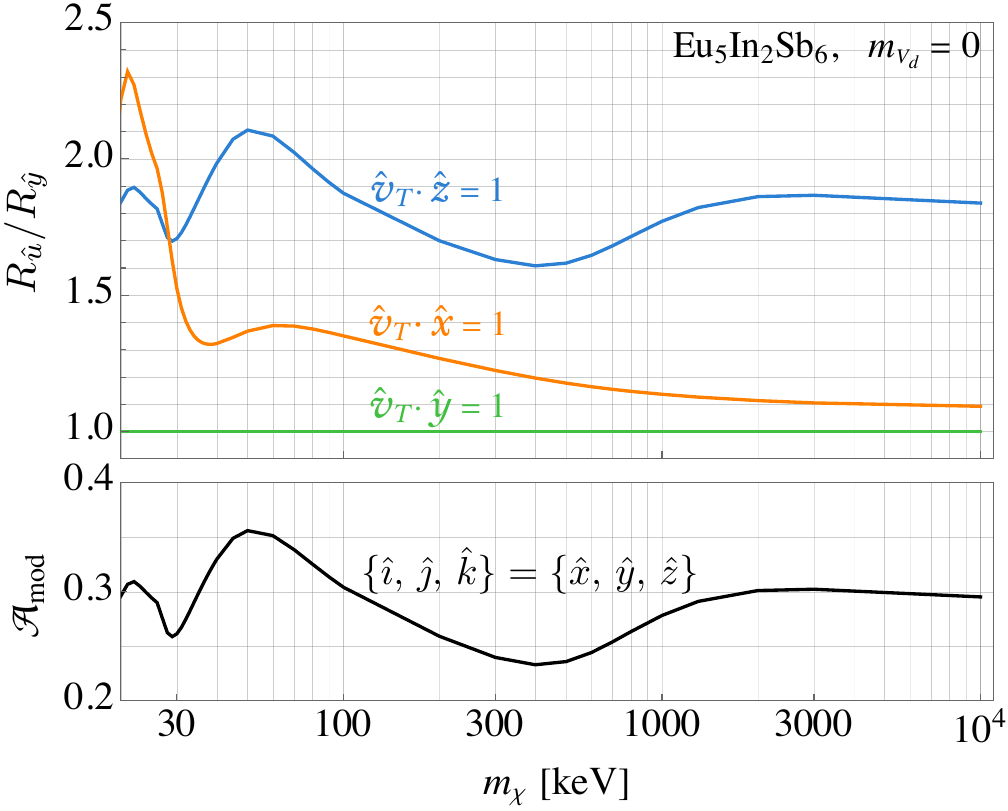}
    \caption{\emph{Top panel:} The instantaneous dark matter scattering rate in \Eu, $R_{\bm{\hat{u}}}$, as a function of dark matter mass, $m_\chi$, when the crystal's ${\bm{\hat{u}}}$-symmetry axis is aligned with the dark matter wind, i.e., when $\hat{v}_T\cdot{\bm{\hat{u}}}=1$. Each of the three curves corresponds to a particular choice of ${\bm{\hat{u}}}$-direction, and are normalized to the instantaneous rate when ${\bm{\hat{u}}}={\bm{\hat{y}}}$.
    The difference between them indicates the non-trivial directional dependence of the crystal's response to dark matter scattering.
    \emph{Bottom panel:} The fractional modulation amplitude of the signal rate, $\mathcal{A}_\text{mod}$, as a function of $m_\chi$, when the target reference frame $\{\bm{\hat{\imath}},\,\bm{\hat{\jmath}},\,\bm{\hat{k}}\}$ is identified with the crystal symmetry axes $\{\bm{\hat{x}},\,\bm{\hat{y}},\,\bm{\hat{z}}\}$. In both panels, an ultralight mediator is assumed.}
    \label{fig:RateAnisotropy}
\end{figure}

We can relate the longitudinal dielectric function $\epsilon_L(\omega,\vec{q\,})$ to our first-principles electronic structure calculation of the electric susceptibility $\chi(\omega,\vec{q\,})$ of \Eu{} in Appx.\,\ref{DFTtheory} via:
\begin{equation}
\epsilon_L(\omega,\vec{q\,})~=~1-\frac{4\pi}{|\vec{q\,}|^2}\,\chi(\omega,\vec{q\,}).
\end{equation}
These quantities exhibit a directional dependence on the momentum transfer $\vec{q\,}$ due to the anisotropic electronic structure of \Eu~(see, e.g., Fig.\;\ref{fig:Chi_q_omega}). In practice, this translates into a dependence of the scattering rate on the incoming direction of the dark matter particles, and, consequently, into a daily modulation of the scattering rate due to the Earth's rotation relative to the dark matter wind.
This effect is captured in \eqref{scatteringRate} through the time-dependent orientation of the target crystal in the Milky Way halo frame, $\hat{v}_T(t)$\footnote{A secondary daily modulation effect may be present if dark matter particles interact with the Earth before reaching the detector, leading to a distortion of the dark matter velocity distribution at the location of the detector. This distortion modulates over the course of a sidereal day, since the distance dark matter particles travel across the Earth before reaching the detector varies as the Earth rotates relative to the dark matter wind \cite{Collar:1992qc,Collar:1993ss,Hasenbalg:1997hs,Kouvaris:2014lpa,Kouvaris:2015xga,Emken:2019tni,Avalos:2021fxm,DAMIC-M:2023hgj,Bertou:2025adb}. In the sub-MeV mass range, this effect can be significant for dark matter-nucleon scattering cross sections $\gtrsim 10^{-35}$\;cm$^2$, and is expected to exacerbate the modulation stemming from the target's anisotropic response. In our studies in Sec.\;\ref{sec:Sensitivity}, we have ignored dark matter propagation through the Earth leading to this secondary modulation effect, and therefore the sensitivity reach estimates we provided are expected to be conservative.}.
Adopting the convention proposed in \cite{Griffin:2018bjn,Coskuner:2019odd}, we have:
\begin{eqnarray}\label{vcrystal}
\vec{v}_T(t)&=&v_\oplus\,\hat{v}_T(t)\\
&=&v_\oplus\Big[\,\text{sin}\,\theta_\oplus\,\text{sin}\phi(t)\,\bm{\hat{\imath}}\nonumber\\
&&~~~~~~~~+\,\text{sin}\theta_\oplus\,\text{cos}\theta_\oplus\,(\text{cos}\phi(t)-1)\,\bm{\hat{\jmath}}\nonumber\\
&&~~~~~~~~+\,\big(1+\text{sin}^2\theta_\oplus\,(\text{cos}\phi(t)-1)\big)\,\bm{\hat{k}}\,\Big],\nonumber
\end{eqnarray}
where $t$ is sidereal time, with $\phi(t)=2\pi (t/\text{23.9345 h})$, and $\theta_\oplus=42\degree$ is the angle between the Earth's rotation axis and the direction of the dark matter wind (i.e., the angle between the North Celestial Pole and the Cygnus constellation). Note that \eqref{vcrystal} neglects the motion of the Earth around the Sun by approximating $v_\oplus$ to a constant.  The directions $\bm{\hat{\imath}}$, $\bm{\hat{\jmath}}$, and $\bm{\hat{k}}$ define a Cartesian reference frame associated with the target crystal---by convention, the crystal will be oriented such that, at $t=0$, $\bm{\hat{k}}$ will point towards the Cygnus constellation
(and therefore will be aligned with the direction of the dark matter wind),
and $\bm{\hat{\jmath}}$ will lie in the plane defined by $\bm{\hat{k}}$ and the Earth's rotation axis, with an angle of $\pi/2+\theta_\oplus$ from the North Celestial Pole. This convention does not specify the orientation of the $\{\bm{\hat{\imath}},\,\bm{\hat{\jmath}},\,\bm{\hat{k}}\}$ frame relative to the symmetry axes of the crystal;
preferably, one would like to determine this relative orientation by maximizing the daily modulation of the signal rate. 
In order to make this determination, in the top panel of Fig.\;\ref{fig:RateAnisotropy} we show the instantaneous rate $R_{\bm{\hat{u}}}$ when the ${\bm{\hat{u}}}$-symmetry axis of the crystal is aligned with the dark matter wind for the three choices of ${\bm{\hat{u}}}$, namely, $\bm{\hat{x}}$, $\bm{\hat{y}}$, and $\bm{\hat{z}}$ (adopting the axis-labeling convention of Appx.\,\ref{DFTtheory}). All curves are normalized to the instantaneous rate $R_{\bm{\hat{y}}}$. It is clear from Fig.\,\ref{fig:RateAnisotropy} that the directional dependence of the signal rate varies as a function of dark matter mass, and, therefore, so will the optimal crystal orientation.

Combined with our chosen convention for the target orientation in \eqref{vcrystal}, the curves in Fig.\,\ref{fig:RateAnisotropy} indicate that the modulation amplitude of the signal rate will be maximized by the following choices:
\begin{equation}
\{\bm{\hat{\imath}},\,\bm{\hat{\jmath}},\,\bm{\hat{k}}\}=\begin{cases}
 \{\bm{\hat{z}},\,\bm{\hat{y}},\,\bm{\hat{x}}\}&\text{for}~~~m_\chi~\lesssim~29~\text{keV,}\vspace{4pt}\\
 \{\bm{\hat{x}},\,\bm{\hat{y}},\,\bm{\hat{z}}\}&\text{for}~~~m_\chi~\gtrsim~29~\text{keV.}
\end{cases}
\end{equation}
While these choices have been informed by the \emph{total} instantaneous rate at different crystal orientations, we note that they remain optimal even when the rate's energy spectrum is taken into account.

In the freeze-in scenario with an ultralight mediator, the mass range $m_\chi\lesssim30$\;keV is excluded by bounds on stellar cooling~\cite{Vogel:2013raa} (see gray-shaded region in Fig.\,\ref{ReachPlot}). Hence, 
for simplicity, in the sensitivity reach estimates discussed in Sec.\;\ref{sec:Sensitivity}, the target reference frame $\{\bm{\hat{\imath}},\,\bm{\hat{\jmath}},\,\bm{\hat{k}}\}$ was identified with the
$\{\bm{\hat{x}},\,\bm{\hat{y}},\,\bm{\hat{z}}\}$ symmetry axes of \Eu~for the entire mass range shown in Fig.\,\ref{ReachPlot}.
Given this choice, we can obtain the fractional modulation amplitude of the signal rate, $\mathcal{A}_\text{mod}$, defined by
\begin{equation}
\mathcal{A}_\text{mod}~\equiv~\frac{1}{2}\,\frac{R_\chi(t_\text{max})-R_\chi(t_\text{min})}{\overline{R}_\chi},\nonumber
\end{equation}
where the instantaneous scattering rate $R_\chi(t)$ is given in \eqref{scatteringRate}, $t_\text{max}$ ($t_\text{min}$) is the time of day at which the instantaneous rate reaches its maximum (minimum), and
$\overline{R}_\chi$
is the sidereal daily average of 
$R_\chi(t)$. 
We note that, due to the orthorhombic crystalline structure of \Eu{}, the modulation of $R_\chi(t)$ is well described by a sinusoidal curve.

In the bottom panel of Fig.\,\ref{fig:RateAnisotropy}, we show the fractional modulation amplitude of the signal rate, $\mathcal{A}_\text{mod}$, as a function of dark matter mass. In the viable mass range $m_\chi\simeq30\;\text{keV}-3\;\text{MeV}$, the total rate displays a modulation amplitude in the range of $\sim$$23-36\%$, showcasing the suitability of \Eu~as a detector target for daily modulation-based searches, which is the key SPLENDOR feature enabling the low-threshold analyses described in Sec.\;\ref{sec:Sensitivity}.


\end{document}